\documentclass[aps,prd,twocolumn,nofootinbib,superscriptaddress,floatfix]{revtex4-1}

\usepackage{amsmath,amssymb,mathtools}
\usepackage{booktabs}
\usepackage{array}
\usepackage{graphicx}
\usepackage{xcolor}
\usepackage{microtype}
\usepackage[pass]{geometry}
\usepackage{etoolbox}
\usepackage[colorlinks=true,linkcolor=blue,citecolor=blue,urlcolor=blue]{hyperref}
\usepackage[nameinlink]{cleveref}
\hypersetup{
  pdftitle={Bootstrapping "black holes" at low impact parameter},
  pdfauthor={Diptarka Das and Aninda Sinha}
}
\crefname{section}{Sec.}{Secs.}\Crefname{section}{Section}{Sections}
\crefname{subsection}{Sec.}{Secs.}\Crefname{subsection}{Section}{Sections}
\crefname{appendix}{Appendix}{Appendices}\Crefname{appendix}{Appendix}{Appendices}
\crefname{figure}{Fig.}{Figs.}\Crefname{figure}{Figure}{Figures}
\crefname{table}{Table}{Tables}\Crefname{table}{Table}{Tables}
\crefname{equation}{Eq.}{Eqs.}\Crefname{equation}{Equation}{Equations}

\newcommand{\dd}{\mathrm d}
\newcommand{\lam}{\lambda}
\newcommand{\sig}{\sigma}
\newcommand{\ME}{M_{\rm EFT}}
\newcommand{\MP}{M_{\rm Pl}}
\newcommand{\GN}{\widehat G_N}
\newcommand{\rphys}{\rho^{\rm phys}_{\rm res}}
\newcommand{\rtot}{\rho^{\rm phys}_{\rm tot}}
\newcommand{\reik}{\rho^{\rm phys}_{\rm eik}}
\newcommand{\rphysi}{\rho^{\rm phys}_{{\rm res},i}}
\newcommand{\rtoti}{\rho^{\rm phys}_{{\rm tot},i}}
\newcommand{\reiki}{\rho^{\rm phys}_{{\rm eik},i}}
\newcommand{\Ktwo}{\mathcal K_2}
\newcommand{\cE}{\mathcal E}

\makeatletter
\let\arxiv@sectionhang\@hangfrom@section
\let\arxiv@sectionformat\@sectioncntformat

\newcommand{\arxivlastbiblabel}{LastBibItem}
\patchcmd{\NAT@bibsetnum}{\ref{LastBibItem}}{\ref{\arxivlastbiblabel}}
  {}{\PackageError{arxiv-manuscript}{Cannot set reference list width}{}}
\patchcmd{\endthebibliography}{\label{LastBibItem}}{\label{\arxivlastbiblabel}}
  {}{\PackageError{arxiv-manuscript}{Cannot label the reference list}{}}

\newcommand{\startarxivsupplement}{%
  \clearpage
  \newgeometry{top=0.65in,bottom=0.66in,left=0.62in,right=0.62in}%
  \setlength{\columnsep}{0.25in}%
  \setlength{\textfloatsep}{13pt plus 2pt minus 3pt}%
  \setlength{\dbltextfloatsep}{13pt plus 2pt minus 3pt}%
  \let\@hangfrom@section\arxiv@sectionhang
  \let\@sectioncntformat\arxiv@sectionformat
  \renewcommand{\arxivlastbiblabel}{SupplementLastBibItem}%
  \@removefromreset{equation}{section}%
  \setcounter{section}{0}%
  \setcounter{subsection}{0}%
  \setcounter{subsubsection}{0}%
  \setcounter{equation}{0}%
  \setcounter{figure}{0}%
  \setcounter{table}{0}%
  \setcounter{footnote}{0}%
  \renewcommand{\thesection}{S\arabic{section}}%
  \renewcommand{\thesubsection}{\Alph{subsection}}%
  \renewcommand{\thesubsubsection}{\arabic{subsubsection}}%
  \renewcommand{\theequation}{S\arabic{equation}}%
  \renewcommand{\thefigure}{S\arabic{figure}}%
  \renewcommand{\thetable}{S\arabic{table}}%
  \renewcommand{\theHsection}{supplement.\arabic{section}}%
  \renewcommand{\theHequation}{supplement.\arabic{equation}}%
  \renewcommand{\theHfigure}{supplement.\arabic{figure}}%
  \renewcommand{\theHtable}{supplement.\arabic{table}}%
  \renewcommand{\theHfootnote}{supplement.\arabic{footnote}}%
  \crefalias{appendix}{section}%
  \crefalias{subappendix}{subsection}%
  \crefalias{subsubappendix}{subsubsection}%
  \onecolumngrid
  \phantomsection
  \pdfbookmark[0]{Supplemental Material}{arxiv.supplement}%
  {\centering
   \large\bfseries Supplemental Material for ``Bootstrapping `black holes' at low impact parameter''\par
   \vspace{0.7em}%
   \normalfont\normalsize Diptarka Das\textsuperscript{1} and
   Aninda Sinha\textsuperscript{2,3}\par
   \vspace{0.3em}%
   {\small\itshape
   \textsuperscript{1}Department of Physics, Indian Institute of Technology Kanpur, Kanpur 208016, India\par
   \textsuperscript{2}Centre for High Energy Physics, Indian Institute of Science, C. V. Raman Avenue, Bangalore 560012, India\par
   \textsuperscript{3}Quantum Horizons Alberta, PHAS, University of Calgary, Canada\par}%
   \vspace{0.4em}%
   {\small(Dated: \today)\par}%
   \vspace{1.4em}}%
  \twocolumngrid
}
\makeatother

\begin{document}
\title{Bootstrapping ``black holes'' at low impact parameter}
	
	\author{Diptarka Das}
	\affiliation{Department of Physics, Indian Institute of Technology Kanpur, Kanpur 208016, India}
	\author{Aninda Sinha}
	\affiliation{Centre for High Energy Physics, Indian Institute of Science, C. V. Raman Avenue, Bangalore 560012, India}
	\affiliation{Quantum Horizons Alberta, PHAS, University of Calgary, Canada}
	
	\date{\today}
	
	\begin{abstract}
We use the stringy dispersion relation to obtain a numerical upper bound on
the radius of a uniformly absorptive disk in a six-dimensional gravitational
effective field theory. Near the boundary of the allowed region of low-energy
couplings, the bound is of order the Schwarzschild radius. The dispersion
relation connects those couplings to the partial-wave spectrum. In the
calculation, we fix the eikonal contribution in a specified high-energy
region at large impact parameter and vary the real parts of the remaining
elastic \(S\)-matrix elements within the bounds imposed by unitarity.
The real part of each element sets an upper limit on its absorption
probability, so the radius bound applies to any unitary amplitude satisfying
the same assumptions. Near the boundary, a strongly absorptive disk of the
classical trapped-surface radius is excluded if it must extend down to low
collision energies. Raising the minimum energy lifts this exclusion.
The angular momentum at the edge of the absorptive core grows approximately
as \(s^{0.7}\), where \(s\) is the squared center-of-mass energy. This is close
to the \(s^{2/3}\) scaling expected for an edge at a fixed multiple of the
Schwarzschild radius in six dimensions. The corresponding dimension-dependent scaling also
describes full-amplitude bootstrap results in nine to eleven dimensions.
 In the extremal spectra, the band's outer edge stays at roughly the same impact parameter as the
collision energy increases, about \(5\)--\(6\) EFT cutoff lengths
\(M_{\rm EFT}^{-1}\) at weak gravity.
\end{abstract}
	
	\maketitle
	
	\section{Introduction}
	\label{sec:introduction}
	
	Causality, analyticity, unitarity and crossing symmetry constrain scattering
	amplitudes far more than one might expect, and dispersion relations are the
	tool that turns these principles into equations.  In a massive effective
	field theory (EFT) the coefficients of the low-energy expansion must come
	from positive high-energy spectral
	data~\cite{Mandelstam,EdenBook,Froissart,
		MartinAnalyticity,LukaszukMartin,RoyEquations,
		AdamsArkaniHamedDubovskyNicolisRattazzi,deRhamMelvilleTolleyZhou,
		BellazziniSoftness,CheungRemmen,deRhamSpin,BellazziniMiroRattazziRiembauRiva,
		ArkaniHamedHuangHuangEFThedron,
		ChiangHuangRodinaWengDeprojecting,
		ChiangHuangLiRodinaWengNonProjective,
		CaronHuotVanDuong,
		CaronHuotMazacRastelliSimmonsDuffin,EliasMiroGuerrieriGumus}.  The
	numerical \(S\)-matrix bootstrap searches for such spectra directly, together
	with the dual functionals that certify the boundary of the allowed region of
	coefficients~\cite{PaulosPenedonesToledoVanReesVieira,
		GuerrieriPenedonesVieira,EliasMiroGuerrieriGumus,deRhamTolleyWangZhou,
		KruczenskiPenedonesVanReesWhitePaper}.
	
	With gravity the question becomes harder and more interesting.  The forward
	amplitude is dominated by the massless graviton pole, the low-energy imprint
	of the long-range force.  At high energy and large impact parameter the pole
	is carried by the eikonal \(S\)-matrix, built from many soft graviton
	exchanges
	\cite{Weinberg, Weinberg2, tHooftGravitonDominance,KabatOrtiz,ACV,
		ACVCollapse,GrossMendeHighEnergy,
		GrossMendeStringBeyond,DAppollonioDiVecchiaRussoVeneziano,
		DiVecchiaHeissenbergRussoVenezianoReview,GiddingsPorto}.  At impact
	parameters of order the Schwarzschild radius \(R_S\) one expects strong
	gravity: large time delays, black-disk absorption, black-hole formation and
	resonances, or more coherent reflective dynamics of the fuzzball
	type~\cite{EardleyGiddingsBH,GiddingsRychkov,YoshinoNambu,
		YoshinoRychkov,GiddingsSrednicki,MathurFuzzballReview}.  Classical
	trapped-surface constructions fix the scale.  In six dimensions the
	Yoshino--Rychkov construction finds an apparent horizon for \(b<1.47R_S\)
	in the Schwarzschild--Tangherlini convention used below~\cite{YoshinoRychkov}.
	String theory adds the expected correspondence between
	highly excited strings and black
	holes~\cite{HorowitzPolchinskiCorrespondence, CeplakEmparanPuhmTomasevic,
		HorowitzPolchinskiSelfGravitating,DamourVenezianoSelfGravitatingStrings,
		SenExposeBlackHole,CMW}.  These scenarios leave different patterns in
	partial-wave space.  Can a dispersive bootstrap see them?
	
\begin{figure*}[t]
		\centering
		\includegraphics[width=0.60\textwidth]{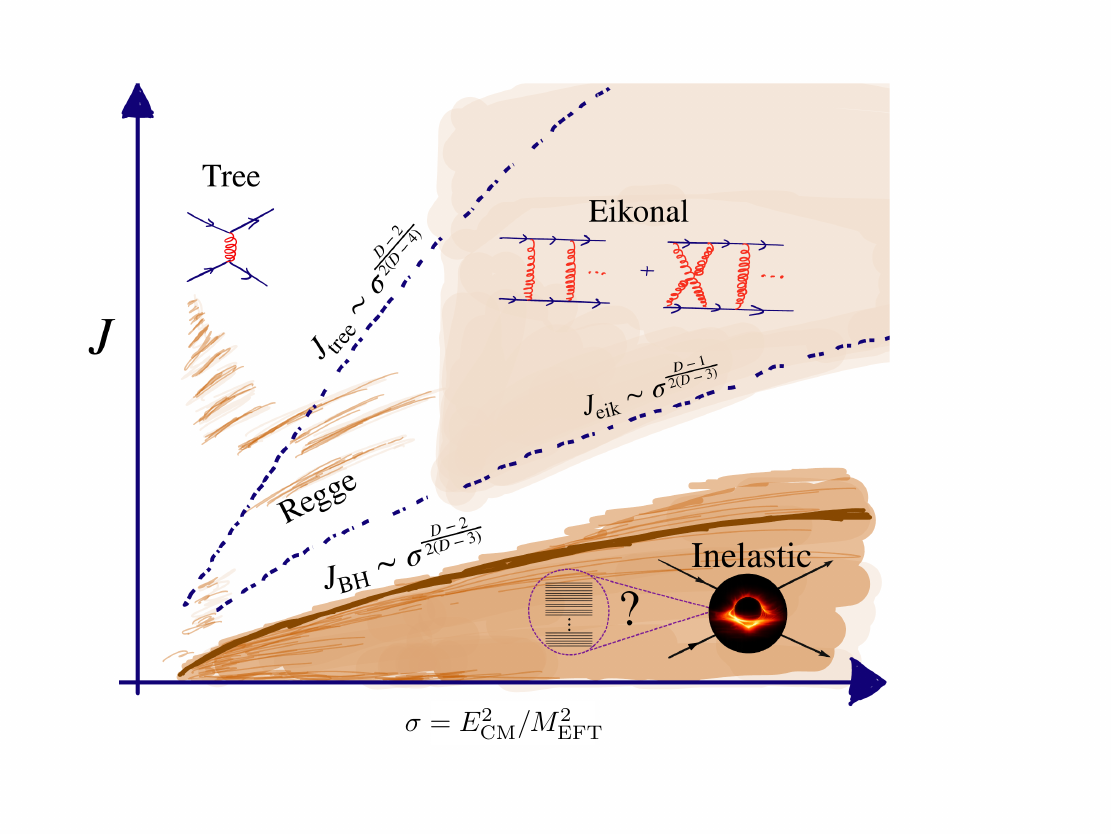}
		\caption{The regimes one expects in \(2\to2\) gravitational scattering for
			\(D>4\), drawn in the plane of energy and spin; this is a map of the
			question, not a result.  The eikonal region (light brown) is prescribed in
			our calculation.  The bootstrap is free to place spectral weight anywhere
			below it.  What it does place there differs from the cartoon in one
			important way: the low-impact band it selects is reflective and sits at a
			fixed impact parameter (\(J\propto\sigma^{1/2}\)), while the part that
			scales like the black-hole curve \(J_{\rm BH}\propto\sigma^{2/3}\) is the
			absorptive core bounded in \cref{sec:absorptive-density-pilot}.}
		\label{fig:schematics}
	\end{figure*}

	A great deal is known about gravitational EFT bounds from dispersion
	relations at fixed momentum transfer, from impact-parameter causality and
	from related sum rules~\cite{CamanhoEdelsteinMaldacenaZhiboedov,
		AlberteDeRhamJaitlyTolleyMasslessSpin2,
		CaronHuotLiParraMartinezSimmonsDuffin,
		CaronHuotLiParraMartinezSimmonsDuffinHigherD,
		HenrikssonMcPeakRussoVichiWGC,ChangParraMartinez, PengRodinaTokarevaXu}.
	A bound on a Wilson coefficient says that some positive spectrum must exist.
	It does not say whether that spectrum lives at large impact parameter, at
	small impact parameter, on a Regge trajectory, or in some mixture of these.
	H\"aring and Zhiboedov answered the question of what the graviton pole is
	``made of'' in an averaged sense: it can be carried by high-energy spectral
	weight at large impact parameter \cite{HaringZhiboedov}.  We ask the
	complementary question.  We supply that eikonal contribution explicitly (the light
	brown region of \cref{fig:schematics}) and ask where the \emph{remaining}
	positive spectrum goes, and what it does there.  \Cref{fig:schematics}
	sketches the regimes one expects in the plane of energy and spin, marks the
	part we prescribe, and summarizes in its caption how our answer differs from
	the expectation.
	
	Our tool is the stringy dispersion relation (SDR) of
	Ref.~\cite{Bhat:2025zex}.  It is crossing symmetric and has an auxiliary
	parameter \(\lambda\). The name SDR records
	its origin in representations of string
	amplitudes~\cite{SahaSinhaFieldTheoryExpansions,WanZhouVeneziano,
		EcknerFigueroaTourkineDualRegge,EcknerFigueroaMetayerTourkine,
		EcknerFigueroaTourkineReggeBootstrap, HaringZhiboedovStringyBootstrap,
		BhatChowdhurySahaSinhaStringBootstrap, CheungHillmanRemmenVirasoroShapiro};
	nothing in it assumes that the completion is a perturbative string theory.
	It is a field-theoretic dispersion relation under stated assumptions on
	analyticity and Regge growth, and a close relative of the fixed-\(a\)
	crossing-symmetric relations and of the three-channel representations of
	Auberson--Khuri and Mahoux--Roy--Wanders
	\cite{ChangParraMartinez,PengRodinaTokarevaXu, AubersonKhuri,MahouxRoyWanders,
		SinhaZahed,EliasMiroGuerrieriGumusZahed}.  We work in \(D=6\), which is free
	of the infrared problems of four dimensions, with identical scalars that
	exchange a graviton.  The SDR becomes a linear program (LP) for the residual
	partial-wave densities
	\[
	\rho_J=1-\operatorname{Re}S_J,
	\qquad
	0\le \rho_J\le2,
	\]
	on a grid in energy and spin, with the eikonal density of Einstein
	gravity prescribed at large impact parameter.  In this variable
	\(\rho_J=1\) is required for a black disk and \(\rho_J=2\) is the perfectly reflecting
	point \(S_J=-1\).  The density tells us where the spectrum is.  It cannot
	tell us by itself the precise absorptive or reflective values of scattering, but can give useful bounds.
	The quotation marks in our title are a reminder of this impreciseness\\
	.
	
	We find three things.
	\begin{enumerate}
		\item \emph{The residual spectrum is organized.}
		The extremal spectra occupy only part of the available region. The upper-boundary extremal solutions exhibit a cap-saturated band at low
		impact parameter (\cref{sec:capped,sec:weak-gravity}). For
		\(M_{\rm EFT}>M_{\rm Pl}\), the examples also show a separate high-spin
		ridge and leave most of the region between the band and the prescribed
		eikonal region empty. The reduced costs in these examples show that
		populating the gap carries a large cost and that the cap on the band
		is an active constraint (\cref{sec:dual-slack}).
		
		\item \emph{The band is not gravitational.}  Its edge sits at a fixed impact
		parameter, \(J_{\rm edge}\propto\sigma^{0.5}\), at every coupling.  In the
		physical ordering \(M_{\rm Pl}>M_{\rm EFT}\) the edge stays at
		\((5\text{--}6)M_{\rm EFT}^{-1}\) while the gravitational radius shrinks,
		the same edge and support appear at \(G_N=0\), and changing the sampled
		range of \(\lambda\) by a factor of three does not move it
		(\cref{sec:weak-gravity}).  The reflective band is a property of the capped
		dispersive problem at the cutoff scale.
		
		\item \emph{The absorptive core is gravitational, and the bootstrap bounds
			its size.}  The density-only problem constrains only
		\(\operatorname{Re}S_J\) and is therefore an outer relaxation of
		partial-wave unitarity.  The largest disk over which a given absorption can
		be sustained near the extremal boundary then bounds the absorptive radius of
		any unitary completion of these spectra from above.  The bound is of order
		\(R_S\).  The Yoshino--Rychkov trapped-surface radius gives a classical
		benchmark for comparison.  Asking for an absorptive disk of that size
		lowers the largest value of the Wilson coefficient \(g_3\) that can be
		reached at fixed \(g_2\).  If at most half a percent of it may be
		given up, a disk of the classical radius \(1.47R_S\) that absorbs with
		probability \(0.9\) is excluded unless it begins above
		\(ER_S\simeq30\)--\(50\).  Above that energy the density constraints admit
		the disk, and the amount of \(g_3\) it requires decreases as a power of the
		energy.  In units of the graviton pole, and compared at the same relative
		position on the boundary of the allowed region, this amount is
		approximately the same function of \(ER_S\) at three couplings
		(\cref{sec:ceiling-floor}).  The edge of
		the absorptive core grows as \(\sigma^{0.7}\), near the gravitational value
		\(2/3\), whatever the shape of the region on which absorption is requested.
		At weak coupling the bound in Schwarzschild units scales approximately as \((\ME/\MP)^{-0.45}\)
		(\cref{sec:absorptive-density-pilot}).  This is our main result.
	\end{enumerate}
	The third result above, connects to a string-inspired full-amplitude graviton bootstrap in
	\(D=9\)--\(11\) \cite{DasMyersSinhaBHScales}, where \(\operatorname{Im}S_J\)
	is available.  There the landmarks of the absorption map scale with energy
	with exponents close to \((D-2)/[2(D-3)]\); the same formula gives \(2/3\) in
	six dimensions.  That calculation also supports the eikonal trust region we
	use, and it shows absorptive and reflective layers side by side inside the
	capture radius (\cref{sec:d10-sky-physical-absorption}).
	
	Our bounds and allowed regions refer to a specified energy and spin grid and
	a specified set of \(\lambda\) values. The calculation prescribes the
	large-impact-parameter eikonal contribution and optimizes the remaining
	density above \(s=\ME^2\). Below that threshold the massless discontinuity
	belongs on the low-energy side of the dispersion relation. We have verified using inclusion of one loop effects that our weak coupling results are robust. Tests at additional \(\lambda\) values are reported
	in Supplemental Sec.~\ref{S-sec:nlambda-drift}.

	\Cref{sec:equation} sets up the dispersion relation, the eikonal contribution and the
	linear program.  \Cref{sec:capped} describes the extremal spectra when
	\(M_{\rm EFT}>M_{\rm Pl}\), and \cref{sec:weak-gravity} the weak-gravity
	baseline.  \Cref{sec:absorptive-density-pilot} contains the bound on the
	absorptive radius and the comparison with \(D=9\)--\(11\).  We end with open
	problems in \cref{sec:discussions}.  Derivations, numerical checks and the
	reproducibility record are in the Supplemental Material, and the code is
	supplied with the paper.

\subsection*{Note added} In versions 1 and 2 of this preprint, we incorrectly attributed the energy dependence of the outer reflective band to gravity. We now find that its outer edge lies at a fixed impact parameter. When we select spectra that favour strong absorption, an inner core emerges whose radius grows with energy approximately as the Schwarzschild radius does.
	
	\section{Dispersion relation, eikonal contribution and optimization}
	\label{sec:equation}
	
	\subsection{The SDR and the coefficient sum rule}
	
	We consider identical massless scalars in \(D=6\) and use the
	crossing-symmetric variables \(x=-(st+su+tu)\) and \(y=-stu\).  Below the
	scale \(M_{\rm EFT}\) the amplitude is
	\begin{equation}
		\begin{split}
			M_{\rm low}
			&=8\pi G_N\left(\frac{st}{u}+\frac{su}{t}+\frac{tu}{s}\right)\\
			&\quad+g_2\,(s^2+t^2+u^2)+g_3\,stu+\cdots\\
			&=-\frac{8\pi G_N\,x^2}{y}+2g_2\,x-g_3\,y+\cdots .
		\end{split}
		\label{eq:eft-expansion}
	\end{equation}
	with \(8\pi G_N=M_{\rm Pl}^{-4}\).  The coefficients \(g_2\) and \(g_3\) are
	normalized as in
	Refs.~\cite{CaronHuotVanDuong,CaronHuotMazacRastelliSimmonsDuffin}; in the
	variables \(x\) and \(y\) this brings a factor of two and a sign.  The dimensionless dispersive energy is \(\sigma=s/\ME^2\).
	In the dispersion kernels below, the arguments written \(x,y,t\) stand
	for \(x_{\rm phys}/\ME^4,y_{\rm phys}/\ME^6,t_{\rm phys}/\ME^2\),
	respectively. The dimensionless \(\lambda\) corresponds to
	momentum transfer \(-\lambda\ME^2\), and the amplitude in those kernels
	is \(\ME^2 M_{\rm phys}\). We retain dimensional \(g_2,g_3,G_N\) and
	define the normalized coefficients
	\begin{equation}
		X=\frac{g_2\ME^2}{8\pi G_N},
		\qquad
		Y=\frac{g_3\ME^4}{8\pi G_N}.
		\label{eq:xy-def}
	\end{equation}
	The strength of gravity in these units is
	\(\GN\equiv G_N\ME^4=(8\pi)^{-1}(\ME/\MP)^4\).  Changing it moves
	the black-hole scale across the spectral grid, so it is a parameter of the
	problem and not a choice of units.
	
	The dispersion relation is the member of the SDR family of
	Ref.~\cite{Bhat:2025zex} with the fewest subtractions that still gives sum
	rules for \(g_2\) and \(g_3\) (higher subtractions are discussed there and in
	Ref.~\cite{AthiraSahaSarenSinha}).  It holds for amplitudes that grow more
	slowly than \(s^2\) at fixed physical momentum transfer \(t_{\rm phys}<0\),
	\begin{equation}
		M_{\rm phys}(s,t_{\rm phys})=O(s^{2-\epsilon}),
		\qquad \epsilon>0 ,
		\label{eq:regge-growth-class}
	\end{equation}
	which includes Regge-behaved string amplitudes.  The gravitational eikonal
	grows faster in the forward direction, but the condition is never imposed at
	\(t=0\), so there is no conflict.  The relation contains an auxiliary
	parameter \(\lambda\).  The amplitude does not depend on it; the kernel does.
	Demanding that every \(\lambda\) reproduce the same low-energy coefficients is
	what generates our constraints.  The exact SDR holds when \(-\lambda\) lies in
	the scaled fixed-\(t\) analyticity domain of the amplitude~\cite{Bhat:2025zex}.  We
	sample \(0<\lambda<1/3\), the interval on which fixed-\(a\) representations
	are normally used (\(\lambda=-a\))
	\cite{ChangParraMartinez,PengRodinaTokarevaXu}.  Larger \(\lambda\) is not
	excluded on physical grounds, but the Gegenbauer polynomials then grow
	rapidly and the linear programs become ill conditioned; constraints from a
	larger valid range could only shrink the regions we find.
	
	The \(s\)-channel absorptive part is expanded in partial waves,
	\begin{equation}
		\mathcal A^{(s)}_\rho(\sig,t)
		=
		\frac1{\sig}
		\sum_{J\in2\mathbb Z_{\ge0}}
		n_J^{(6)}\rho_J(\sig)\,
		\frac{C_J^{(3/2)}
			\left(1+\frac{2t}{\sig}\right)}{C_J^{(3/2)}(1)}.
		\label{eq:absorptive-amplitude}
	\end{equation}
	with \(C_J^{(3/2)}\) a Gegenbauer polynomial, even \(J\) only, and
	\(n_J^{(6)}>0\) a normalization given in the Supplemental Material
	(Sec.~\ref{S-app:eikonal-background}).  The SDR evaluates it at a momentum
	transfer \(t_\lambda(x,y;\sigma)=s_2-\lambda\), where
	\begin{equation}
		s_2^\pm(\sig;x,y)
		=
		-\frac{\sig-2\lam}{2}
		\left[
		1\pm
		\sqrt{
			1+
			\frac{4(y+\lam x-\lam^3)}{(\sig+\lam)(\sig-2\lam)^2}
		}
		\right],
		\label{eq:s2pm-main}
	\end{equation}
	and \(z_\lambda=1+2t_\lambda/\sigma\) is the corresponding scattering angle.
	The two roots are related by \(z\to-z\) and give the same relation for even
	spins.  We use the near-forward one, for which \(t_\lambda\to-\lambda\) at
	large \(\sigma\).  This is what makes the SDR suited to gravity.  At increasing energy the physical momentum transfer approaches
	\(-\lambda\ME^2\), where the eikonal description applies. The full
	relation still samples an energy-dependent transfer and maintains crossing
	symmetry.
	
	With the kernel
	\begin{equation}
		K_0(\sig;x,y;\lam)
		=
		\frac{\bigl(x-3\sig^2\bigr)}{\bigl(y-x\sig+\sig^3\bigr)}
		+\frac{1}{\sig+\lam},
		\label{eq:pq0-kernel-main}
	\end{equation}
	the SDR reads
	\begin{widetext}
		\begin{align}
			M(x,y)
			&=
			W_{00}(3\lam^2,-2\lam^3)
			-\frac1\pi
			\int_0^\infty\dd\sig\,
			K_0(\sig;x,y;\lam)\,
			\mathcal A^{(s)}
			\bigl(\sig,t_\lam(x,y;\sig)\bigr)
			\nonumber\\
			&\hspace{15mm}
			+\frac1\pi
			\int_0^\infty\dd\sig\,
			K_0(\sig;3\lam^2,-2\lam^3;\lam)\,
			\mathcal A^{(s)}
			\bigl(\sig,t_\lam(3\lam^2,-2\lam^3;\sig)\bigr).
			\label{eq:pq0-full-sdr-main}
		\end{align}
	\end{widetext}
	Here \(W_{00}\) is the subtraction constant at the crossing-symmetric point.
	The first and last terms do not depend on \(x\) and \(y\), so only the middle
	term contributes to the sum rules for the Wilson coefficients.  Its part
	above the first massive threshold is
	\begin{equation}
		\mathfrak M_\rho(x,y;\lam)
		=
		-\frac1\pi
		\int_1^\infty\dd\sig\,
		K_0(\sig;x,y;\lam)\,
		\mathcal A^{(s)}_\rho
		\bigl(\sig,t_\lam\bigr),
		\label{eq:pq0-sdr-main}
	\end{equation}
	and the contribution of the massless cut below \(\sigma=1\) is left on the
	low-energy side (we come back to it in \cref{sec:discussions}).  Projecting
	\cref{eq:pq0-sdr-main} onto the coefficients of \(x\) and \(y\) gives two
	one-parameter families of sum rules with the pole structure
	\begin{equation}
		\begin{split}
			\mathcal C_x[\rho](\lam)&=\frac{16\pi \GN}{\lam}+2g_2\ME^6+\cdots,\\
			\mathcal C_y[\rho](\lam)&=\frac{8\pi \GN}{\lam^2}-g_3\ME^8+\cdots .
		\end{split}
		\label{eq:split-rows-main}
	\end{equation}
	The \(1/\lambda\) and \(1/\lambda^2\) terms are the image of the graviton
	pole.  In the combination
	\(\Ktwo=\mathcal C_x-\lam\,\mathcal C_y\) the $W_{00}$ term, as well as the derivative of the partial waves
	with respect to the scattering angle cancels, because \(t_\lam\) depends on
	\(x\) and \(y\) only through \(y+\lam x\) (Supplemental Material,
	Sec.~\ref{S-app:algebra}).  It obeys
	\begin{equation}
		\frac{\lam\Ktwo[\rho](\lam)}{8\pi \GN}
		=
		1+2\lam X+\lam^2Y .
		\label{eq:k2-split-convention}
	\end{equation}
	with the explicit kernel
	\begin{equation}
		\begin{split}
			\Ktwo[\rho](\lam)
			&=\sum_{J\in2\mathbb Z_{\ge0}} n_J^{(6)}
			\int_1^\infty \frac{\dd\sig}{\pi\sig^2}\,
			\frac{2\sig+3\lam}{\sig^3}\\
			&\quad\times
			\frac{C_J^{(3/2)}\!\left(
				\sqrt{(\sig-3\lam)/(\sig+\lam)}\right)}{C_J^{(3/2)}(1)}\,
			\rho_J(\sig).
		\end{split}
		\label{eq:k2-explicit-kernel}
	\end{equation}
	Every sampled \(\lambda\) must return the same \(X\) and \(Y\) from one and
	the same density.  That requirement is a set of bootstrap constraints.
	
	\subsection{Density, unitarity cap and the eikonal contribution}
	
	The variable in \cref{eq:k2-explicit-kernel} is the absorptive partial-wave
	density.  Writing \(S_J=1+if_J\),
	\begin{equation}
		\rho_J(\sig)=1-\operatorname{Re}S_J(\sig)
		=\operatorname{Im}f_J(\sig),
		\qquad
		0\le\rho_J(\sig)\le2 .
		\label{eq:rho-def-main}
	\end{equation}
	where the bounds follow from \(|S_J|\le1\). A black disk requires \(S_J=0\), hence \(\rho_J=1\).
	The upper endpoint \(\rho_J=2\) is \(S_J=-1\), perfect reflection with a
	phase of \(\pi\).  The factor \(e^{-S_{\rm BH}}\)
	suppresses the \emph{exclusive} return to the two-particle state: with
	\(S_J^{\rm el}=\eta_Je^{2i\delta_J}\) and \(\eta_J\sim e^{-S_{\rm BH}/2}\)
	\cite{GiddingsSrednicki,GiddingsPorto},
	\begin{equation}
		\rho_J^{\rm BH}
		=1-\operatorname{Re}S_J^{\rm el}
		=1-e^{-S_{\rm BH}/2}\cos(2\delta_J),
		\label{eq:black-hole-entropy-partial-wave}
	\end{equation}
	which is one up to exponentially small terms.  Saturation near \(\rho_J=2\),
	which we will meet often, is a different and reflective feature.
	
	The tree-level graviton pole gives a real Born partial wave.  Its imaginary
	part arises through unitarity and, at high energy and large spin, is resummed
	by the eikonal \(S_J\simeq e^{i\chi_J}\)~\cite{tHooftGravitonDominance,ACV},
	whose density is \(1-\cos\chi_J\).  This starts at order \(G_N^2\) in any
	fixed bin, but its dispersive integral over the continuum at large impact
	parameter is not uniform and reproduces the order-\(G_N\) pole coefficient
	(Supplemental Material, Secs.~\ref{S-app:eikonal-background}
	and~\ref{S-app:algebra}).  Reference~\cite{HaringZhiboedov} established this
	in an averaged sense.  We make a stronger assumption: on a trust region of
	large impact parameter and spin we insert the eikonal density pointwise.
	
	In six dimensions the impact parameter and eikonal phase of a bin are
	\begin{equation}
		\nu=\frac32,
		\qquad
		b=\frac{2(J+\nu)}{\ME\sqrt\sig}.
		\label{eq:b-def}
	\end{equation}
	\begin{equation}
		\chi(\sig,J)
		=
		\frac{G_N\ME^2\sig}{\pi b^2}
		=
		\frac{\GN\sig^2}{4\pi(J+\nu)^2},
		\label{eq:chi-def}
	\end{equation}
	and the prescribed density is
	\begin{equation}
		\reik(\sig,J)=1-\cos\chi(\sig,J)
		\label{eq:rho-eik}
	\end{equation}
	on the set
	\begin{widetext}
		\begin{equation}
			\begin{aligned}
				\cE(\chi_{\max})=
				\{(\sig,J):\;&
				\sqrt\sig\ge4,\quad J\ge20,\quad b\ge2\,\ME^{-1},\quad
				b/R_S(\sig)\ge r_{\min},\\
				&0\le\chi(\sig,J)<\chi_{\max}\},
				\qquad
				r_{\min}=3,\quad \chi_{\max}=30 .
			\end{aligned}
			\label{eq:eikonal-active}
		\end{equation}
	\end{widetext}
	where
	\begin{equation}
		R_S(\sig)=\left(\frac{3G_N\ME}{2\pi}\right)^{1/3}\sig^{1/6}.
		\label{eq:rs-def}
	\end{equation}
	is the Schwarzschild--Tangherlini radius.  The energy cut keeps the eikonal contribution
	away from the low-energy expansion, the spin cut keeps it semiclassical, and
	the cut at \(b/R_S\ge r_{\min}\) keeps it out of the strong-gravity region.
	The value \(r_{\min}=3\) is a choice.  It is supported after the fact by the
	full-amplitude bootstrap of Ref.~\cite{DasMyersSinhaBHScales}, in which the
	large-spin phase agrees with the tree-level eikonal to within three percent
	for \(b\ge4R_S\). Moving inward, the phase departs from that prediction
	around \(b=(2.5\text{--}3)R_S\), where inelastic scattering sets in.  On \(\cE\) the density
	is prescribed and there is no residual variable.  Everywhere else
	\(\reik=0\) and the residual density \(\rphys\) is free, with the total
	\(\rtot=\reik+\rphys\) subject to
	\begin{equation}
		0\le \rtoti\le2,
		\qquad\hbox{or}\qquad
		0\le\rphysi\le2-\reiki .
		\label{eq:physical-cap}
	\end{equation}
	The region between the low-impact band and the eikonal layer in the figures
	below is therefore not excluded by hand.  It is full of variables, and when
	it comes out empty that is a result.
	
	\subsection{Scales and hierarchies}
	\label{sec:scales}
	
	A partial wave \(J\) at energy \(E=\sqrt s\) probes the impact parameter
	\(b\simeq2(J+\nu)/E\).  Comparing it with the gravitational scales gives the
	guides drawn in \cref{fig:schematics},
	\[
	J_{\rm Born/eik}\sim \sigma^{\frac{D-2}{2(D-4)}},\qquad
	J_{\rm BH}\sim \sigma^{\frac{D-2}{2(D-3)}} ,
	\]
	up to constants: above $J_{\rm Born/eik}$ the single-graviton Born approximation
	holds, and $J_{\rm BH}$ is the black-hole scale \(b\sim R_S(E)\)
	\cite{GiddingsPorto,GiddingsSrednicki}.  Between them the partial waves are
	eikonal.  In \(D=6\) the black-hole curve grows as \(\GN^{1/3}\sigma^{2/3}\),
	a Regge trajectory would grow like \(\sigma\), and a fixed impact parameter
	corresponds to \(\sigma^{1/2}\).  Much of what follows is about telling
	these three powers apart.  As a reference curve in the figures we use the
	rotating black-hole estimate of Giddings and Porto~\cite{GiddingsPorto} with
	\(b=\kappa R_J\) and \(\kappa=3\), the nonnegative root of
	\begin{equation}
		(1+\kappa^2)J^3
		+\frac32(3+\kappa^2)J^2
		+\frac{27}{4}J+\frac{27}{8}
		-\frac{3\GN}{16\pi}\kappa^3 \sig^2=0
		\label{eq:rotating-bh-main}
	\end{equation}
	(\cref{app:rotating-bh}).  It is not a constraint.
	
	In terms of the reduced Planck mass,
	\begin{equation}
		\frac{b}{R_S}
		=
		2\left(J+\frac32\right)
		\left(\frac{16\pi^2}{3}\right)^{1/3}
		\sigma^{-2/3}
		\left(\frac{M_{\rm Pl}}{M_{\rm EFT}}\right)^{4/3},
		\label{eq:intro-planck-impact-sigma}
	\end{equation}
	so the window \(b\lesssim3R_S\) opens only above \(E\simeq2.7M_{\rm Pl}\),
	and in the physical hierarchy \(M_{\rm Pl}>M_{\rm EFT}\) it sits at
	\(\sigma\gg1\).  We therefore use two orderings and keep them apart.  In
	\cref{sec:capped} we take \(M_{\rm EFT}>M_{\rm Pl}\), with \(\GN=4\pi^2\) and
	\(M_{\rm Pl}/M_{\rm EFT}\simeq0.18\), which brings the strong-gravity window
	down to \(\sigma\) of order ten.  This is a numerical device.  The variables
	here are the positive spectral density.  In \cref{sec:weak-gravity} and in the
	weak-coupling part of \cref{sec:absorptive-density-pilot} we use the physical
	ordering, \(M_{\rm Pl}/M_{\rm EFT}=1.4\)--\(2.4\).  Partial waves remain
	well defined at trans-Planckian energies; what is lost near \(b\sim R_S\) is
	control of the weak-field expansion, not unitarity
	\cite{CaronHuotLiParraMartinezSimmonsDuffin,
		CaronHuotLiParraMartinezSimmonsDuffinHigherD,HaringZhiboedov}.  At
	\(J=O(1)\) the label \(b=2(J+3/2)/E\) is a WKB assignment and not a sharply
	resolved impact parameter.
	
	\subsection{Grid, eikonal source and linear program}
	\label{sec:finite-grid-defs}
	\label{sec:eikonal-implementation}
	\label{sec:optimization}
	
	The residual density lives on bins \(i=(r,J)\) with
	\begin{equation}
		\sig_r=\frac{N_\sigma}{r},
		\qquad
		r=1,\ldots,N_\sigma .
		\label{eq:sigma-grid}
	\end{equation}
	and \(J=0,2,\ldots,J_{\max}\).  The grid is uniform in \(1/\sigma\), which
	is the natural measure of the kernel, and starts at the threshold
	\(\sigma=1\).  The values of \(\lambda\) are
	\begin{equation}
		\theta_\alpha=\frac{(2\alpha-1)\pi}{4N_\lambda},
		\qquad
		\lam_\alpha=\frac{\sin^2\theta_\alpha}{4-\sin^2\theta_\alpha},
		.
		\label{eq:lambda-grid}
	\end{equation}
	for \(\alpha=1,\ldots,N_\lambda\). This grid is uniform in the threshold angle and resolves the region where the
	Gegenbauer argument varies fastest.  For \(N_\lambda=120\) they run from
	\(1.1\times10^{-5}\) to \(0.33331\).  We label a grid by the triple
	\((N_\sigma,J_{\max},N_\lambda)\): \(N_\sigma\) energies, even spins up to
	\(J_{\max}\), and \(N_\lambda\) sampled values of \(\lambda\).  Where the
	number of values of \(\lambda\) does not matter we write only
	\((N_\sigma,J_{\max})\).
	
	Separating the prescribed density from the residual one in
	\cref{eq:k2-split-convention} gives the equation we solve,
	\begin{widetext}
		\begin{equation}
			\frac{\lam\Ktwo[\rphys](\lam)}{8\pi \GN}
			+F_{\rm eik}(\lam;\GN)
			=1+2\lam X+\lam^2Y,
			\qquad
			F_{\rm eik}\equiv
			\frac{\lam\Ktwo[\reik](\lam)}{8\pi \GN}.
			\label{eq:k2-main}
		\end{equation}
	\end{widetext}
	The source \(F_{\rm eik}\) appears once and must contain the whole trust
	region, which is unbounded in energy and spin even though the residual grid
	is finite.  A finite phase window retains the fraction
	\begin{widetext}
		\begin{equation}
			\begin{aligned}
				\alpha_E(\chi_{\min},\chi_{\max})
				&=\frac{2}{\pi}\left[P(\chi_{\max})-P(\chi_{\min})\right],\\
				P(\chi)&=\operatorname{Si}(\chi)-\frac{1-\cos\chi}{\chi},
				\qquad
				\operatorname{Si}(\chi)=\int_0^\chi\frac{\sin t}{t}\,dt .
			\end{aligned}
			\label{eq:alpha-window-main}
		\end{equation}
	\end{widetext}
	of the forward pole; \([0,30]\) keeps \(\alpha_E=0.9795\).  We split the
	source at an even spin \(J_\ast\),
	\begin{equation}
		F_{\rm eik}^{\rm hyb}(\lam;J_\ast)
		=F_{\rm eik}^{J\le J_\ast,\,\mathrm{grid}}(\lam)
		+F_{\rm eik}^{J>J_\ast,\,\mathrm{cont}}(\lam).
		\label{eq:hybrid-eikonal-source}
	\end{equation}
	The first term uses the exact Gegenbauer kernel on its own fine energy
	quadrature.  In the second the normalized Gegenbauer polynomial is replaced
	by its large-spin limit \(2J_1(b\ME\sqrt\lambda)/(b\ME\sqrt\lambda)\) at fixed
	\(b\), the spin sum becomes an integral, and the variables \((\sigma,J)\)
	are traded for \((\chi,b)\).  The \(2\sigma\) and \(3\lambda\) terms of the
	kernel then give
	\(F_{\rm eik}^{\rm cont}=F_{\rm eik}^{(0)}+F_{\rm eik}^{(1)}\) with
	\begin{widetext}
		\begin{align}
			F_{\rm eik}^{(0)}(\lambda)
			&=\frac{2}{\pi}
			\int_{\chi_{\min}}^{\chi_{\max}}\!d\chi\,
			\frac{1-\cos\chi}{\chi^2}
			\left[J_0\!\left(b_{\min}\ME\sqrt\lambda\right)
			-J_0\!\left(b_{\max}\ME\sqrt\lambda\right)\right],
			\label{eq:continuum-eikonal-leading}\\
			F_{\rm eik}^{(1)}(\lambda)
			&=\frac{3\GN\lambda^2}{\pi^2}
			\int_{\chi_{\min}}^{\chi_{\max}}\!d\chi\,
			\frac{1-\cos\chi}{\chi^3}
			\left[H\!\left(b_{\min}\ME\sqrt\lambda\right)
			-H\!\left(b_{\max}\ME\sqrt\lambda\right)\right],
			\label{eq:continuum-eikonal-regular}\\
			H(x)&\equiv\int_x^\infty\frac{J_1(t)}{t^2}\,dt .
			\label{eq:continuum-eikonal-h}
		\end{align}
	\end{widetext}
	Here \(b_{\min}(\chi)\) and \(b_{\max}(\chi)\) are the cuts of
	\cref{eq:eikonal-active}, together with \(J>J_\ast\), rewritten at fixed
	\(\chi\).  The derivation, a stable form of \(H\), and the quadrature checks
	are in the Supplemental Material (Secs.~\ref{S-sec:supp-grids}
	and~\ref{S-app:algebra}).  We take \(J_\ast=320\) when
	\(M_{\rm EFT}>M_{\rm Pl}\).  At weak coupling the phase window moves to
	energies and spins far beyond any grid, and we integrate the whole eikonal contribution
	in the continuum.
	
	With the residual density on the grid, \cref{eq:k2-main} is one equality for each
	\(\lambda_\alpha\),
	\begin{equation}
		\begin{aligned}
			\sum_i A_{\alpha i}\rphysi-\lam_\alpha^2Y
			&=r_{X,\alpha},\\
			r_{X,\alpha}\equiv r_X(\lam_\alpha)
			&=1+2\lam_\alpha X-F_{\rm eik}^{\rm hyb}(\lam_\alpha;J_\ast),\\
			A_{\alpha i}
			&=\frac{\lam_\alpha\mathcal K_{2,i}(\lam_\alpha)}{8\pi \GN}.
		\end{aligned}
		\label{eq:sampled-row}
	\end{equation}
	For fixed \(X\) the right-hand side is known.  The unknowns are the weights
	\(\rphysi\), one for each bin outside \(\cE\), and the single number \(Y\).
	We maximize or minimize \(Y\) subject to these equalities and to the box
	\begin{equation}
		0\le\rphysi\le u_i,
		\qquad
		u_i=2-\reiki,
		\label{eq:upper-bound-ui}
	\end{equation}
	This traces the upper and lower boundaries of the allowed region in the
	\((X,Y)\) plane and, more usefully for us, returns the density that realizes
	each boundary point.  We solve the primal problem with
	HiGHS~\cite{HuangfuHallHiGHS}, which also returns the reduced costs used in
	\cref{sec:dual-slack}; the dual problem is written out in the Supplemental
	Material (Sec.~\ref{S-app:reduced-cost}).  The grids
	\((N_\sigma,J_{\max},N_\lambda)\) range from \((300,160,60)\) to
	\((900,320,120)\), and each solution is tested on up to \(1601\) further
	values of \(\lambda\) that were not imposed.
	
\begin{figure}[t]
		\centering
		\includegraphics[width=0.92\columnwidth]{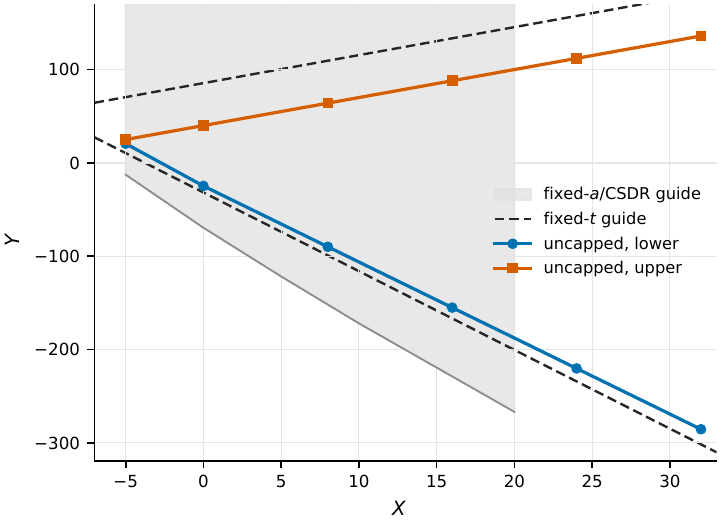}
		\caption{What the eikonal contribution does before unitarity is imposed bin by bin:
			the allowed region in \((X,Y)\) with the eikonal contribution supplied but no
			upper bound on the residual density (\(\GN=0.4\pi^2\), \((900,256,120)\)
			grid).  It is an open wedge, close to the fixed-\(t\) guide lines (dashed)
			and unlike the fixed-\(a\) region (gray); neither guide is imposed.
			Compare the closed leaf of \cref{fig:g6p5-leaf}, obtained with the cap.}
		\label{fig:uncapped-wedge}
	\end{figure}

	Without the upper bound in \cref{eq:upper-bound-ui} the same problem gives the
	wedge-shaped cone of \cref{fig:uncapped-wedge}.  It lies close to the region
	allowed by fixed-\(t\) sum rules, and its spectra are sparse, but their
	densities reach \(\rho\simeq10^3\) (Supplemental Material,
	Sec.~\ref{S-sec:supp-uncapped}).  The cap is what makes the spectra
	physical, and it changes them qualitatively.
	
	\section{Extremal spectra with the eikonal contribution supplied}
	\label{sec:capped}
	
	We begin with \(\GN=4\pi^2\), that is \(M_{\rm Pl}/M_{\rm EFT}\simeq0.18\),
	where the region \(b\sim R_S\) falls inside the sampled energies.  The leaf
	is traced on the grid \((N_\sigma,J_{\max},N_\lambda)=(300,160,80)\) of
	\cref{sec:finite-grid-defs}, the spectra and reduced costs are computed on
	\((600,240,80)\), and convergence is checked up to \((900,320,120)\); the eikonal source has its own 6400-point energy
	quadrature and a 4000-point continuum tail, so its accuracy does not depend
	on the residual grid.
	
	One fact about linear programs should be kept in mind throughout.  A basic
	optimum with \(N_\lambda\) equality constraints has at most \(N_\lambda\)
	variables strictly between their bounds. Sparse support and many bins at
	the cap are generic features of these solutions, including at weak gravity
	(\cref{sec:weak-gravity}).  What carries information is where the
	support lies, how it differs between the two boundaries, and how it moves
	with the coupling.
	
	\subsection{The leaf}
	\label{sec:g6p5-leaf}
	
	\begin{figure}[t]
		\centering
		\includegraphics[width=\columnwidth]{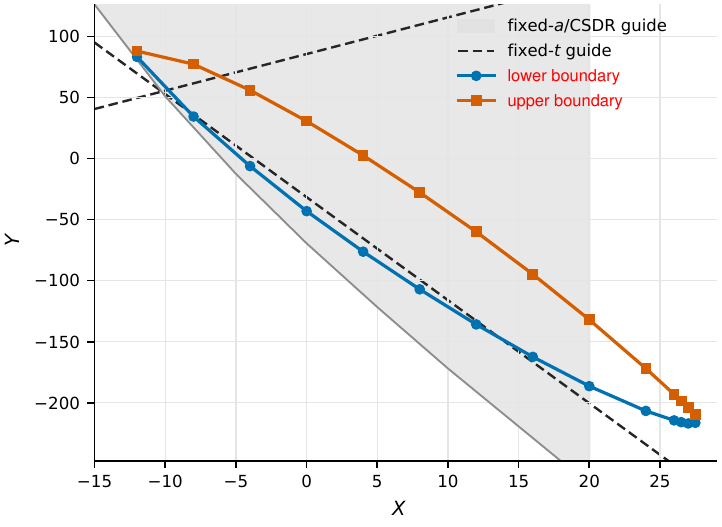}
		\caption{Allowed region in the normalized coefficients \((X,Y)\) at
			\(\GN=4\pi^2\) on the \((300,160,80)\) grid, with the eikonal contribution
			supplied and the unitarity cap imposed.  The orange and blue curves are the
			upper and lower boundaries; every spectrum discussed in
			\cref{sec:capped} is the density at a point on one of them.  The dashed
			lines (fixed-\(t\) sum rules) and the gray region (fixed-\(a\) relation) are
			shown for orientation and are not imposed.  The region closes between
			\(X=-12.5\) and \(X=28\).}
		\label{fig:g6p5-leaf}
	\end{figure}
	
	With the cap the allowed region closes into the leaf of
	\cref{fig:g6p5-leaf}, as in the fixed-\(a\) analysis of
	Ref.~\cite{PengRodinaTokarevaXu} (their published example is in \(D=5\), so
	the comparison is qualitative).  The boundary values are tabulated in the
	Supplemental Material (Table~\ref{S-tab:g6p5-leaf-table}); the larger grids
	reproduce them at the percent level or better.
	
	\subsection{A band, a ridge and a gap}
	\label{sec:spectra}
	
	\begin{figure*}[t]
		\centering
		\includegraphics[width=0.80\textwidth]{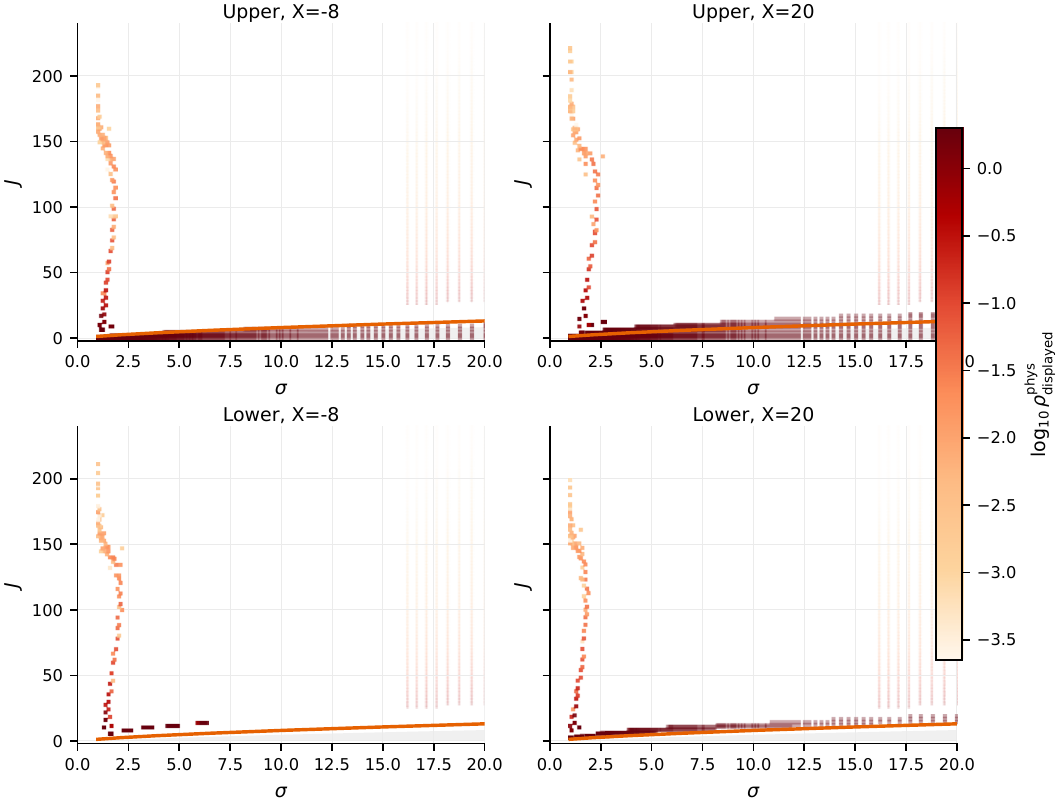}
		\caption{The densities behind four boundary points of
			\cref{fig:g6p5-leaf} (\(\GN=4\pi^2\), \((600,240,80)\) grid, shown for
			\(\sigma\le20\)): upper boundary (top) and lower boundary (bottom) at
			\(X=-8\) and \(X=20\).  Squares are the optimized residual density, small
			translucent circles sample the prescribed eikonal density, and the orange
			curve is the rotating black-hole reference \(J_{\rm BH}^{(\kappa=3)}\).
			Three structures recur: a band of bins at the cap along the bottom, a
			steep ridge at low energy and high spin, and an empty gap between the
			band and the eikonal layer.  The band is absent only on the lower
			boundary at negative \(X\).}
		\label{fig:g6p5-four-witness}
	\end{figure*}
	
	\Cref{fig:g6p5-four-witness} shows the densities at four boundary points.
	Residual variables exist everywhere below the eikonal layer, and the white
	regions are where the optimum sets them to zero.  Three structures stand
	out.  Along the bottom of each panel there is a band of bins at the cap,
	\(\rho\simeq2\), which starts at \(J=0\) and ends near the black-hole
	reference curve.  At low energy a steep ridge climbs to high spin; it grows
	from \(J=12\)--\(20\) to \(14\)--\(28\) when the point at which the eikonal contribution
	bins begin is moved from \(\sigma=80\) to \(160\), so it is not produced by
	that boundary (Supplemental Material, Sec.~\ref{S-sec:supp-strong}).
	Between the band and the eikonal layer the spectrum is almost empty.
	
	\begin{table}[t]
		\centering
		\caption{Support of the four solutions of \cref{fig:g6p5-four-witness}:
			number of occupied residual bins, number within ten percent of the cap,
			fraction of the residual weight at \(b/R_S<3\), and the weighted mean of
			\(b/R_S\).}
		\small
		\begin{tabular}{ccrrrr}
			\toprule
			\(X\)&boundary
			&\(N_{\rm res}\)
			&\(N_{\rho\ge1.8}\)
			&\(f_{b/R_S<3}\)
			&\(\langle b/R_S\rangle_\rho\)\\
			\midrule
			\(-8\)&upper&1491&1412&0.961&1.04\\
			20&upper&2604&2525&0.978&1.45\\
			\(-8\)&lower&198&120&0&4.60\\
			20&lower&1034&955&0.866&2.46\\
			\bottomrule
		\end{tabular}
		\label{tab:g6p5-support-ledger}
	\end{table}
	
	\Cref{tab:g6p5-support-ledger} makes this quantitative.  On the upper
	boundary more than \(96\%\) of the residual weight lies at \(b/R_S<3\), and
	nearly all occupied bins are at the cap.  On the lower boundary the band is
	present at \(X=20\) but absent at \(X=-8\), where the weight sits in the gap
	instead; it turns on between \(X=0\) and \(X=8\).  The ridge is long but
	light: it occupies many bins on the lower boundary at negative \(X\) and
	carries little of the sum rule (Supplemental Material,
	Sec.~\ref{S-sec:boundary-morphology}).  The band is thus a property of
	particular extremal solutions and not something the cap produces
	automatically.
	
	These solutions are numerically stable.  Raising \(N_\lambda\) from 40 to
	120 and the spectral grid from \((300,160)\) to \((1200,400)\) moves
	\(Y_{\max}\) at \(X=20\) by less than \(2.5\times10^{-5}\) once
	\(N_\lambda\ge60\).  On 1601 values of \(\lambda\) that were not imposed the
	median residual is below \(10^{-3}\) and the maximum is \(1.3\)--\(1.9\%\).
	Dual simplex and interior point return the same optimum.  Imposing eight
	finite amplitude differences at three boundary points, which tests the SDR
	away from \(x=y=0\), changes \(Y\) by at most \(0.14\%\) and leaves the
	support in place (Supplemental Material, Sec.~\ref{S-sec:supp-strong}).
	
	\subsection{The band ends at a fixed impact parameter}
	\label{sec:band-edge}
	
	The band ends near the black-hole reference curve in
	\cref{fig:g6p5-four-witness}, and it is tempting to call it a black-hole
	band.  The energy dependence of its edge says otherwise.  We follow the
	contiguous run of bins with \(\rtot\ge1.8\) upward from \(J=0\) at each
	energy and fit \(J_{\rm edge}+3/2\propto\sigma^p\).  On the \(X=20\)
	upper-boundary solution this gives \(p=0.50\)--\(0.51\) on every window
	between \(\sigma=3\) and \(600\), with a scatter of \(0.02\)--\(0.05\) in
	\(\log J\), and \(p=0.49\)--\(0.52\) at \(\GN=2\pi^2\) and \(\pi^2\)
	[\cref{fig:core-vs-band}(b)].  A radius proportional to
	\(R_S(E)\propto\sigma^{1/6}\) would give \(p=2/3\).  The band ends instead
	at a fixed impact parameter,
	\[
	b_{\rm edge}\simeq9.3,\ 7.4,\ 6.3\ M_{\rm EFT}^{-1}
	\quad\text{at}\quad \GN=4\pi^2,\ 2\pi^2,\ \pi^2 ,
	\]
	so that at \(\GN=4\pi^2\) the ratio \(b_{\rm edge}/R_S\) falls from \(2.7\)
	at \(\sigma=5\) to \(1.7\) at \(\sigma=75\) and the edge crosses the
	black-hole curve.  The two coincide only in the window \(\sigma\lesssim20\)
	of \cref{fig:g6p5-four-witness}\footnote{To probe further we need to investigate how absorptive this spectra is \ref{sec:absorptive-density-pilot}.}.  The next two subsections explain why the
	gap is empty and why the band sits where it does; \cref{sec:weak-gravity}
	shows that the band survives when gravity is switched off.
	
	\subsection{Why the gap stays empty}
	\label{sec:dual-slack}
	
	\begin{figure*}[t]
		\centering
		\includegraphics[width=0.95\textwidth]{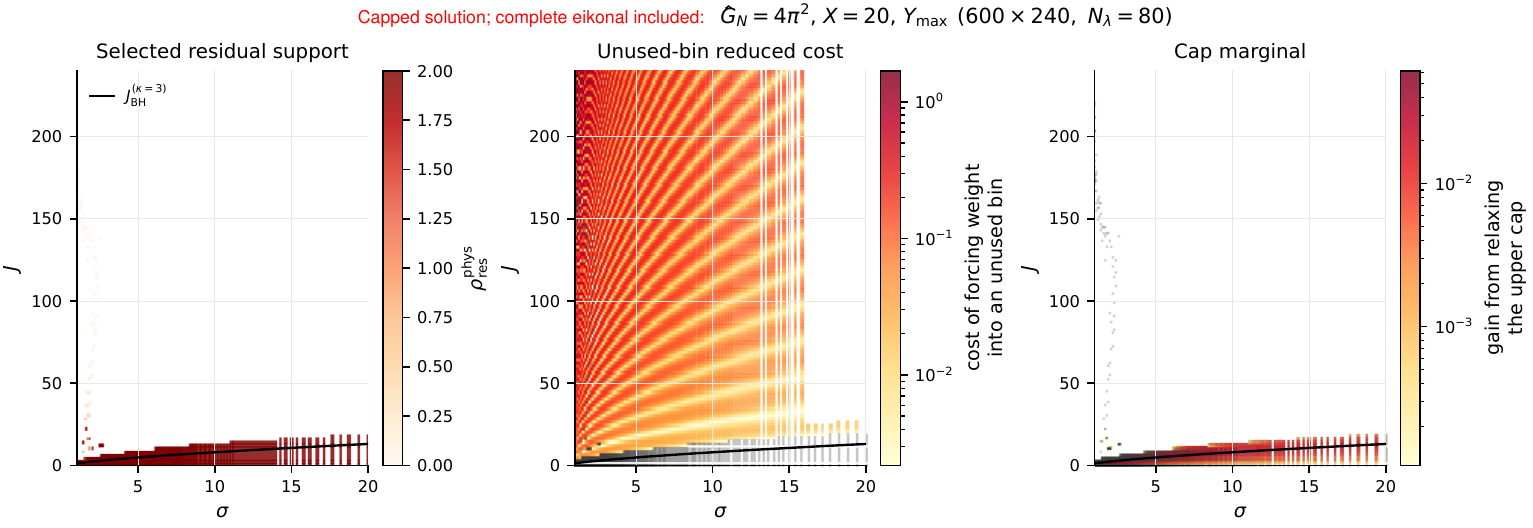}
		\caption{Why the optimum looks the way it does, at \(\GN=4\pi^2\),
			\(X=20\), upper boundary, \((600,240,80)\) grid, \(\sigma\le20\).  Left:
			the density.  Middle: for each empty bin, the first-order loss in
			\(Y_{\max}\) if that bin is forced to carry weight and everything else is
			reoptimized; the gap between the band and the eikonal layer is expensive.
			Right: for each bin at the cap, the first-order gain in \(Y_{\max}\) if the
			cap is relaxed there; the band is held back by unitarity.  The black curve
			is the rotating black-hole reference.}
		\label{fig:x20-dual-slack-sigma20}
	\end{figure*}
	
	The linear program has variables throughout the gap.  Why does it not use
	them?  The reduced costs answer this directly.  For an empty bin the reduced
	cost is the rate at which \(Y_{\max}\) falls if the bin is forced to carry
	density while all other variables readjust.  For a bin at the cap the
	corresponding marginal is the rate at which \(Y_{\max}\) would rise if the
	cap were relaxed there.  Both are properties of the whole program, not of a
	single kernel (conventions in the Supplemental Material,
	Sec.~\ref{S-app:reduced-cost}).
	
	\Cref{fig:x20-dual-slack-sigma20} shows them at \(X=20\).  On the full grid
	the median cost of an empty bin above the reference curve is \(0.173\),
	thirty times the median \(0.0056\) of the empty bins at low impact
	parameter, and the bins of the band have a positive median marginal of
	\(0.019\).  The gap is therefore not a reservoir of equally good unused
	bins, and the band is not merely full: it would take more weight if
	unitarity allowed.  
	
	\subsection{Why the band sits at the cutoff scale}
	\label{sec:demystifying-bh-band}
	
	\begin{figure*}[t]
		\centering
		\includegraphics[width=0.95\textwidth]{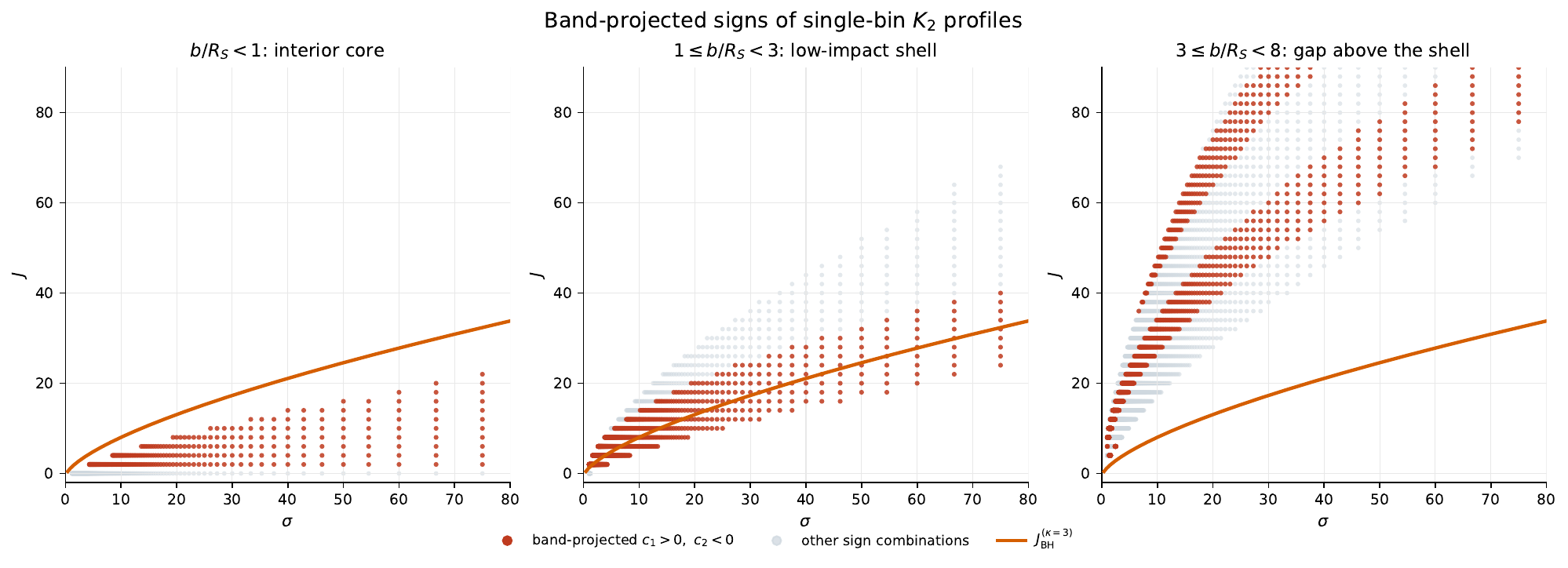}
		\caption{Which bins have a kernel of the right shape.  Each point is a bin
			\((\sigma,J)\) of the \((600,80)\) grid, whether or not it is occupied in
			any solution.  Red: the profile \(\lambda\Ktwo\) of the bin, projected on
			\(\{1,\lambda,\lambda^2\}\) over the sampled window, has a positive linear
			and a negative quadratic coefficient, the signs required on the upper
			boundary at \(X>0\).  Gray: any other combination.  The classification
			does not depend on \(G_N\).  The panels sort the bins by \(b/R_S\) at
			\(\GN=4\pi^2\), for which the orange reference curve is drawn; in
			absolute units the red region ends at
			\(b\simeq9.8\,M_{\rm EFT}^{-1}\) at every energy.}
		\label{fig:k2-column-band-projected-signs}
	\end{figure*}
	
	The location of the band can be understood before any optimization.  Each
	bin enters every constraint through its kernel profile
	\begin{equation}
		v_{\sig,J}(\lambda_\alpha)
		=
		\lambda_\alpha\mathcal K_{2,(\sig,J)}(\lambda_\alpha),
		\label{eq:band-column-definition}
	\end{equation}
	a function of \(\lambda\) on the sampled window.  We project it on
	\(\{1,\lambda,\lambda^2\}\) by least squares,
	\begin{equation}
		v_{\sig,J}(\lambda_\alpha)
		\simeq
		c_0^{\rm band}(\sig,J)
		+c_1^{\rm band}(\sig,J)\lambda_\alpha
		+c_2^{\rm band}(\sig,J)\lambda_\alpha^2 .
		\label{eq:band-projected-coefficients}
	\end{equation}
	and compare with the right-hand side \(1+2\lambda X+\lambda^2Y\) of
	\cref{eq:k2-main}.  On the upper boundary at positive \(X\) the linear term
	is positive and the quadratic term negative, so the useful bins are those
	with
	\begin{equation}
		c_1^{\rm band}>0,\qquad c_2^{\rm band}<0 .
		\label{eq:useful-band-projected-sign}
	\end{equation}
	This test knows nothing about how bins cooperate, which is the business of
	the reduced costs, but it shows which bins have the right shape.
	
	\Cref{fig:k2-column-band-projected-signs} displays the result.  The bins
	with the signs of \cref{eq:useful-band-projected-sign} form a coherent
	region at low impact parameter, and beyond it the signs are mixed.  The
	figure is organized by \(b/R_S\) at the reference coupling, where the
	coherent region ends near \(b/R_S\simeq2.8\).  But the kernel does not
	contain \(G_N\), and at large spin it depends on energy and spin only
	through the combination \(b\ME\sqrt\lambda\), since the normalized Gegenbauer
	polynomial tends to \(2J_1(b\ME\sqrt\lambda)/(b\ME\sqrt\lambda)\).  The boundary of
	the coherent region is therefore a fixed impact parameter.  Evaluating it
	bin by bin we find
	\[
	b_\ast\simeq9.8\,M_{\rm EFT}^{-1}
	\qquad(\lambda_{\max}=1/3),
	\]
	the same to three percent at \(\sigma=20\), \(75\) and \(300\), and
	\(b_\ast\ME\sqrt{\lambda_{\max}}\simeq5.6\) to within three percent when the
	upper end of the sampled window is lowered to \(0.25\), \(0.2\) and
	\(0.1\).  Measured in units of
	\(R_S\) this boundary moves as \(G_N^{-1/3}\), which is all there is to the
	coupling dependence of the number \(2.8\).
	
	This is the origin of the band.  The bins that can help the objective are
	those below a fixed impact parameter of a few cutoff lengths, and the
	optimizer fills them from the bottom up until the sum rule is saturated.
	At the strong coupling used here, reducing the upper end of the sampled
	\(\lambda\) range from \(1/3\) to \(0.2\) leaves the observed edge at
	\(9.3\,\ME^{-1}\). The outer limit inferred from the kernel signs moves
	from \(9.8\,\ME^{-1}\) to \(12.6\,\ME^{-1}\), so those signs alone
	do not fix where the band ends (Supplemental Sec.~\ref{S-sec:lambda-window}).
	The same argument accounts for the
	missing band on the lower boundary at negative \(X\).  There the target has
	a negative linear term, the low-impact bins have the wrong sign, and using
	them would require compensating weight elsewhere, which a positive and
	capped spectrum can supply only by lowering \(Y\).
	
	\begin{figure*}[t]
		\centering
		\includegraphics[width=0.80\textwidth]
		{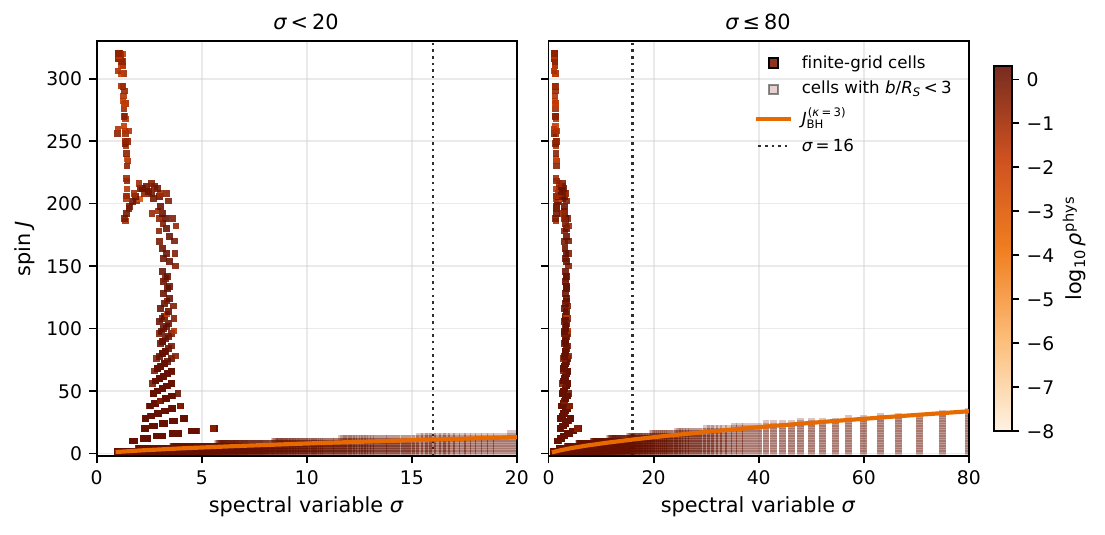}
		\caption{What the spectrum does when the eikonal contribution is not supplied
			(\(\GN=4\pi^2\), \(X=6.5\), upper boundary, \((1200,320,120)\) grid; left
			\(\sigma<20\), right \(\sigma\le80\)).  The pole sum rule is still
			imposed, and the grid satisfies it with a branch at low energy and spin up
			to \(300\), far outside the eikonal regime.  The low-impact band along the
			bottom is still there but carries little of the pole.  Compare
			\cref{fig:g6p5-four-witness}, where the eikonal contribution is supplied and this
			branch is absent.}
		\label{fig:noeik-control-x6p5}
	\end{figure*}
	
	\subsection{What the eikonal contribution does}
	\label{sec:no-eikonal-controls}
	\label{sec:oscillation}
	\label{sec:ext-eik-results}
	
	The eikonal density is an input.  Two controls show what it contributes
	(Supplemental Material, Secs.~\ref{S-sec:noeik-scan}
	and~\ref{S-sec:supp-inward}).
	
	If the eikonal contribution is removed, the pole sum rule is still imposed but the grid
	itself has to supply it.  At \(\GN=4\pi^2\) and \(X=6.5\), with the spectral grid and sampled \(\lambda\) set refined together up to \((1200,320,120)\), \(Y_{\max}\)
	reaches \(9.7\), still drifting by three percent in the last step, and the
	weight that carries the pole moves to a branch at low energy and high spin; the low-impact band remains but contributes
	little to the pole (\cref{fig:noeik-control-x6p5}).  The sampled sum rules
	thus admit a quite different
	spectral origin for the pole, one that lies outside the eikonal regime.  Supplying
	the eikonal density selects the standard realization before one asks how
	the spectrum completes at small impact parameter.  Without it the boundary
	value can look unremarkable while the equations are badly violated between
	the sampled points: at \(\GN=\pi^2/4000\) removing the eikonal contribution changes
	\(Y_{\max}\) by \(0.4\%\) and raises the worst residual from \(0.02\) to
	\(0.47\).
	
	If instead the eikonal contribution is continued inward, to \(b/R_S\ge0.9\), \(0.75\),
	\(0.5\) and \(0.25\), with residual variables kept in every bin and the cap
	applied to the total density, the set of occupied bins hardly changes
	(overlap above \(0.999\)).  The density in those bins shifts by \(8\) to
	\(28\%\) and develops rapid modulation at low spin ( See Sec.\ref{S-sec:supp-inward} in Supplemental).
	
	\begin{figure*}[t]
		\centering
		\includegraphics[width=0.88\textwidth]
		{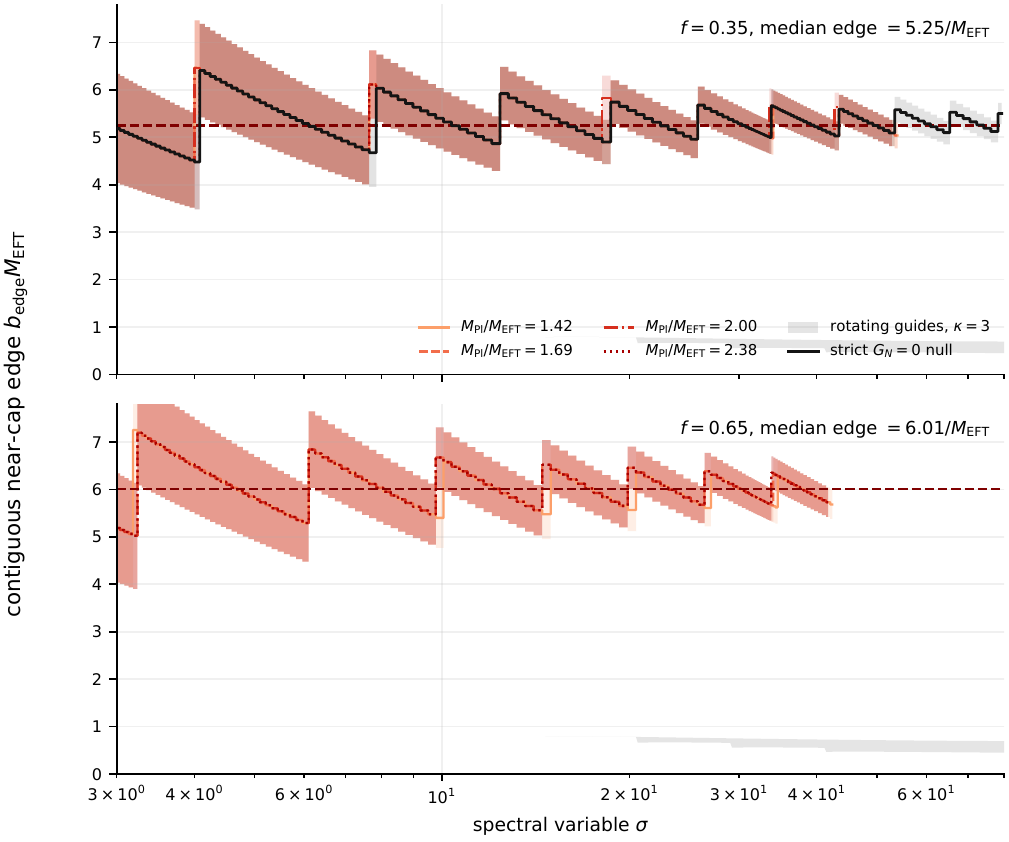}
		\caption{The edge of the cap-saturated band in units of the cutoff length,
			against energy, at two matched positions on the upper boundary
			(\(f=0.35\) top, \(f=0.65\) bottom).  Colored bands: the four weak
			couplings \(\GN=\pi^2/r\), \(r=1000,2000,4000,8000\), each drawn as the
			interval between the last occupied spin and the next.  Dashed lines:
			medians.  Black staircase (top): the same problem at \(G_N=0\) with the
			physical \(g_2\) matched.  Gray: the rotating black-hole reference curves
			for the four couplings, which lie an order of magnitude lower and move
			with \(G_N\).  The edge does neither.  This is the evidence that the band
			is the non-gravitational baseline of the capped problem.}
		\label{fig:continuum-weak-edge}
	\end{figure*}
	
	\section{Weak gravity: the baseline of the capped problem}
	\label{sec:weak-gravity}
	
	We now restore the physical hierarchy \(M_{\rm EFT}<M_{\rm Pl}\).  The
	eikonal phase window then lies at energies and spins,
	\(\chi\sim \GN\sigma^2/J^2\), that no fixed grid reaches.  At
	\(\GN=\pi^2/300\) an eikonal quadrature grid of the size used in \cref{sec:capped} holds
	\(8\times10^{-4}\) of the pole-sensitive source at \(\lambda=10^{-4}\),
	where the full eikonal phase window holds \(0.979\).  We therefore integrate the
	eikonal contribution in the continuum, \cref{eq:continuum-eikonal-leading}, and
	discretize only the residual spectrum.  The couplings are
	\begin{equation}
		\begin{split}
			\GN=\frac{\pi^2}{r},
			\qquad r&=1000,2000,4000,8000,\\
			\frac{M_{\rm Pl}}{M_{\rm EFT}}&=\left(\frac{r}{8\pi^3}\right)^{1/4}.
		\end{split}
		\label{eq:continuum-weak-ladder}
	\end{equation}
	with the trust region of \cref{eq:eikonal-active}. The residual grid samples
	high energies logarithmically and then compactifies the tail. Both this grid
	and the eikonal integration extend beyond the energies displayed in
	\cref{fig:continuum-weak-edge}.
	
	As \(G_N\) decreases the leaves grow strongly in the normalized variables
	\((X,Y)\), but in the physical coefficients \((g_2,g_3)\) they nearly
	coincide (Supplemental Material, Sec.~\ref{S-sec:supp-weak}): most of the
	apparent motion is the factor \(1/G_N\) in the normalization. The leftmost allowed value of \(g_2\) is
	\(g_{2,\min}\simeq-(0.82\text{--}0.91)\,8\pi G_N/\ME^2\) and tends linearly to zero as
	gravity is switched off, in line with Ref.~\cite{ChangParraMartinez}.
	Because the leaves differ, we compare spectra at matched positions and not
	at equal \(X\).  Let \(X_{\rm L}\) and \(X_{\rm R}\) be the smallest and
	largest values of \(X\) on the leaf at a given coupling, its two tips.  We
	take the solution on the upper boundary at
	\begin{equation}
		X_f=X_{\rm L}+f\,(X_{\rm R}-X_{\rm L}),
		\qquad f=0.35,\ 0.65 ,
		\label{eq:leaf-fraction}
	\end{equation}
	so that \(f\) is the fractional position along the leaf, with the tips
	recomputed at each \(G_N\).  All solutions are obtained on
	\((600,240,96)\) and have residuals of at most two percent on 801 unsampled
	values of \(\lambda\).
	
	\subsection{The edge does not follow the gravitational radius}

	At each energy we follow the sequence of successive spin bins as long as their densities remain within ten percent of
	the cap upward from \(J=0\), as in \cref{sec:band-edge}, and convert its end
	to an impact parameter.  \Cref{fig:continuum-weak-edge} shows the result.
	The median edge is
	\begin{equation}
		\begin{split}
			\ell_{\rm micro}&=5.25M_{\rm EFT}^{-1}\quad(f=0.35),\\
			\ell_{\rm micro}&=6.01M_{\rm EFT}^{-1}\quad(f=0.65).
		\end{split}
		\label{eq:continuum-weak-micro-radii}
	\end{equation}
	Across the four couplings the median varies from \(5.25\,\ME^{-1}\) to
	\(5.30\,\ME^{-1}\) at \(f=0.35\) and from \(5.94\,\ME^{-1}\) to
	\(6.01\,\ME^{-1}\) at \(f=0.65\). The sets of occupied bins at neighboring
	couplings have a Jaccard overlap above \(0.98\).\footnote{The Jaccard
	overlap of two supports is the number of bins occupied in both divided by
	the number occupied in either. It equals one for identical supports.}
	Over the same range of couplings the Schwarzschild radius at fixed
	\(\sigma\) changes by a factor of two.  The edge depends on where one sits
	on the boundary, and it does not depend on \(G_N\).
	
	\subsection{The same band at \texorpdfstring{\(G_N=0\)}{GN=0}}
	\label{sec:gn0}
	
	If the edge does not move with the coupling, does it need gravity at all?
	At \(G_N=0\) there is no pole and no eikonal contribution, the normalized variables are
	singular, and the problem is posed in the physical coefficients,
	\begin{equation}
		\lambda\Ktwo[\rho^{\rm phys}](\lambda)-\lambda^2g_3\ME^8
		=2\lambda g_2\ME^6 .
		\label{eq:gn0-null-sum-rule}
	\end{equation}
	We compare the \(f=0.35\) solution at \(\GN=\pi^2/4000\), for which
	\(g_2\ME^6=5.56\times10^3\), with the \(G_N=0\) solution at
	\(g_2\ME^6=5.60\times10^3\) on the same grid, with \(0\leq\rho^{\rm phys}\leq2\).  The median edges are
	\(5.30\,\ME^{-1}\) and \(5.27\,\ME^{-1}\), the edges differ by \(4.9\%\) rms
	over 149 common energies, the sets of occupied bins overlap with a Jaccard
	index of \(0.988\), and the
	values of \(g_3\) agree to \(1.1\%\).  The \(G_N=0\) solution satisfies the
	unsampled equations to \(1.3\times10^{-5}\).  This is the black staircase in
	\cref{fig:continuum-weak-edge}.
	
	One might still worry that a radius of five cutoff lengths is set by the
	sampled window, \(b\ME\sqrt{\lambda_{\max}}\simeq3\), and not by the threshold.
	It is not.  Repeating the \(G_N=0\) calculation with \(\lambda\) sampled up
	to \(0.25\), \(0.20\) and \(0.10\) in place of \(1/3\) gives median edges of
	\(5.26\), \(5.27\) and \(5.29\,M_{\rm EFT}^{-1}\) for \(3\le\sigma\le80\),
	against \(5.27\), while \(g_3\) changes by one part in a thousand
	(Supplemental Material, Sec.~\ref{S-sec:lambda-window}).  Above
	\(\sigma\simeq80\) the windows with $\lambda_{\max} = 1/3, 0.25$, and $0.20$ still agree, at \(5.30\)--\(5.31\);
	with \(\lambda_{\max}=0.10\) the band no longer extends that far.  The kernel boundary \(b_\ast\) of
	\cref{sec:demystifying-bh-band} moves from \(9.8\,\ME^{-1}\) to \(18\,\ME^{-1}\) over the same
	range and plays no role here, because at weak coupling the band ends well
	inside it.
	
	There is a simple reason for the size of the band.  At \(\lambda\to0\),
	\cref{eq:gn0-null-sum-rule} fixes \(\Ktwo[\rho](0)=2g_2\ME^6\). A density at the cap
	inside a fixed impact parameter \(b_0\) gives
	\(\Ktwo(0)\simeq2\pi\rho_{\max}(b_0\ME)^4\) in the large-spin
	approximation, so
	\begin{equation}
		b_0\simeq\left(\frac{g_2\ME^6}{\pi\rho_{\max}}\right)^{1/4}
		M_{\rm EFT}^{-1}
		=5.5\,M_{\rm EFT}^{-1}
		\label{eq:forward-moment-radius}
	\end{equation}
	for \(g_2\ME^6=5600\) and \(\rho_{\max}=2\), against the measured
	\(5.27\).  The \(\lambda\to0\) sum rule fixes the total weight needed to reproduce
	\(g_2\). With a cap on the density, filling from the smallest impact
	parameters takes a disk of about the size in \cref{eq:forward-moment-radius}.
	About two thirds of that weight comes from \(\sigma<3\) and \(99\%\) from
	\(\sigma<80\). Changing the upper end of the \(\lambda\) range leaves the
	forward moment fixed and barely moves the edge in that energy range. The
	same estimate accounts for the dependence on position in
	\cref{eq:continuum-weak-micro-radii}: \(g_2\) is approximately proportional
	to \(f\) at weak coupling, and
	\((0.65/0.35)^{1/4}=1.17\) against the measured ratio \(1.14\).  At strong
	coupling the estimate does not apply as it stands, because the forward
	residual term in the small-\(\lambda\) sum rule is \(30\) to \(70\) times
	\(2g_2\ME^6\). Almost all of that term comes from \(\sigma<3\), where
	the adjustable density compensates for the eikonal contribution at the
	smallest sampled \(\lambda\).
	
	The reflective band is therefore a feature of the capped dispersive problem
	at the scale of the first threshold, with a radius fixed by \(g_2\) and the
	cap.  Gravity displaces its edge outward,
	from \(5.3\,\ME^{-1}\) at \(G_N=0\) to \(6.3\,\ME^{-1}\),
	\(7.4\,\ME^{-1}\) and \(9.3\,\ME^{-1}\) at
	\(\GN=\pi^2\), \(2\pi^2\) and \(4\pi^2\), but does not create it and does
	not give it a gravitational energy dependence.
	
	\subsection{High energy}
	\label{sec:matched-high-energy-null}
	
	At \(\sigma\sim10^4\)--\(10^5\) the empty interval between the outer
	edge of the occupied, adjustable density and the inner edge of the
	prescribed eikonal region can close.  On an \((800,400,120)\) grid at fixed
	\(g_2\ME^6\simeq5.6\times10^3\) and \(\GN=\pi^2/r\) with
	\(r=1000\), \(500\), \(100\), \(50\), the bins at the cap fill a wedge that
	reaches all the way out to the prescribed eikonal region.  For \(r=1000\) and \(500\) its
	outer edge is the inner boundary of the prescribed eikonal region \(b=r_{\min}R_S\), to within one even
	spin, for \(10^4\le\sigma\le5.5\times10^4\); at higher energy it is pinned
	at the spin cutoff, and for \(r=100\) and \(50\) the cut on the phase
	competes and the edge falls below the boundary.  When \(r_{\min}\) is
	changed from \(3\) to \(4\) and \(6\) at \(r=1000\) the outer edge moves
	with it, and nothing else
	does (\(Y\) changes by less than \(3\times10^{-4}\)).  The scaling
	\(J\propto \GN^{1/3}\sigma^{2/3}\) of this edge is thus inherited from the
	definition of the eikonal cut.  The lower edge of the wedge is free.  It is linear,
	\begin{equation}
		J_{\rm low}+\frac32\simeq a(\GN)\sigma+\beta,
		\qquad a(\GN)\propto \GN^{-0.35\pm0.02}.
		\label{eq:matched-high-energy-lower-edge}
	\end{equation}
	compatible with \(G_N^{-1/3}\), and in that sense Regge-like, with an
	effective slope \(\alpha'_{\rm eff}M_{\rm EFT}^2\propto
	(M_{\rm Pl}/M_{\rm EFT})^{4/3}\).  A linear lower edge together with support
	that reaches the prescribed eikonal region is reminiscent of the
	string/black-hole
	correspondence~\cite{HorowitzPolchinskiCorrespondence,CeplakEmparanPuhmTomasevic},
	but a test of that idea needs a string scale that can be varied
	independently of \(G_N\), which this setup does not have.  Details are in
	the Supplemental Material (Sec.~\ref{S-sec:supp-wedge}).
	
	\section{An upper bound on the absorptive radius}
	\label{sec:absorptive-density-pilot}
	
	So far we have located the spectrum.  We now ask what it does.  The density
	fixes only the real part of the partial wave.  With
	\(q_J=\operatorname{Im}S_J\), the absorption probability is
	\begin{equation}
		\begin{split}
			\mathcal A_J&=1-|S_J|^2
			=2\rho_J-\rho_J^2-q_J^2
			=\mathcal A_{0,J}-q_J^2,\\
			\mathcal A_{0,J}&=2\rho_J-\rho_J^2.
		\end{split}
		\label{eq:zero-q-absorption-proxy}
	\end{equation}
	so \(\mathcal A_{0,J}\) is the largest absorption compatible with a given
	density, reached when \(q_J=0\).  When a bin at the cap has
	\(\mathcal A_{0,J}=0\): the band of \cref{sec:capped} is a reflector,
	\(S_J\simeq-1\), and cannot be a black disk whatever its location.
	
	A filled black disk absorbs strongly from its center to its edge.
	The inelastic cross section gives more weight to larger impact parameters,
	so even a ring around a reflective center can give a large cross section.
	To look for a filled core, we minimize the weighted mean of
	\((\rho_J-1)^2\) inside a chosen region. This favors \(\rho_J=1\) in
	each bin, where complete absorption is possible.
	
	For the radius bound we require high absorption throughout a trial disk.
	The unitarity disk
	\((1-\rho_J)^2+q_J^2\le1\) projects onto the interval \(0\le\rho_J\le2\).
	Any spectrum that satisfies full partial-wave unitarity and the sampled sum
	rules is therefore a feasible point of our density-only program, and the
	program is an outer relaxation of the unitary problem.  Suppose we ask the
	relaxed program to make \(\mathcal A_{0,J}\ge\eta\) everywhere inside some
	region while keeping \(Y\) within a fraction \(\epsilon\) of its optimum,
	and it cannot.  Then no unitary completion can either, because its
	absorption is no larger than \(\mathcal A_0\) and its density is a feasible
	point.  The largest region for which the relaxed program succeeds is an
	upper bound on the absorptive region of every unitary completion of these
	spectra within \(\epsilon\) of the boundary.  We don't need \(q_J\) for this statement.
	
	The bound uses the eikonal density prescribed in \cref{eq:eikonal-active}
	and the adjustable spectrum above \(s=\ME^2\). The discontinuity from
	massless intermediate states below that threshold is part of the low-energy
	matching. At \(\lambda\to0\) the prescribed phase range supplies
	\(F_{\rm eik}(0^+)=\alpha_E=0.97948\) of the forward pole. The contribution
	of any fixed residual grid to \(\lambda\Ktwo\) vanishes in this limit,
	which explains the endpoint residual in the numerical checks.

	\subsection{The absorptive core and its energy dependence}
	\label{sec:absorptive-radius-bound}
	
	\begin{figure*}[t]
		\centering
		\includegraphics[width=0.90\textwidth]{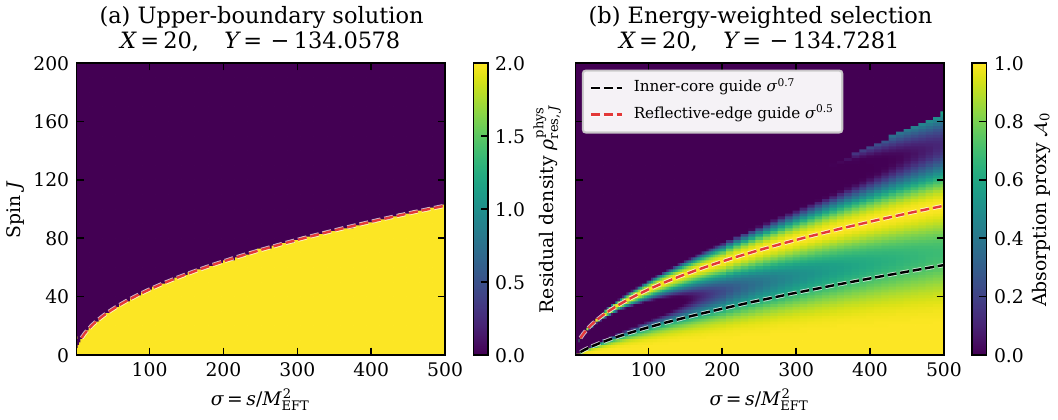}
		\par\smallskip
		\includegraphics[width=0.90\textwidth]{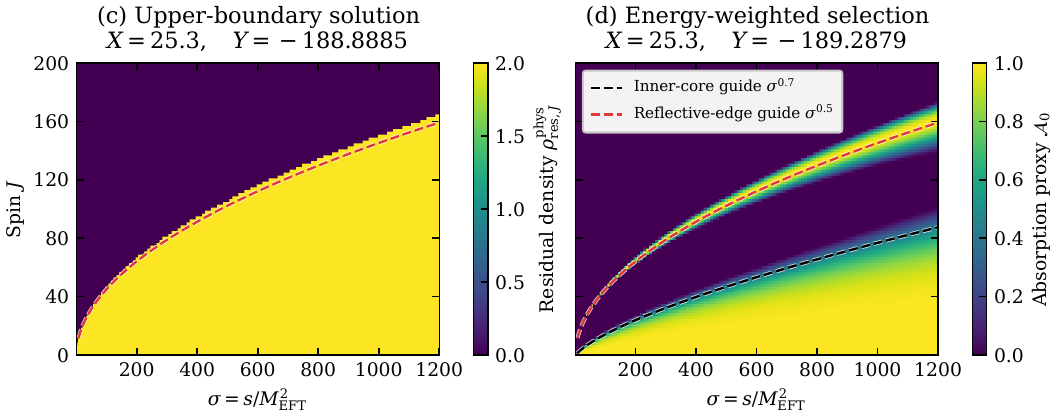}
		\caption{Spectra at \(\GN=4\pi^2\) on the refined grid. Bins with a
			prescribed eikonal contribution are masked in the background color.
			(a) Residual density on the upper boundary at \(X=20\).
			(b) Absorption proxy \(\mathcal A_0\) after the energy-weighted \(L^2\) selection at
			the same \(X\), with \(Y\) lowered by \(0.5\%\) of \(|Y_{\max}|\).
			(c) Residual density on the upper boundary at \(X=25.3\), near the
			tip of the allowed region. (d) The absorption proxy \(\mathcal A_0\)
			after the energy-weighted selection at the same \(X\). The bright core
			and the thin outer rim are separated by a reflective strip.
			The black dashed guides in (b) and (d), \(J+3/2\propto\sigma^{0.7}\),
			are anchored at the first outward crossing of \(\mathcal A_0=1/2\)
			at \(\sigma=102\) in each selected spectrum. The red dashed guide
			in all four panels, \(J+3/2\propto\sigma^{1/2}\), is at fixed impact
			parameter \(b\simeq9.28\,M_{\rm EFT}^{-1}\).}
		\label{fig:absorptive-core-selection}
	\end{figure*}
	
	\begin{figure*}[t]
		\centering
		\includegraphics[width=\textwidth]{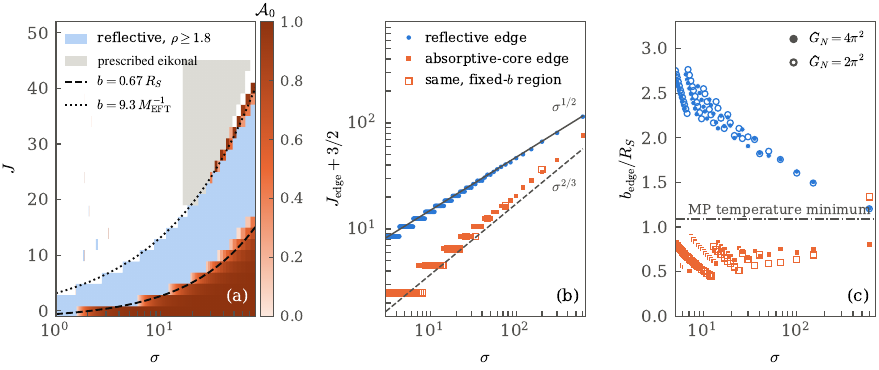}
		\caption{The main result.  (a) The original \(L^2\) selection on the
			\((600,240,80)\) grid, on a logarithmic energy axis.  Blue: bins that stay reflective
			(\(\rho\ge1.8\)).  Orange: the absorption proxy
			\(\mathcal A_0=2\rho-\rho^2\) of the remaining bins.  Gray: prescribed
			eikonal contribution.  An absorptive core forms inside the reflective band.  The thin line of
			orange bins at the outer tip of the band is not a second absorptive
			structure but the softened edge of the band: there the density
			passes from \(2\) to \(0\) through \(\rho=1\), at the fixed impact
			parameter \(b=9.55\,M_{\rm EFT}^{-1}\) for all \(\sigma\ge35\).  The
			extremal solution has a sharp step at this edge, and the step survives
			when the objective region does not reach the tip.  The dashed
			curve is \(b=0.67R_S\); the dotted curve is the fixed impact parameter
			\(b=9.3\,M_{\rm EFT}^{-1}\).  (b) The two edges against energy.  The
			reflective edge follows \(\sigma^{1/2}\), a fixed impact parameter; the
			edge of the absorptive core follows a steeper law close to
			\(\sigma^{2/3}\), the Schwarzschild radius.  Open squares: the core edge
			when absorption is requested on a region of fixed impact parameter, whose
			own shape is \(\sigma^{1/2}\).  (c) The same edges in units of \(R_S\), at
			\(\GN=4\pi^2\) (filled) and \(2\pi^2\) (open): the core edge is flat near
			\(0.6\)--\(0.7\), the reflective edge falls.  The line marks the impact
			parameter at which a Myers--Perry black hole of the same mass has its
			lowest temperature (\cref{sec:d10-sky-physical-absorption}).}
		\label{fig:core-vs-band}
	\end{figure*}
	
	Panels (a) and (b) of \cref{fig:absorptive-core-selection} compare the
	upper-boundary and absorption-selected spectra on a refined grid at
	\(X=20\). We give each even spin the same weight at fixed energy before
	tapering the objective smoothly from \(2.5R_S\) to \(3R_S\), and weight each
	energy by its cell width in \(1/\sigma\). Up to \(\sigma=20\), the first
	outward crossings of \(\mathcal A_0=0.5\) and \(0.9\) agree with those of
	the original selection to within one even-spin bin. The contrast between
	the core and the outer transition weakens at higher energies.
	
	Panels (c) and (d) of \cref{fig:absorptive-core-selection} compare the
	two solutions at \(X=25.3\). The selected spectrum keeps a reflective
	strip to higher energies. Its thin outer rim lies near the sharp drop
	in the upper-boundary density. Across the rim the selected density falls
	from near the cap to near zero, passing through \(\rho_J=1\).
	Over \(610\le\sigma\le1200\), the two sides of this rim at
	\(\mathcal A_0=0.9\) give exponents \(p=0.49\) and \(0.51\) in
	\(J+3/2\propto\sigma^p\), consistent with a fixed impact parameter.
	The core edge at the same threshold gives \(p=0.68\), close to the
	Schwarzschild value \(2/3\).
	
	For the fits in \cref{fig:core-vs-band} we use the original \(L^2\)
	selection on the \((600,240,80)\) grid. At \(X=20\) and \(\GN=4\pi^2\)
	we allow \(Y\) to drop by \(0.5\%\) below
	\(Y_{\max}\) and minimize the mean squared distance of the density from
	\(\rho=1\) over the 1985 bins with \(b/R_S<2\), with all other bins held
	fixed.\footnote{This calculation starts from an upper-boundary solution in
		which the eikonal density is switched on smoothly between \(2R_S\) and \(3R_S\)
		(\(Y_{\max}=-133.99\), against \(-131.96\) for the sharp cut at
		\(3R_S\)), and allows a slack of \(10^{-2}\) in the normalized sum rules.
		The controls in \cref{tab:core-controls} impose them exactly.}
	\Cref{fig:core-vs-band}(a) shows this selection in terms of the
	absorption proxy \(\mathcal A_0\). An absorptive core, \(\rho\simeq1\),
	forms inside the reflective band, whose outer part stays at the cap.
	
	The core has an edge, which we define at each energy as the spin at which
	\(\mathcal A_0\) first drops below one half on the way out from \(J=0\).
	Its energy dependence is shown in \cref{fig:core-vs-band}(b), next to that
	of the reflective edge of \cref{sec:band-edge}.  The two are different.  The
	reflective edge has \(p=0.50\), a fixed impact parameter.  The core edge
	has \(p=0.70\) for \(10\le\sigma\le80\) and \(0.69\) for
	\(20\le\sigma\le200\), with a scatter of \(0.06\)--\(0.08\) in \(\log J\); in
	units of the Schwarzschild radius it stays between \(0.64\) and \(0.70\)
	from \(\sigma=5\) to \(75\) while the reflective edge falls from \(2.7\) to
	\(1.7\) [\cref{fig:core-vs-band}(c)].  Below \(\sigma\simeq10\) the core is
	one to three spins wide and the fitted exponent is dominated by the
	staircase; we do not use those energies.
	
	\begin{table}[t]
		\centering
		\caption{Exponent \(p\) of the absorptive-core edge,
			\(J_{\rm edge}+3/2\propto\sigma^p\), and its position in units of \(R_S\),
			for different couplings, norms and shapes of the region on which
			absorption is requested (\(\epsilon=0.5\%\) throughout).  The last line
			is the reflective band.  A gravitational radius has \(p=2/3\), a fixed
			impact parameter gives \(p=1/2\). The coupling column is
			\(\GN=G_N\ME^4\). In the \(L^1\) rows at \(\GN=4\pi^2\), the bins
			between the region and \(10R_S\) are free to readjust; in all other rows
			the exterior is frozen.}
		\small
		\begin{tabular}{llccc}
			\toprule
			\(\GN\) & norm, region & \(p_{10\text{--}80}\) & \(p_{20\text{--}200}\)
			& \(b_{\rm edge}/R_S\) at \(\sigma=20,\,75\)\\
			\midrule
			\(4\pi^2\) & \(L^2\), \(b<2R_S\) & 0.70 & 0.69 & 0.66,\ 0.70\\
			\(4\pi^2\) & \(L^1\), \(b<2R_S\) & 0.65 & 0.72 & 0.66,\ 0.61\\
			\(4\pi^2\) & \(L^1\), \(b\ME<5.3\) & 0.70 & 0.75 & 0.66,\ 0.70\\
			\(2\pi^2\) & \(L^2\), \(b<10R_S\) & 0.77 & 0.73 & 0.58,\ 0.67\\
			\(2\pi^2\) & \(L^1\), \(b<10R_S\) & 0.79 & 0.73 & 0.58,\ 0.56\\
			\(2\pi^2\) & \(L^1\), \(b\ME<12\) & 0.77 & 0.72 & 0.58,\ 0.67\\
			\midrule
			\(4\pi^2\) & reflective band & 0.51 & 0.51 & 2.09,\ 1.71\\
			\bottomrule
		\end{tabular}
		\label{tab:core-controls}
	\end{table}
	
	An obvious objection is that the region on which we asked for absorption,
	\(b<2R_S\), itself scales with \(R_S\), so that the exponent might have been
	put in by hand.  \Cref{tab:core-controls} answers it.  The exponent and the
	position of the edge are the same when the region is five times larger,
	\(b<10R_S\); when it is a region of fixed impact parameter, \(b\ME<5.3\) or
	\(b\ME<12\), whose own boundary scales as
	\(\sigma^{1/2}\); when the squared distance from \(\rho=1\) is replaced by
	the absolute distance, which turns the selection into a linear program;
	when the sum rules are imposed exactly; and at half the coupling.  In every
	case \(p\) lies between \(0.65\) and \(0.79\) and the edge between \(0.56\)
	and \(0.70\) Schwarzschild radii, against \(p=0.51\) for the reflective
	band.  The edge therefore follows an energy dependence close to that of the
	gravitational radius.\footnote{Enlarging the objective region from
	\(b<2R_S\) to \(b<10R_S\) leaves the edge near
	\((0.6\text{--}0.7)R_S\). The measured exponent is about \(0.70\)--\(0.75\)
	on windows above \(\sigma=20\). The core occupies at most fifteen spins
	below \(\sigma=75\), so these fits do not resolve a departure from
	\(2/3\).}

	\subsection{The bound}
	
	The selection above shows where absorption is cheap.  To turn it into a
	bound we fix a shape and maximize its size.  For a given loss in \(Y\),
	the fraction of its maximum that is given up,
	\begin{equation}
		\epsilon=\frac{Y_{\max}-Y_{\rm constrained}}{|Y_{\max}|}
		\label{eq:absorptive-radius-price}
	\end{equation}
	and a threshold \(\eta\) we look for the largest radius \(B_*\) such that
	the program remains feasible with
	\begin{equation}
		\begin{split}
			\mathcal A_{0,J}\geq\eta
			\;\Longleftrightarrow\;
			1-\sqrt{1-\eta}&\leq\rho_J\leq1+\sqrt{1-\eta},\\
			b/R_S&<B_*/R_S.
		\end{split}
		\label{eq:inverse-absorptive-radius}
	\end{equation}
	For fixed \(B_*/R_S\) this is a linear program.  By the argument at the start of
	this section, a unitary completion whose \(Y\) lies within \(\epsilon\) of
	\(Y_{\max}\) cannot absorb with probability \(\eta\) or more throughout a
	disk larger than \(B_*(\epsilon,\eta)\).
	
	\begin{figure*}[t]
		\centering
		\includegraphics[width=0.86\textwidth]{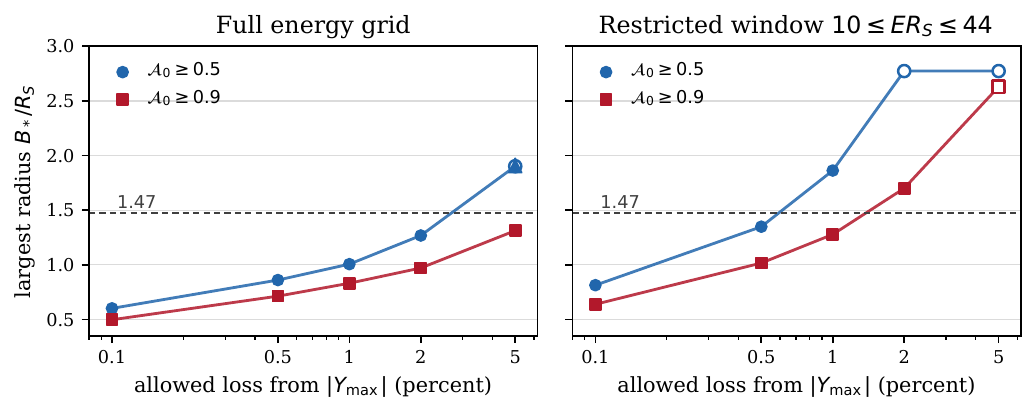}
		\caption{The bound: the largest radius \(B_*\), in units of \(R_S\), of a
			disk in which the absorption can reach \(\eta=0.5\) (blue) or \(0.9\) (red)
			everywhere, as a function of the fraction of \(Y_{\max}\) given up, at
			\(\GN=4\pi^2\), \(X=20\).  Left: the condition is imposed at all energies
			of the grid.  Right: only for \(10\le ER_S\le44\).  A unitary completion
			within that distance of the boundary cannot have a larger uniformly
			absorptive disk.  The dashed line is the classical apparent-horizon value
			\(1.47\)~\cite{YoshinoRychkov}, which is not an input; the
			Myers--Perry value discussed in \cref{sec:d10-sky-physical-absorption}
			is \(1.09\).  Open symbols: the density box prevents extending the
			disk further; the triangle is a lower bound.}
		\label{fig:inverse-absorptive-radius}
	\end{figure*}
	
	\Cref{fig:inverse-absorptive-radius} shows the result at \(\GN=4\pi^2\).
	At \(\epsilon=0.5\%\) and with the condition imposed at all energies,
	\(B_*=0.86R_S\) for \(\eta=0.5\) and \(0.71R_S\) for \(\eta=0.9\).  The bound
	loosens as more of \(Y\) is given up: reaching the classical apparent-horizon radius \(1.47R_S\) at
	every sampled energy of the original grid (\(\ME\le E\le\sqrt{600}\,\ME\))
	costs \(2.7\%\) at \(\eta=0.5\) and \(6.3\%\) at \(\eta=0.9\).  We can also readily see the energy dependence.  If
	absorption is required only for \(10\le ER_S\le44\), the radii at
	\(\epsilon=0.5\%\) are \(1.35R_S\) and \(1.02R_S\), and \(1.47R_S\) costs
	\(0.6\%\) and \(1.5\%\).  The reason is that a uniform disk is constrained
	mainly by its lowest energies, where each bin weighs most in the sum rule.
	We have examined this at \(\GN=2\pi^2\) by replacing the disk with regions
	\(J+3/2\le C\sigma^p\) (Supplemental Material,
	Sec.~\ref{S-sec:supp-absorb}).  For \(p\le2/3\) the largest admissible
	regions all pass close to \(b\simeq2\,M_{\rm EFT}^{-1}\) at
	\(\sigma\simeq2\): they are pinned at the lowest energies.  The shape that
	holds the most bins for a given loss is neither the fixed impact parameter
	(\(p=1/2\)) nor a Regge-like region (\(p=1\)), each of which holds about
	\(17\%\) fewer, but \(p\simeq0.75\)--\(0.85\), close to the exponent of the
	core selected above.  When the disk is imposed only in a window of energies
	a factor of three wide, the bound rises from \(1.2R_S\) around \(\sigma=7\)
	to \(2.3R_S\) above \(\sigma\simeq40\).  The bound is therefore a family.  For losses of a percent or less and any
	reasonable threshold, a disk imposed at all energies has a radius between
	\(0.7\) and \(1.5\) Schwarzschild radii, and this scale is set at the lowest
	energies of the grid.
	
	We also vary \(X\) over the entire leaf at \(\GN=4\pi^2\).
	At each \(X\) we use the same absolute allowance,
	\(\Delta Y=Y_{\max}(X)-Y\le0.66994\), equal to the \(0.5\%\)
	allowance at \(X=20\). With \(\mathcal A_0\ge0.9\) throughout the disk
	at every sampled energy, the largest radius admitted by the density
	program increases from \(0.71R_S\) at \(X=20\) to \(0.9782R_S\)
	over the whole leaf. A spectrum at \(X\simeq-10.87\) attains the limiting radius in the
	density program, and the certificate in Supplemental
	Sec.~\ref{S-sec:whole-leaf-bound} excludes any larger disk across the
	full range of \(X\) under the same allowance.

	\subsection{A disk of the classical trapped-surface radius}
	\label{sec:ceiling-floor}
	
	\begin{figure*}[t]
		\centering
		\includegraphics[width=0.92\textwidth]{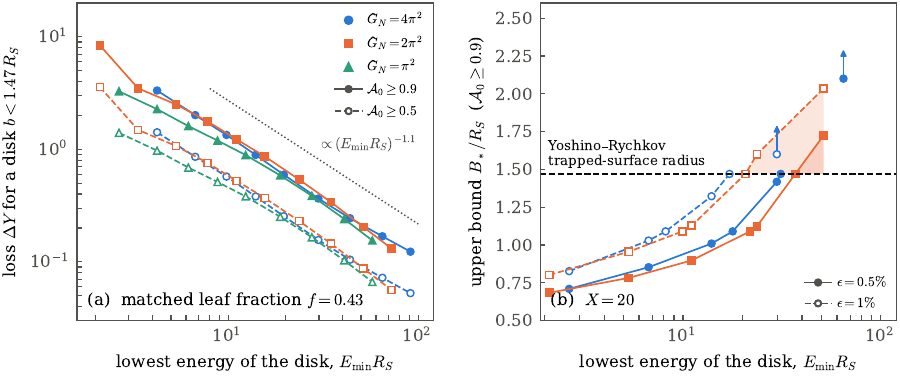}
		\caption{A disk of the classical trapped-surface radius as a test.
			(a) The loss \(\Delta Y=Y_{\max}-Y_{\rm constrained}\) required by a disk
			\(b<1.47R_S\), the classical trapped-surface radius in \(D=6\), in which
			the absorption reaches \(0.9\) (filled symbols) or \(0.5\) (open) at all
			energies of the grid above \(E_{\min}\).  The three couplings are compared
			at the same leaf fraction, \(f=0.43\) (\cref{eq:leaf-fraction}), and
			approximately collapse onto one curve in \(E_{\min}R_S\) (rms scatter
			\(9\%\)); the dotted line is the slope of
			\cref{eq:price-of-classical-disk}.
			(b) The upper bound \(B_*/R_S\) for a disk with \(90\%\) absorption
			above \(E_{\min}\), at \(X=20\) and losses of \(0.5\%\) (filled) and
			\(1\%\) (open) of \(|Y_{\max}|\), for \(\GN=4\pi^2\) (circles) and
			\(2\pi^2\) (squares).  Arrows mark scans that reached only a lower
			limit on \(B_*\).  At impact parameters below the dashed line the
			Yoshino--Rychkov construction gives a classical trapped
			surface~\cite{YoshinoRychkov}.  The shading shows the interval
			\(1.47\le B_{\rm abs}/R_S\le B_*/R_S\) for the physical disk radius
			\(B_{\rm abs}\), assuming trapped-surface formation implies at least
			\(90\%\) absorption throughout the sampled energy range.  Shading
			stops at the last resolved point of each curve.  Where
			\(B_*<1.47R_S\), a uniformly
			\(90\%\)-absorptive disk of the classical radius is excluded.}
		\label{fig:ceiling-floor}
	\end{figure*}
	
	In \(D=6\), the construction of Yoshino and
	Rychkov~\cite{YoshinoRychkov} gives a classical apparent horizon for
	\(b<1.47R_S\).  The construction thus places a lower bound on the
	largest impact parameter for classical horizon formation.
	
	For a unitary amplitude within the allowed loss in \(Y\), let
	\(B_{\rm abs}/R_S\) be the largest radius in Schwarzschild units for which
	\(1-|S_J|^2\ge0.9\) throughout the disk at every sampled energy above
	\(E_{\min}\).  The bootstrap gives
	\(B_{\rm abs}/R_S\le B_*/R_S\) because it tests the largest absorption
	compatible with each \(\operatorname{Re}S_J\). We make the additional
	semiclassical assumption that trapped-surface formation reduces the
	elastic return probability to \(|S_J|^2\leq 0.1\) throughout the energy
	range considered. Under this assumption, the radius of a uniformly
	\(90\%\)-absorptive disk satisfies \(B_{\rm abs}/R_S\geq 1.47\).
	This is a semiclassical assumption motivated by
	\(ER_S\gg1\).  The shading in \cref{fig:ceiling-floor}(b) shows the
	resulting interval along the resolved curves.  Whether a unitary amplitude
	realizes a radius in that interval also depends on \(\operatorname{Im}S_J\).
	In \cref{fig:ceiling-floor}(a) we impose the necessary condition
	\(\mathcal A_0\ge0.9\) out to \(1.47R_S\) at all sampled energies above
	\(E_{\min}\) and compute the loss in \(Y\).
	
	The loss falls as a power of the lowest energy.  At \(\GN=4\pi^2\) and
	\(X=20\), the point of \cref{fig:inverse-absorptive-radius}, it is
	\(1.8\%\) of \(Y_{\max}\) for \(E_{\min}R_S=10\), \(0.53\%\) for \(30\) and
	\(0.12\%\) for \(91\).
	
	A percentage of \(Y_{\max}\) cannot be compared between couplings, or
	between points of the leaf, because \(Y_{\max}\) changes sign along the
	upper boundary.  What can be compared is the loss itself,
	\(\Delta Y=Y_{\max}-Y_{\rm constrained}\), in the units of
	\cref{eq:k2-main} in which the graviton pole is one, taken at the same
	position on the leaf.  We located the two tips of the leaf at
	\(\GN=4\pi^2\), \(2\pi^2\) and \(\pi^2\) on the same grid.  The point
	\(X=20\) lies at leaf fractions \(f=0.85\), \(0.43\) and \(0.28\)
	respectively (\cref{eq:leaf-fraction}), three quite different places, so a
	comparison at \(X=20\) does not compare like with like.  We therefore
	compare the three couplings at the same leaf fraction, \(f=0.43\), which
	means different values of \(X\): \(X=4.25\), \(20\) and \(39.27\) at
	\(\GN=4\pi^2\), \(2\pi^2\) and \(\pi^2\).  At these three points the losses
	approximately collapse for \(8\le E_{\min}R_S\le91\), although the coupling
	changes by a factor of four (\cref{fig:ceiling-floor}a).  A single power law fits them with an rms
	scatter of \(9\%\) (extremes \(-16\%\) and \(+14\%\)):
	\begin{equation}
		\Delta Y\simeq0.58\left(\frac{20}{E_{\min}R_S}\right)^{1.10}.
		\label{eq:price-of-classical-disk}
	\end{equation}
	With the exponent fixed to one the scatter is \(11\%\).  With the lower
	threshold \(\mathcal A_0\ge0.5\) the loss is smaller by a factor of
	\(2.33\)--\(2.34\).  This is the ratio of the amounts \(1-\sqrt{1-\eta}\) by
	which the two thresholds tighten the density box at each end, so the loss responds almost linearly to the tightening.  At \(f=0.28\) the prefactor is
	\(0.46\) and the three couplings agree only within \(40\%\), with exponents
	between \(0.85\) and \(1.15\), and the three references do not share the
	same eikonal cut (Supplemental Material, Sec.~\ref{S-sec:supp-matched}).  We
	therefore claim an approximate collapse at this leaf fraction over the
	sampled range, in which the coupling enters mainly through
	\(E_{\min}R_S\).
	
	Read the other way, the largest black disk admitted for a fixed loss grows
	with \(E_{\min}\).  At \(X=20\) and half a percent of \(Y_{\max}\) from
	the boundary, the disk of the classical radius is excluded at
	\(E_{\min}R_S=29.8\) and admitted at \(43.8\) for \(\GN=4\pi^2\), and
	excluded at \(34.7\) and admitted at \(51.4\) for \(2\pi^2\); interpolating
	the loss between these trials puts the crossings at about \(31\) and
	\(37\), and at one percent at about \(17\) and \(21\)
	(\cref{fig:ceiling-floor}b).  The exclusions at the
	lower energy of each pair have been confirmed with a dual functional
	evaluated on every bin of the grid (Supplemental Material,
	Sec.~\ref{S-sec:supp-floor}).  Below these energies a completion
	that close to the boundary cannot absorb like a classical black hole out to
	the trapped-surface radius.  Above them the density constraints no longer
	forbid it, and the largest admitted disk keeps growing, to \(1.7R_S\) at
	\(E_{\min}R_S=51\) for \(2\pi^2\) and beyond \(2.1R_S\) at \(65\) for
	\(4\pi^2\).  Whether such a disk is realized is not decided by the density.
	
	Two remarks.  First, the loss is nonzero for every \(E_{\min}\) we have
	tried: a uniformly absorptive disk of this size lies strictly inside the
	boundary, at a distance that decreases with energy over the sampled range.
	This fits the earlier sections, where the
	extremal spectra were the reflective baseline of the capped problem.
	Second, in \(D=6\) the Bekenstein--Hawking entropy is
	\(S_{\rm BH}=\pi ER_S\), so \cref{eq:price-of-classical-disk} is close to
	an inverse power of the entropy of the smallest black hole that is asked
	for.  For a disk of the Myers--Perry radius \(1.09R_S\) of
	\cref{app:rotating-bh} and \(X=20\) the exponent is \(1.01\) and \(0.96\) at the two
	strong couplings, with \(\epsilon\,E_{\min}R_S\simeq0.09\) and \(0.10\).  We record this and do not
	interpret it.
	
	\subsection{Weak coupling}
	\label{sec:weak-absorptive-scaling}
	
	In the physical hierarchy the core at low impact parameter is empty in the
	reference solutions, not reflective, and the bound constrains how far it
	can be filled.  We repeat the construction at fixed \(\epsilon=0.5\%\) for
	\begin{equation}
		\GN=\frac{\pi^2}{r},\qquad
		r\in\{1000,5000,6000,6500,7500,8000\},
		\label{eq:weak-gravity-coupling-set}
	\end{equation}
	that is \(0.42\le M_{\rm EFT}/M_{\rm Pl}\le0.71\), on a \((600,240)\) grid
	adapted to cover the shell \(b\sim R_S\) (a fifth of the nodes below
	\(\sigma=32\), most of the rest logarithmically spaced up to
	\(2.5\times10^5\)), with the eikonal contribution at \(b\ge4R_S\).  Moving the eikonal contribution to
	\(3R_S\) changes the radius by \(0.1\%\), and a \((900,320)\) grid changes it
	by \(0.4\%\).
	
	\begin{table}[t]
		\centering
		\caption{Fits of the bound at weak coupling,
			\(B_\eta/R_S=F_\eta(M_{\rm EFT}/M_{\rm Pl})^{a_\eta}\), to the six
			couplings of \cref{eq:weak-gravity-coupling-set} at
			\(\epsilon=0.5\%\).}
		\small
		\begin{tabular}{cccc}
			\toprule
			\(\eta\) & \(F_\eta\) & \(a_\eta\) & \(R^2_{\log}\)\\
			\midrule
			\(0.4\) & \(2.155\pm0.004\) & \(-0.430\pm0.003\) & \(0.99985\)\\
			\(0.5\) & \(1.998\pm0.004\) & \(-0.452\pm0.003\) & \(0.99985\)\\
			\(0.6\) & \(1.901\pm0.005\) & \(-0.450\pm0.004\) & \(0.99973\)\\
			\(0.7\) & \(1.830\pm0.006\) & \(-0.439\pm0.005\) & \(0.99958\)\\
			\(0.8\) & \(1.763\pm0.005\) & \(-0.431\pm0.004\) & \(0.99965\)\\
			\(0.9\) & \(1.693\pm0.006\) & \(-0.423\pm0.005\) & \(0.99955\)\\
			\bottomrule
		\end{tabular}
		\label{tab:weak-fits}
	\end{table}
	
	\begin{figure}[t]
		\centering
		\includegraphics[width=0.92\columnwidth]{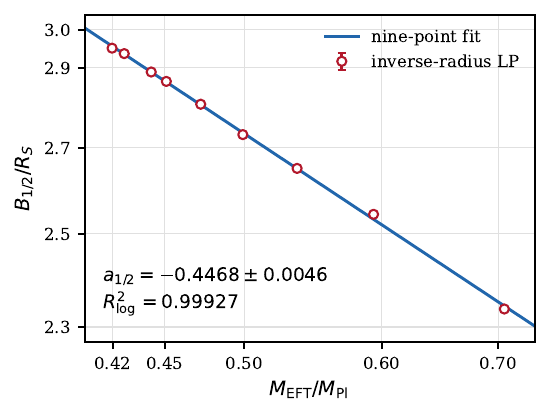}
		\caption{The bound at weak coupling: \(B_{1/2}/R_S\) at
			\(\epsilon=0.5\%\) for nine couplings \(\GN=\pi^2/r\),
			\(1000\le r\le8000\), against \(M_{\rm EFT}/M_{\rm Pl}\), with the
			power-law fit of \cref{eq:weak-gravity-nine-point-midpoint-fit}.  A purely
			gravitational radius would be a horizontal line and a fixed impact
			parameter a line of slope \(-4/3\).  The slope found is \(-0.45\).}
		\label{fig:weak-nine-eta05-power-fit}
	\end{figure}
	
	The radii follow a power law,
	\begin{equation}
		\frac{B_\eta}{R_S}
		=F_\eta
		\left(\frac{M_{\rm EFT}}{M_{\rm Pl}}\right)^{a_\eta}.
		\label{eq:weak-gravity-absorption-scaling}
	\end{equation}
	to better than \(0.3\%\) at every point (\cref{tab:weak-fits}). In
	physical units at fixed \(\sigma\), the same fit reads
	\[
	  B_\eta(\sigma)
	  =F_\eta\left(\frac{3}{16\pi^2}\right)^{1/3}
	    \frac{\sigma^{1/6}}{\ME}
	    \left(\frac{\ME}{\MP}\right)^{4/3+a_\eta}.
	\]
	Here \(8\pi G_N=\MP^{-4}\). The exponent \(4/3+a_\eta\) is positive for every fitted threshold. The
	fit therefore extrapolates to \(B_\eta\to0\) as \(G_N\to0\) at fixed
	\(\ME\) and \(\sigma\), even as \(B_\eta/R_S\) increases. With three
	more couplings at \(\eta=0.5\), \cref{fig:weak-nine-eta05-power-fit},
	\begin{equation}
		a_{1/2}=-0.4468\pm0.0046,
		\qquad R^2_{\log}=0.99927,
		\label{eq:weak-gravity-nine-point-midpoint-fit}
	\end{equation}

	The exponents for all thresholds lie between \(-0.45\) and \(-0.42\).  Two
	reference values frame this number.  A radius set by gravity alone would
	make \(B_\eta/R_S\) independent of the coupling, exponent \(0\).  A fixed
	impact parameter in cutoff units, like the edge of the reflective band,
	would give \(B_\eta/R_S\propto(M_{\rm EFT}/M_{\rm Pl})^{-4/3}\).  The
	measured exponent is one third of the way from the first to the second.
	It is close to \(-4/9\), which is what a radius of the form
	\(R_S^{2/3}b_0^{1/3}\), with \(b_0\) a fixed length in cutoff units, would
	give.  The errors quoted are regression errors only: the exponent moves by
	\(0.03\) between thresholds (\cref{tab:weak-fits}) and also depends on the
	grid and on the reference solution at each coupling.  We have no derivation
	of this form and record
	it as an empirical regularity of the fitted interval; classical gravity at
	fixed dimension predicts a coupling-independent ratio, so the exponent
	measures how far these weak-coupling spectra are from that regime.  The
	half-absorption radius of the density selection of
	\cref{sec:absorptive-radius-bound}, followed from \(\GN=\pi^2/20\) to
	\(\pi^2/300\), moves in the same direction with exponent \(-0.50\).
	
	\subsection*{A check with graviton loops}
	\label{sec:weak-gravity-loops}
	As a check of the weak-coupling radius bound, we repeated the
	\(\eta=1/2\) calculation at \(X=20\) and \(\GN=\pi^2/r\) for
	\(r=1000\) and \(8000\), including the pure-gravity one-loop term
	in Eq.~(B.4) of Ref.~\cite{ChangParraMartinez} in the low-energy
	sum rule, with matching and renormalization scales both set to
	\(M_{\rm EFT}\). The loop term shifted the unconstrained and
	disk-constrained maxima of \(Y\) by the same amount within
	numerical accuracy, so the loss \(\Delta Y\) was unchanged.
	With the loss allowance fixed at \(0.5\%\) of the tree-level
	\(|Y_{\max}|\), the radius bound stayed at the same grid point
	in each case, with \(B_{1/2}/R_S\simeq2.34\) and \(2.95\), respectively.
	
	\subsection{Comparison with a full-amplitude bootstrap in \(D=9\)--\(11\)}
	\label{sec:d10-sky-physical-absorption}
	
	The absorption proxy in \cref{fig:absorptive-core-selection}(d) shows where
	the density allows strong absorption. To see how much is actually absorbed,
	we need the full partial wave \(S_J\). The string-inspired bootstrap of
	the maximally supersymmetric graviton amplitude of
	Refs.~\cite{GuerrieriPenedonesVieiraWhereString,
		GuerrieriMuraliPenedonesVieiraWhereM}, carried to high degree and spin
	cutoffs of several hundred in \(D=9\), \(10\) and \(11\)
	\cite{DasMyersSinhaBHScales}.  There the full amplitude is constructed, the
	leading higher-derivative coefficient is minimized, and both
	\(\operatorname{Re}S_J\) and \(\operatorname{Im}S_J\) are available.  That
	calculation imposes full partial-wave unitarity, \(|S_J|\le1\) for the
	complex \(S_J\).  The present paper imposes only positivity of the density
	with its cap, \(0\le\rho_J\le2\), which is the projection of that condition
	onto \(\operatorname{Re}S_J\).  The
	dimension, the external states and the objective all differ from ours, so
	only locations and scalings can be compared, not heights.  Four points of
	contact are worth recording.
	
	\begin{table}[t]
		\centering
		\caption{Energy exponent \(p\) of structures at low impact parameter,
			\(J\propto s^{p}\).  A length proportional to the Schwarzschild radius
			gives \((D-2)/[2(D-3)]\); a fixed impact parameter gives \(1/2\) in any
			dimension.  The \(D=9\)--\(11\) values are from the collapse of the
			absorption map in Ref.~\cite{DasMyersSinhaBHScales} and are estimates
			in the basis and energy windows used there, with a systematic uncertainty
			of about \(0.1\) from the truncation degree.}
		\small
		\begin{tabular}{cccc}
			\toprule
			\(D\) & \(\dfrac{D-2}{2(D-3)}\) & measured & structure\\
			\midrule
			6 & 0.667 & 0.70--0.75 & absorptive core\\
			6 & 0.667 & 0.50--0.51 & reflective band\\
			9 & 0.583 & 0.58 & absorption map\\
			10 & 0.571 & 0.64 & absorption map\\
			11 & 0.563 & 0.60 & absorption map\\
			\bottomrule
		\end{tabular}
		\label{tab:exponents-D}
	\end{table}
	
	\emph{The eikonal region.}  In \(D=9\)--\(11\) the phase of the extremal
	amplitude at large spin follows the tree-level eikonal to three percent for
	\(b\ge4R_S\) and departs from it at \((2.5\text{--}3)R_S\), where the
	inelasticity sets in.  The cut \(r_{\min}=3\) of \cref{eq:eikonal-active} is
	a conservative edge of the eikonal regime and does not hide structure.
	
	\emph{The energy exponent.}  The landmarks of the \(D=9\)--\(11\) absorption
	maps collapse when the spin is scaled as \(s^{p}\) with \(p\) close to
	\((D-2)/[2(D-3)]\), the exponent of \(ER_S(E)\); a fixed spin and a fixed
	length are both disfavored.  \Cref{tab:exponents-D} puts our numbers next
	to these.  The same test, applied in \(D=6\), separates the absorptive core,
	which follows the gravitational exponent, from the reflective band, which
	follows \(1/2\).  The natural variable in both calculations is the energy in
	units of the Schwarzschild--Tangherlini temperature,
	\(E/T_0=4\pi ER_S/(D-3)\).  Our window \(10\le ER_S\le44\) is
	\(42\le E/T_0\le184\); the \(D=9\)--\(11\) maps are analyzed at
	\(E/T_0\simeq200\)--\(350\).
	
	\emph{The radius.}  In \(D=9\)--\(11\) the first absorptive band inside
	the principal elastic recovery sits at \(b\simeq0.30/T_0\) in all three
	dimensions, that is \(0.63\), \(0.54\) and \(0.47\,R_S\).  This is close to
	the impact
	parameter at which a singly spinning Myers--Perry black hole with the mass
	and angular momentum of the collision has its lowest temperature
	(\(bT_0=0.287\), \(0.291\), \(0.294\); \cref{app:rotating-bh}).  The
	Myers--Perry black hole is a reference for the scale here, not a claim
	that a stationary black hole forms, and the agreement, a few percent in
	the amplitudes of Ref.~\cite{DasMyersSinhaBHScales}, is sensitive at the
	ten-percent level to the basis in which the amplitude is expanded.  The same rule in \(D=6\) gives
	\(bT_0=0.26\), or \(b=1.09R_S\).  Our bound allows a uniformly absorptive
	disk of \(0.7\)--\(0.9R_S\) at all energies and \(1.0\)--\(1.35R_S\) in the
	window \(10\le ER_S\le44\), at half a percent from the boundary, and the
	absorptive core selected in \cref{sec:absorptive-radius-bound} ends at
	\(0.6\)--\(0.7R_S\).  All of these, and the trapped-surface value
	\(1.47R_S\), are within a factor of two of one another.  The dispersive
	bound, the classical construction and the rule abstracted from the
	higher-dimensional amplitudes single out the same scale.
	
	\emph{The layering.} Inside the capture radius of the \(D=9\)--\(11\)
	amplitudes, moving outward takes us through an absorptive band, a region of
	little absorption where \(S_J\) returns close to \(-1\), and a broad outer
	absorptive band before the eikonal region. In the full nonlinear unitarity
	condition, \(q_J=\operatorname{Im}S_J\) can lower the absorption below
	\(\mathcal A_{0,J}\), even where \(\rho_J\) is close to one.
	This lets a region of weak absorption separate two strongly absorptive
	bands. We will discuss the physics of these bands in
	Ref.~\cite{DasMyersSinhaBHScales}.
	
	The reflective region in the \(D=9\)--\(11\) amplitudes follows a
	gravitational length scale. The cap-saturated band in our density
	calculation has an outer edge at fixed impact parameter and survives at
	\(G_N=0\). The thin rim in \cref{fig:absorptive-core-selection}(d) follows
	that edge. Our radius bound limits the size of a uniformly absorptive disk.
	A separate absorptive band beyond the disk can coexist with the bound.
	
	\section{Discussion}
	\label{sec:discussions}
	
	We set out to learn where the positive spectrum of a gravitational EFT goes
	once the eikonal contribution of the graviton pole is in place.  The answer has
	two parts.  The spectrum that the capped dispersive problem selects by
	itself is a reflective band at low impact parameter.  Its edge
	sits at a fixed impact parameter, the kernel of the dispersion relation
	explains why, and the same band is there at \(G_N=0\).  The spectrum that
	scales with the gravitational radius has to be asked for.  When we request
	absorption near the extremal boundary, an absorptive core forms inside the
	band; its edge grows with energy with an exponent of \(0.70\)--\(0.75\),
	against \(2/3\) for the Schwarzschild radius and \(1/2\) for the band, and
	the largest uniformly absorptive disk that the constraints admit has a
	radius of \(0.7\)--\(1.5R_S\).  Because the density-only problem is a
	relaxation of unitarity, this radius is an upper bound for any unitary
	completion of these spectra.  The classical trapped-surface radius, the
	Myers--Perry scale suggested by the amplitudes in \(D=9\)--\(11\), and the
	dispersive bound all fall within a factor of two of each other.
	The Yoshino--Rychkov radius is a lower bound on the largest impact
	parameter for classical horizon formation.  Our dispersive result is an
	upper bound on the radius of a uniformly absorptive disk.  Near the boundary
	a disk of the classical size is excluded when it must absorb uniformly
	down to low energy.  The exclusion
	is lifted above \(E_{\min}R_S\simeq20\)--\(50\), and the loss in \(Y\) that
	the disk requires falls roughly as \((E_{\min}R_S)^{-1.1}\) and, in units of the graviton
	pole and at the matched position we tested, depends on the coupling
	approximately only through \(E_{\min}R_S\).  Whether a unitary amplitude
	realizes such a disk is a question involving the phase as well.
	
	Several things would sharpen this picture.
	
	\emph{The imaginary part.}  The bound becomes a statement about the
	absorptive region itself only when \(q_J=\operatorname{Im}S_J\) is
	reconstructed and the full disk \((1-\rho_J)^2+q_J^2\le1\) is imposed.  The
	dispersive real part is linear in the densities, so each sampled disk is a
	second-order cone and the problem remains convex.  We have tried this on a
	\((120,60,16)\) grid at \(\GN=4\pi^2\), \(X=20\), with the disk imposed at
	five energies for \(J=2\).  \(Y\) changes by \(3\times10^{-6}\), the support
	and the 474 cap-saturated bins are unchanged, and \(q_2=0\) is forced
	wherever \(\rho_2=2\), which confirms that the band is reflective.  Between
	the sampled energies, however, the reconstructed \(q_J\) receives
	contributions of order \(10^2\) from high-spin densities that are
	individually below \(10^{-8}\).  A usable constraint will need to be smeared
	in energy and to control the high-spin tail.  With a complex amplitude in
	hand one could also look for Ericson-type fluctuations
	\cite{EricsonFluctuations,EricsonTheoryFluctuations} at the rotating
	black-hole scale.
	
	\emph{The massless cut.}  The free spectrum here begins at \(\sigma=1\),
	while massless states contribute to the cut below it.  The weak-gravity
	numbers are therefore statements about the capped problem above the EFT
	threshold and not yet EFT bounds.  The correct treatment keeps the computed
	low-energy discontinuity fixed, bounds its unresolved remainder by a
	perturbative envelope, and reserves the cap for \(\sigma\ge1\).  The
	dispersive split is set up in the Supplemental Material
	(Sec.~\ref{S-app:eft-matching}). The pure-gravity
	\hyperref[sec:weak-gravity-loops]{one-loop check} tests one contribution to
	this matching. An optimization including the unresolved low-energy
	remainder will be reported elsewhere~\cite{AthiraSahaSarenSinha}.
	
	\emph{Other dimensions.}  The exponent test is cheap and makes a prediction:
	in \(D=5\) and \(D=7\) the absorptive core should follow \(p=3/4\) and
	\(5/8\), while the reflective band stays at \(1/2\).  Together with
	\cref{tab:exponents-D} this would show whether the formula
	\((D-2)/[2(D-3)]\) holds across dimensions and across two very different
	bootstraps.  A string scale that can be varied independently of \(G_N\)
	would, in the same spirit, turn the linear lower edge of
	\cref{sec:matched-high-energy-null} into a test of the string/black-hole
	correspondence.
	
	The lesson we draw is that positivity and unitarity of the density alone
	already know about the Schwarzschild radius, but only as a bound.  Whether
	the bound is saturated, and by what, is a question for the full unitarity
	disk.
	
	\begin{acknowledgments}
		
		We thank Arhum Ansari, P. Athira, Faizan Bhat, Sumit R. Das, Saptaswa Ghosh, Apratim Kaviraj, Arnab P. Saha, Soumen Saren and Ahmadullah Zahed for discussions.
		The numerical coding and draft-preparation workflow made use of AI coding and
		reasoning assistants, including OpenAI Codex and ChatGPT and Anthropic's Claude, under human supervision.  All
		physics assumptions, numerical choices, and conclusions are the responsibility
		of the authors.  We are supported by ANRF grant ANRF/ARG/2025/001338/PS.  AS
		also acknowledges support from a Quantum Horizons Alberta senior fellowship.
		DD acknowledges support from the P.~K.~Kelkar Fellowship at IIT Kanpur.
	\end{acknowledgments}
	
	\appendix
	
	\section{Rotating black-hole reference scales}
	\label{app:rotating-bh}
	
	The curve \(J_{\rm BH}^{(\kappa)}\) drawn in the figures is the estimate of
	Giddings and Porto~\cite{GiddingsPorto} for the impact parameter below which
	a rotating black hole of the collision's mass and angular momentum can
	form.  With the rotating radius \(R_J\) defined by
	\begin{equation}
		R_J^{D-5}
		\left[
		R_J^2+\frac{(D-2)^2J^2}{4\ME^2\sig}
		\right]
		=
		\frac{16\pi G_N\ME\sqrt{\sig}}{(D-2)\Omega_{D-2}} .
		\label{eq:gp-radius}
	\end{equation}
	and \(b=2(J+3/2)/(\ME\sqrt\sigma)\) in \(D=6\), the condition \(b=\kappa R_J\) is
	\begin{equation}
		(J+\tfrac32)
		\left[
		(J+\tfrac32)^2+\kappa^2J^2
		\right]
		=
		\frac{3\GN}{16\pi}\kappa^3\sig^2,
		\label{eq:rotating-bh-factorized}
	\end{equation}
	which is the cubic \cref{eq:rotating-bh-main}.  At large spin
	\begin{equation}
		J_{\rm BH}^{(\kappa)}(\sig)
		\sim
		\left[
		\frac{3\GN\kappa^3}{16\pi(1+\kappa^2)}
		\right]^{1/3}\sig^{2/3}.
		\label{eq:rotating-bh-asymptotic}
	\end{equation}
	We use \(\kappa=3\).  Neither this curve nor the simpler ruler \(b/R_S\) is
	a threshold: trapped-surface constructions give lower bounds on black-hole
	formation with order-one
	uncertainties~\cite{EardleyGiddingsBH,GiddingsRychkov,YoshinoNambu,YoshinoRychkov},
	and rotation adds another factor of order one.
	
	The second reference scale, used in
	\cref{sec:d10-sky-physical-absorption}, comes from the Myers--Perry
	solution with a single spin~\cite{MyersPerry}.  With horizon radius
	\(r_+\) and rotation parameter \(a\),
	\begin{equation}
		\begin{split}
			r_+^{D-5}\left(r_+^2+a^2\right)&=R_S^{D-3},\\
			T&=\frac{1}{4\pi}\left[\frac{2r_+}{r_+^2+a^2}+\frac{D-5}{r_+}\right],
		\end{split}
		\label{eq:mp-temperature}
	\end{equation}
	and \(J=2Ea/(D-2)\), so that the impact parameter of a collision that
	deposits all of its energy and angular momentum is \(b=2J/E=4a/(D-2)\).  At
	fixed mass \(T\) first decreases with \(a\) and, for \(D\ge6\), increases
	again at large spin.  Its minimum lies at \(a/R_S=1.09\), \(1.05\), \(1.04\)
	and \(1.04\) for \(D=6\), \(9\), \(10\) and \(11\), that is at
	\(b/R_S=1.09\), \(0.60\), \(0.52\), \(0.46\), or
	\(bT_0=0.26\), \(0.287\), \(0.291\), \(0.294\) with
	\(T_0=(D-3)/(4\pi R_S)\).

\startarxivsupplement

The numerical conventions and scales are those of the main text.
In particular, \(\sig=s/\ME^2\), \(\lambda\) is dimensionless,
\(\GN=G_N\ME^4\), and \(b\) is a physical length. In the SDR
kernels, \(x,y,t\) denote invariants divided by \(\ME^4,\ME^6,\ME^2\),
respectively. The physical Born formulas in Sec.~\ref{S-sec:supp-born}
use dimensional Mandelstam invariants.

\section{Grids, sampled \(\lambda\) constraints and source checks}
\label{S-sec:supp-grids}

	We checked this source in three independent ways.  A Mathematica calculation
	verifies the partial-wave normalization, both Bessel primitives, and the Jacobian from
	\((\sigma,J)\) to \((\chi,b)\).  Direct quadrature agrees with
	\cref{S-eq:continuum-eikonal-h-stable}.  Finally, moving the split in
	Eq.~(\ref{eq:hybrid-eikonal-source}) of the main text from \(J_\ast=240\) to \(320\) and then \(400\)
	on the \(N_{\sigma}^{\rm source}=6400\) source quadrature changes the total
	source by at most \(0.08\%\) and \(0.07\%\), respectively, over the sampled
	\(\lambda\) range.  Increasing the source quadrature from 6400 to 9600 nodes
	changes it by at most \(3.1\times10^{-4}\) on the pole-normalized scale, while
	doubling the continuum-tail quadrature from 4000 to 8000 nodes changes it by
	less than \(3.6\times10^{-5}\).  The exact-grid and continuum pieces move
	separately by much larger amounts, making these nontrivial matching checks.
	At \(\GN=4\pi^2\), the regular term is less than
	\(2\times10^{-3}\) of the pole scale in absolute size, although
	it is retained in every  calculation including the complete eikonal contribution reported below.

	\label{S-app:chimax}

	The cuts \(\sqrt\sigma\ge4\), \(J\ge20\), \(b\ge2\ME^{-1}\)
	and \(b/R_S\ge3\) jointly set the lower edge of the prescribed eikonal
	region. Increasing \(\chi_{\max}\) admits smaller spins at fixed \(\sigma\):
	\[
	\chi<\chi_{\max}\quad\Longleftrightarrow\quad
	J+\tfrac32>\sigma\sqrt{\frac{\GN}{4\pi\chi_{\max}}}.
	\]
	It also changes the eikonal term in each sampled sum rule through
	Eq.~\eqref{eq:alpha-window-main}.

\section{Uncapped residual supports}
\label{S-sec:supp-uncapped}

	Before the cap is imposed the eikonal contribution already reshapes the allowed region.  With only
	\(\rho^{\rm phys}_{\rm res}\ge0\) on the residual bins and the complete eikonal contribution inserted at \(\GN=0.4\pi^2\)
	(trust window \(\sqrt\sigma\ge4\), \(J\ge20\), \(b\ge2\), \(0<\chi<30\)), the sampled SDR cone on the
	\((900,256,120)\) grid lies close to the fixed-\(t\) guide lines and differs visibly from the fixed-\(a\)
	region (Fig.~\ref{fig:uncapped-wedge} of the main text).

	At \(X=8\) the upper and lower solutions occupy sparse curved sets of
	residual bins, with the full high-spin eikonal contribution included
	analytically in the sum rule.

	The nonnegative-\(X\) boundary values on this grid are listed in
	\cref{S-tab:uncapped-table}.  The upper edge is the line
	\(Y_{\max}=40.0749+3X\) to the displayed precision, while a fit to the
	nonnegative-\(X\) lower points gives \(Y_{\min}=-24.6701-8.14723X\), close to
	but distinct from the fixed-\(t\) lower guide.  At \(X=-5\) the finite
	interval is \(20.687<Y<25.075\); the attempted points \(X=-6,-7\) are
	infeasible on this grid.  The
	imposed equality residuals are below \(3.2\times10^{-8}\) relatively, and the
	largest residual on additional \(\lambda\) values in the resolved interval
	among the feasible points is \(2.4\%\).  At \(X=8\) the uncapped upper
	boundary has \(\max\rphys\simeq981\) and the lower boundary
	\(\max\rphys\simeq49.8\).
\begin{table}[t]
		\centering
		\small\setlength{\tabcolsep}{3pt}
\caption{Sampled boundaries of the uncapped residual cone including the complete eikonal contribution
			on the \((900,256,120)\) grid.}
		\begin{tabular}{crr}
			\toprule
			\(X\)&\(Y_{\min}^{\rm uncapped}\)&\(Y_{\max}^{\rm uncapped}\)\\
			\midrule
			0& -24.670& 40.075\\
			8& -89.848& 64.075\\
			16&-155.026& 88.075\\
			24&-220.204&112.075\\
			32&-285.381&136.075\\
			\bottomrule
		\end{tabular}
		\label{S-tab:uncapped-table}
	\end{table}

\section{Weak gravity: spectra, cap scan, sampled window and the high-energy wedge}
\label{S-sec:supp-weak}

	For the four weak couplings in Sec.~\ref{sec:weak-gravity}, the allowed
	regions grow as gravity weakens in \((X,Y)\), while they nearly coincide
	in \((g_2\ME^6,g_3\ME^8)\). The tips of the leaves were located on the \((300,160,64)\) grid
	and matched upper- and lower-boundary solutions promoted to \((600,240,96)\);
	\cref{S-tab:continuum-weak-leaves} lists the brackets and the largest residual on 801 unsampled values of
	\(\lambda\).
\begin{table*}[t]
		\centering
		\small\setlength{\tabcolsep}{3pt}
		\caption{Weak-gravity leaves with the eikonal integral evaluated analytically in spin.  ``Feasible/fail'' gives the
			last feasible value of \(X\) and the adjacent infeasible trial on the tip
			scan.  The final column is the worst dense relative residual among the
			promoted upper- and lower-boundary  solutions at that coupling.}
		
		\begin{tabular}{ccccc}
			\toprule
			\(\GN\) & \(\frac{M_{\rm Pl}}{M_{\rm EFT}}\) & left feas./fail &
			right feas./fail & dense max\\
			\midrule
			\(\pi^2/1000\) & 1.417 & \(-0.9141/-0.9219\) & \(64000/64500\) & 0.01990\\
			\(\pi^2/2000\) & 1.685 & \(-0.8594/-0.8672\) & \(128000/129000\) & 0.01980\\
			\(\pi^2/4000\) & 2.004 & \(-0.8359/-0.8438\) & \(256000/258000\) & 0.01959\\
			\(\pi^2/8000\) & 2.383 & \(-0.8203/-0.8281\) & \(508000/512000\) & 0.01920\\
			\bottomrule
		\end{tabular}
		\label{S-tab:continuum-weak-leaves}
	\end{table*}

	\subsection{Spectrum and finite-grid checks}

	\begin{figure*}[t]
	\centering
	\includegraphics[width=0.93\textwidth]{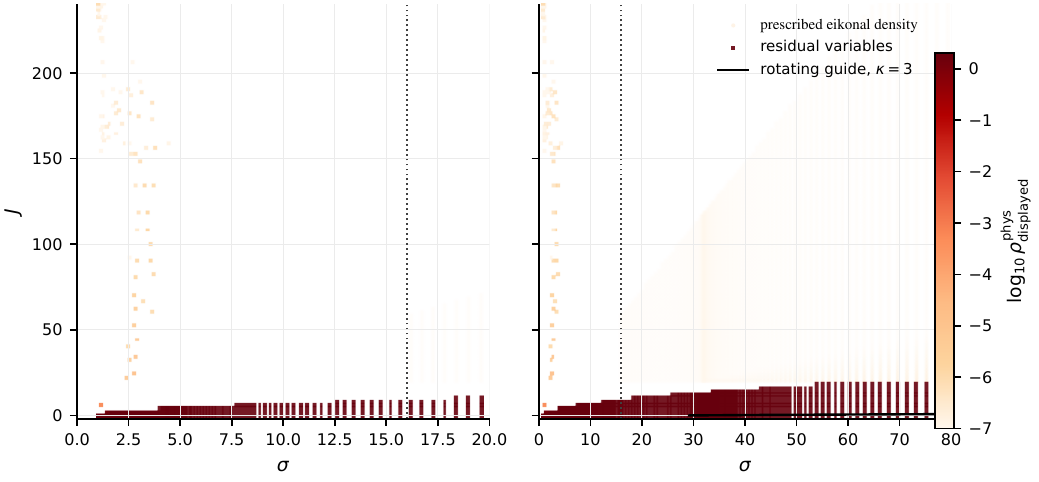}
	\caption{An upper-boundary spectrum at \(\widehat G_N=\pi^2/4000\),
	\(f=0.35\), on the \((600,240,96)\) grid. The prescribed eikonal density
	is integrated over all energies and spins in the sum rule. The sampled
	values shown locate it relative to the adjustable density.}
	\end{figure*}

	Changing the cap exposes a second feature of this weak band: the solution
	remains essentially saturated to the imposed cap.  At \(\GN=\pi^2/4000\) and fixed
	\(X=5.12\times10^4\), lowering \(\rho_{\max}\) from \(2\) to \(1\) and
	\(0.5\) moves \(Y_{\max}\) from \(-8.99\times10^4\) to
	\(-1.70\times10^5\) and \(-2.49\times10^5\).  Roughly \(99\%\) of the
	occupied bins remain within ten percent of the relevant cap.  The median edge
	moves outward from \(4.77\) to \(5.48\) and \(6.29\), in units of \(\ME^{-1}\), so the broad low-spin organization survives but its
	location is cap dependent. This is a useful warning against identifying the
			height or exact edge of one extremal  solution with a physical cross section.
			Together with the  \(G_N=0\) comparison above, the cap  scan shows that the present
		\(\ell_{\rm micro}\) is a property of the capped extremal problem---its kernels,
		threshold and cap---until an independent completion dial demonstrates
		otherwise.

	The upper/lower asymmetry also survives  in the weak-coupling scan.  At the stored
	\(f=0.50\) lower-boundary  solutions, none of the four couplings has a
	contiguous near-cap sequence beginning at \(J=0\) anywhere in
	\(3\leq\sigma\leq80\), and these  solutions carry no residual weight at
	\(b/R_S<3\).  By contrast, the upper-boundary  solutions at the bracketing
	positions \(f=0.35\) and \(0.65\) have such a contiguous edge in
	\(136\)--\(150\) sampled energy bins.  Since a fixed \(b/R_S\) cut is a poor
	scalar diagnostic when \(G_N\) is small, the robust statement is the absence
	of the low-spin saturated edge on the lower boundary, not an exclusive claim
	about black-hole production.

		We do not resolve a residual Regge ridge with appreciable stored bin density for
		\(\sigma\leq80\).  Extending the diagnostic to \(\sigma\leq500\) finds
		\(J\geq30\) bins whose raw sum of \(\rho_{\rm res}^{\rm phys}\) is
		\(5.9\times10^{-5}\), only \(3.8\times10^{-9}\) of the corresponding sum over
		all residual bins.  These raw bin-density sums  describe the support pattern; they
		do not include the energy quadrature, the partial-wave normalization, or the
		\(\Ktwo\) kernel.  The conclusion is therefore only that no visually resolved
		ridge with appreciable pointwise density appears on this weak-coupling family,
		while the outer near-cap component at \(\sigma\geq10^4\) appears in
		Sec.~\ref{sec:matched-high-energy-null} of the main text.

	\subsection{Dependence on the sampled \texorpdfstring{\(\lambda\)}{lambda} window}
	\label{S-sec:lambda-window}

	The band edge at weak coupling, \(b\simeq5.3\,M_{\rm EFT}^{-1}\), satisfies
	\(b\ME\sqrt{\lambda_{\max}}\simeq3\) for \(\lambda_{\max}=1/3\), which raises the
	question whether it is set by the sampled window.  \Cref{S-tab:lambda-window}
	answers it.  At \(G_N=0\) and \(g_2\ME^6=5600\) on the
	\((600,240,96)\) grid of Sec.~\ref{sec:gn0} of the main text, we replaced
	the threshold-angle nodes by Chebyshev nodes on \((0,\lambda_{\max})\) and
	lowered \(\lambda_{\max}\).  The median edge over \(3\le\sigma\le80\) does not
	move.  For comparison the table lists the boundary \(b_\ast\) of the region in
	which the single-bin kernel profile has the useful signs
	(Sec.~\ref{sec:demystifying-bh-band} of the main text), which does scale as
	\(\lambda_{\max}^{-1/2}\).  The lower part of the table repeats the test at \(\GN=4\pi^2\), \(X=20\), on the
	\((600,240)\) grid with the complete eikonal contribution.  With 80 nodes on the shorter windows the solver fails
	numerically, because the rows become too nearly dependent; 48 nodes on \((0,0.2)\) converge, as do 60
	nodes on \((0,1/3)\), included for comparison.  The value of \(Y_{\max}\) depends on the window and on
	the number of nodes, as it must for a relaxation.  The edge does not.

	\begin{table*}[t]
		\centering
		\small\setlength{\tabcolsep}{3pt}
		\caption{The cap-saturated band edge for different sampled windows \(0<\lambda<\lambda_{\max}\).  Upper part:
			\(G_N=0\), \(g_2\ME^6=5600\), \((600,240,96)\) grid; the edge is the median over
			\(3\le\sigma\le80\), and the coefficient column is \(g_3\ME^8\).  Lower part: \(\GN=4\pi^2\),
			\(X=20\), \((600,240)\) grid, edge taken as the median over \(10\le\sigma\le80\), coefficient \(Y_{\max}\).
			The fifth column is the boundary \(b_\ast\) of the kernel region with the useful signs at \(\sigma=75\).
			The last column is the largest relative residual on 801 unsampled values of \(\lambda\) inside the
			window.  Lengths in units of \(M_{\rm EFT}^{-1}\).}
		
		\begin{tabular}{clccccc}
			\toprule
			\(\GN\) & nodes & \(\lambda_{\max}\) & \(b_{\rm edge}\ME\) & \(b_\ast\ME\) & coefficient & residual\\
			\midrule
			0 & threshold angle & 0.333 & 5.27 & 9.8 & \(-17045.3\) & \(1.3\times10^{-5}\)\\
			0 & Chebyshev & 0.333 & 5.27 & 9.8 & \(-17045.3\) & \(1.3\times10^{-5}\)\\
			0 & Chebyshev & 0.250 & 5.26 & 11.2 & \(-17044.3\) & \(1.3\times10^{-5}\)\\
			0 & Chebyshev & 0.200 & 5.27 & 12.6 & \(-17042.7\) & \(1.3\times10^{-5}\)\\
			0 & Chebyshev & 0.100 & 5.29 & 17.7 & \(-17026.9\) & \(3.5\times10^{-6}\)\\
			\midrule
			\(4\pi^2\) & Chebyshev & 0.333 & 9.34 & 9.8 & \(-131.980\) & 0.020\\
			\(4\pi^2\) & threshold angle & 0.333 & 9.34 & 9.8 & \(-131.957\) & 0.020\\
			\(4\pi^2\) & Chebyshev, 60 nodes & 0.333 & 9.19 & 9.8 & \(-127.157\) & 0.119\\
			\(4\pi^2\) & Chebyshev, 48 nodes & 0.200 & 9.33 & 12.6 & \(-109.418\) & 0.060\\
			\bottomrule
		\end{tabular}
		\label{S-tab:lambda-window}
	\end{table*}

		\subsection{The high-energy wedge}
	\label{S-sec:supp-wedge}

	At \(\sigma\sim10^4\)--\(10^5\) the bins at the cap fill a wedge in the \((\sigma,J)\) plane
	(Sec.~\ref{sec:matched-high-energy-null} of the main text).  At each energy we take the contiguous run
	of spins with \(\rho\ge1.8\) that reaches the largest spin and call its ends \(J_{\rm low}\) and
	\(J_{\rm out}\).  The four solutions use the \((800,400,120)\) grid, \(g_2\ME^6\simeq5.6\times10^3\)
	and \(\GN=\pi^2/r\) with \(r=1000\), \(500\), \(100\), \(50\), that is
	\(M_{\rm Pl}/M_{\rm EFT}=1.42\), \(1.19\), \(0.80\), \(0.67\); the scan crosses \(M_{\rm Pl}=M_{\rm EFT}\)
	on purpose.

	\begin{figure*}[t]
		\centering
		\includegraphics[width=0.86\textwidth]
		{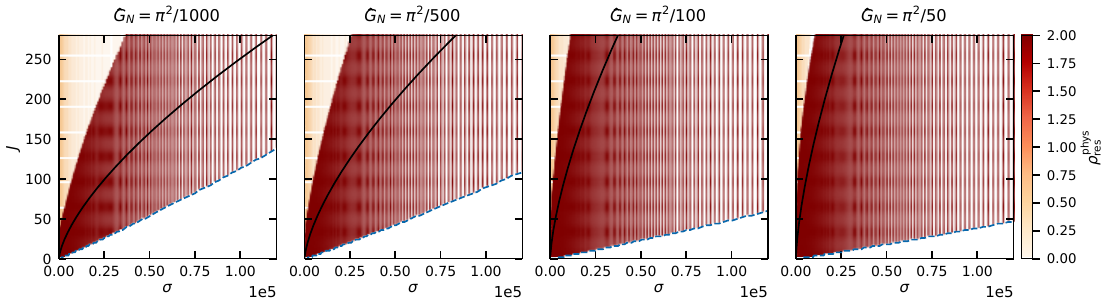}
		\caption{Total density at high energy for \(\GN=\pi^2/r\), \(r=1000\), \(500\), \(100\), \(50\) (left to
			right), on a linear energy axis.  The bins at the cap fill the region between the lower edge
			\(J_{\rm low}\) (dashed blue), which is linear in \(\sigma\), and the prescribed eikonal contribution.  The black
			curve is the \(\kappa=3\) reference.}
		\label{S-fig:matched-high-energy-null}
	\end{figure*}

	The outer edge is the inner edge of the prescribed eikonal region \(b=R_{\min}R_S\),
\begin{equation}
		\begin{split}
		J_{\rm out}+\frac32
		&\simeq\frac{R_{\min}}{2}\left(\frac{3}{2\pi}\right)^{1/3}
		\GN^{1/3}\sigma^{2/3}\\
		&=1.17\,\GN^{1/3}\sigma^{2/3}
		\quad(R_{\min}=3),
		\end{split}
		\label{S-eq:matched-high-energy-edge-powers}
	\end{equation}
	to within one even spin for \(r=500\) and \(1000\) and \(10^4\le\sigma\le5.5\times10^4\)
	(\(J_{\rm out}=116\), \(156\), \(208\), \(274\), \(362\) against \(117\), \(157\), \(209\), \(274\), \(362\)
	at \(r=1000\)); at higher energy it is pinned at \(J_{\max}=400\).  Repeating the \(r=1000\) solve with
	\(R_{\min}=4\) and \(6\) moves \(J_{\rm out}\) to the new boundary (\(156\), \(208\), \(278\), \(364\)
	against \(157\), \(210\), \(279\), \(366\), and \(234\), \(314\) against \(236\), \(316\)), leaves
	\(J_{\rm low}\) identical, and changes \(Y\) by \(1.5\times10^{-5}\) and \(2.9\times10^{-4}\).  The
	scaling \(\GN^{1/3}\sigma^{2/3}\) of this edge is thus inherited from the trust region.  At \(r=100\) and
	\(50\) the cut on \(\chi_{\max}\) competes with the cut on \(b/R_S\) and the edge falls below the
	boundary.

	The lower edge is free and linear,
\begin{equation}
		J_{\rm low}+\frac32\simeq a(\GN)\sigma+\beta,
		\qquad a(\GN)\propto \GN^{-0.35\pm0.02}.
		\label{S-eq:matched-high-energy-lower-edge}
	\end{equation}
	Fits with a free intercept on four energy windows give exponents between \(-0.360\) and \(-0.365\); fits
	through the origin give \(-0.343\) to \(-0.348\).  The intercepts are below one spin.  The data are
	compatible with \(G_N^{-1/3}\) and do not resolve the exponent further.  The result has been obtained on
	one slice of fixed \(g_2\).

\section{Strong coupling: numerical checks}
\label{S-sec:supp-strong}

	\subsection{ Grid convergence of the coefficient bounds and  spectra}
	\label{S-sec:nlambda-drift}

	There are two separate numerical questions.  First, does the optimized
	coefficient stabilize as we impose more values of \(\lambda\)?  Second, does
	the resulting spectrum satisfy the SDR between those values?  More precisely,
	sampling at additional values asks
	whether the optimized coefficient stabilizes as more sampled values of the
	auxiliary parameter are imposed exactly.   We then evaluate the returned
	solution at additional values that were not used in the optimization.
	The latter is a test of interpolation between sampled values of \(\lambda\); it is not an
	additional LP constraint.

	For an  additional value of \(\lambda\), we use
\begin{widetext}
	\begin{equation}
		\Delta_{\rm rel}(\lambda)=
		\frac{|A(\lambda)\rphys-\lambda^2Y-r_X(\lambda)|}{1+|A(\lambda)\rphys|+|\lambda^2Y|+|r_X(\lambda)|},
		\qquad
		\Delta_{\rm dense}=\max_{\lambda\in\Lambda_{\rm dense}}
		\Delta_{\rm rel}(\lambda).
		\label{S-eq:dense-residual-definition}
	\end{equation}
\end{widetext}
	Because its denominator contains the additive reference scale \(1\),
	\(\Delta_{\rm rel}\) is a bounded normalized residual rather than a
	conventional relative error.  This definition remains finite when one term in
	the sum rule crosses zero.
	The  convergence study including the complete eikonal contribution at \(\GN=4\pi^2\), \(X=20\),
	\(Y_{\max}\) is summarized in \cref{S-tab:dense-residual-status}.  Here
	\(\Lambda_{\rm dense}\) contains 1601 threshold-angle points.  The column
	\(\Delta_{\rm resolved}\) restricts this test to the interval between the
	first and last sampled values of \(\lambda\) of the corresponding solve; the final
	column also includes the still smaller-\(\lambda\) points present only in the
	dense grid.
	\begin{table*}[t]
		\centering
		\small\setlength{\tabcolsep}{3pt}
		\caption{Grid refinement and residuals at additional \(\lambda\) values, including the complete eikonal contribution, at
			\(\GN=4\pi^2\), \(X=20\), \(Y_{\max}\).  The equality residual is evaluated
			on the imposed constraints; \(q_{50}\) and \(q_{95}\) use the full dense
			grid.}
		\begin{tabular*}{\textwidth}{@{\extracolsep{\fill}}ccrrrrrr}
				\toprule
				\((N_\sigma,J_{\max})\)&\(N_\lambda\)&\(Y_{\max}\)&
				\(\Delta_{\rm eq}\)&\(q_{50}\)&\(q_{95}\)&
				\(\Delta_{\rm resolved}\)&\(\Delta_{\rm dense}\)\\
				\midrule
				\((300,160)\)&40 &-123.6759&\(7.2\times10^{-16}\)&0.0226&0.0660&0.1054&0.1054\\
				\((300,160)\)&60 &-131.9975&\(1.1\times10^{-15}\)&0.00055&0.00303&0.01597&0.02011\\
				\((300,160)\)&80 &-131.9652&\(7.3\times10^{-8}\)&0.00058&0.00920&0.01906&0.02008\\
				\((600,240)\)&80 &-131.9565&\(7.3\times10^{-8}\)&0.00078&0.00353&0.01481&0.02008\\
				\((600,240)\)&100&-131.9329&\(3.0\times10^{-8}\)&0.00018&0.00420&0.01639&0.02004\\
				\((600,240)\)&120&-131.9186&\(1.4\times10^{-8}\)&0.00061&0.00902&0.01924&0.02000\\
				\((900,320)\)&120&-131.9153&\(1.4\times10^{-8}\)&0.00074&0.00414&0.01319&0.01999\\
				\((1200,400)\)&120&-131.9159&\(1.4\times10^{-8}\)&0.00072&0.00340&0.01144&0.02001\\
				\bottomrule
		\end{tabular*}
		\label{S-tab:dense-residual-status}
	\end{table*}
	The coarse \((300,160)\), \(N_\lambda=40\) result has not reached the plateau.
	From \(N_\lambda=60\) onward the target moves toward a stable value.  At
	\(N_\lambda=120\), the \((600,240)\) and \((900,320)\) results differ by
	\(2.5\times10^{-5}\) relatively, while the \((900,320)\) and
	\((1200,400)\) values differ by only \(5.2\times10^{-6}\).  The median  residual at additional \(\lambda\) values is below
	\(10^{-3}\), and the resolved-interval maximum is \(1.3\)--\(1.9\%\).  The
	remaining \(2\%\) full-grid maximum occurs below the first sampled value of \(\lambda\),
	where the finite phase window leaves the known endpoint floor.  Raising
	\(N_\lambda\) without also increasing the spectral resolution eventually
	overconstrains the coarse problem: at \((300,160)\), \(N_\lambda=100\), phase
	I requires a nonzero equality slack.  This is why both sides of the finite
	matrix are enriched together.  We do not insert isolated constraints below
	the first sampled value of \(\lambda\) into the quoted runs: at fixed spectral resolution
	they probe the same unresolved endpoint and can manufacture infeasibility,
	whereas a controlled extension requires enriching the endpoint spectral basis
	at the same time.

	As an independent solver check, HiGHS dual simplex and HiGHS interior point
	return exactly the same value
	\(Y=-131.9565048401233\) on the \((600,240)\), \(N_\lambda=80\) problem,
	with identical equality and dense-residual diagnostics
	\cite{S-HuangfuHallHiGHS}.  The source itself is tested separately: moving the
	exact-spin/continuum split and increasing both source quadratures changes it
	by less than \(10^{-3}\) on the pole-normalized scale.

		\subsection{Sampled leaf and support pattern along the boundary}
	\label{S-sec:boundary-morphology}

	\Cref{S-tab:g6p5-leaf-table} lists the sampled capped boundary at \(\GN=4\pi^2\)
	on the \((300,160,80)\) leaf grid; every entry is a direct primal optimum.
\begin{table}[t]
		\centering
		\small\setlength{\tabcolsep}{3pt}
		\caption{ Sampled capped boundary including the complete eikonal contribution at \(\GN=4\pi^2\) on
			the \((300,160,80)\) leaf grid.}
		\begin{tabular}{crr}
			\toprule
			\(X\)&\(Y_{\min}\)&\(Y_{\max}\)\\
			\midrule
			-12& 83.208& 87.971\\
			-8& 34.369& 77.223\\
			-4& -6.235& 55.915\\
			0& -43.087& 30.612\\
			4& -76.268& 2.512\\
			8& -107.073& -27.779\\
			12& -135.818& -60.003\\
			16& -162.385& -94.722\\
			20& -186.378& -131.965\\
			24& -206.649& -171.888\\
			26& -214.391& -193.048\\
			27& -216.919& -203.975\\
			27.5& -216.342& -209.610\\
			\bottomrule
		\end{tabular}
		\label{S-tab:g6p5-leaf-table}
	\end{table}

	 Across the eight \((600,240,80)\) solutions at \(X=-8,0,8,20\), the
	upper boundary and positive-\(X\) lower boundary concentrate their residual
	weight at \(b<3R_S\) and \(\rphys\ge1.8\). The negative-\(X\) lower
	boundary instead populates \(3\le b/R_S<6\) and a long, light high-spin
	ridge (\(J\ge35\), \(\sigma\le20\)).

		\subsection{Amplitude-level SDR check}
	\label{S-sec:fad-validation}

	The coefficient sum rule uses the SDR only through derivatives at \(x=y=0\).  The redundancy in
	\(\lambda\) can also be tested at finite kinematics.  For two low-energy points \((x_1,y_1)\) and
	\((x_2,y_2)\) the difference \(\Delta M_{12}=M(x_1,y_1)-M(x_2,y_2)\) does not contain the subtraction
	constant, and
\begin{widetext}
	\begin{align}
		\Delta M_{12}
		&=
		-\frac1\pi\int_1^\infty\dd\sig\,
		\Big[
		K_0(\sig;x_1,y_1;\lam)\,
		\mathcal A^{(s)}_\rho\bigl(\sig,t_\lam(x_1,y_1;\sig)\bigr)
		\nonumber\\
		&\hspace{38mm}
		-
		K_0(\sig;x_2,y_2;\lam)\,
		\mathcal A^{(s)}_\rho\bigl(\sig,t_\lam(x_2,y_2;\sig)\bigr)
		\Big],
		\label{S-eq:fad-row-main}
	\end{align}
\end{widetext}
	The left-hand side is a physical amplitude difference and cannot depend on \(\lambda\).  We take
	\((s_1,t_1)=(0.20,-0.08)\) and \((s_2,t_2)=(0.24,-0.10)\), sample nine Chebyshev values of \(\lambda\) in
	\([0.01,0.30]\), and project out the one \(\lambda\)-independent component, which leaves eight linear
	constraints on the density.  The eikonal contribution contributes through an oscillatory integral, which we evaluate
	for the difference directly; changing its quadratures and the finite-spin/continuum split moves the
	projected source by less than \(0.9\%\).

\begin{table*}[t]
	\centering
		\small\setlength{\tabcolsep}{3pt}
\caption{Matched \((600,240)\), \(N_\lambda=80\) solves including the complete eikonal contribution
			with and without the eight amplitude-difference constraints.  The last
			column is the weighted support overlap
			\(\mathcal J_\rho=\sum_i\min(\rho_i,\rho_i')/
			\sum_i\max(\rho_i,\rho_i')\), evaluated on the union of residual bins.}
		\begin{tabular}{crrrr}
			\toprule
			boundary point & \(Y_{K_2}\) & \(Y_{K_2+\Delta M}\) &
			\(\lvert\Delta Y\rvert/\lvert Y\rvert\) & \(\mathcal J_\rho\)\\
			\midrule
			\(X=20\), upper & \(-131.9565\) & \(-131.9646\) &
			\(6.2\times10^{-5}\) & 0.9991\\
			\(X=8\), upper & \(-27.7272\) & \(-27.7655\) &
			\(1.38\times10^{-3}\) & 0.9985\\
			\(X=8\), lower & \(-107.1103\) & \(-107.0330\) &
			\(7.22\times10^{-4}\) & 0.9837\\
			\bottomrule
		\end{tabular}
		\label{S-tab:fad-validation}
	\end{table*}

	\Cref{S-tab:fad-validation} compares three boundary solutions with and without these constraints.  The
	constraints are met to \(2\times10^{-15}\), the coefficient moves by at most \(0.14\%\), the residual of
	the \(\Ktwo\) rule on unsampled \(\lambda\) stays at \(1.1\)--\(1.4\%\), and the bins at the cap overlap to
	\(0.989\) or better.  The spectra are therefore not artifacts of imposing only the differentiated sum
	rule.  These constraints are a check; they are not used to trace the leaves.

	\subsection{Where the bins with prescribed eikonal density begin}

	The high-spin ridge of Fig.~\ref{fig:g6p5-four-witness} of the main text starts near the energy at
	which the gridded bins with prescribed eikonal density begin, so we repeated the \(X=20\), \(Y_{\max}\) solve on the
	\((300,160,80)\) grid with that energy moved from \(\sigma_*=80\) to \(160\), keeping the complete
	analytic tail in both runs.  Among residual bins with \(J\ge12\), \(b/R_S\ge3\)
	and \(\sigma\le80\), the largest connected high-spin component grows from \(J=12\)--\(20\) to
	\(J=14\)--\(28\).  The ridge is not produced by the handover.

	At \(X=20\), \(Y_{\max}\), the single-bin score \(c_1^{\rm band}(-c_2^{\rm band})\) of
	Sec.~\ref{sec:demystifying-bh-band} of the main text has a Spearman coefficient of \(-0.47\) with the
	cost of forcing weight into an empty bin (\(\sigma\le80\), \(J\le90\)).  For \(1\le b/R_S<3\) the empty
	bins with the useful signs have a median cost of \(1.19\), against \(7.21\) for the others; for
	\(3\le b/R_S<8\) both medians are about \(3.65\).  The signs locate the inexpensive bins at low impact
	parameter but do not account for the whole optimum.

\subsection{From single-bin signs to the full optimization}

	The single-bin signs identify a favorable region of the \(\Ktwo\)
		kernel, but by themselves they cannot show that gravity fixes the position selected
		by the LP.  Outside the prescribed eikonal region, changing \(G_N\) merely
		rescales a cap-saturated single-bin profile by a positive constant.  A genuine
		coupling test must therefore recompute the allowed boundary and compare
		corresponding points using the complete prescribed source.  The weak-gravity
		scan in Sec.~\ref{sec:weak-gravity} of the main text does this in the hierarchy
		\(M_{\rm EFT}<M_{\rm Pl}\).  A fixed-\(X\),
		grid-sampled eikonal \(G_N\)  scan would instead mix different positions on
		different leaves and would inherit the source truncation described above.

	\label{S-app:support-drift}

	The heat maps are finite-grid primal  solutions, not unique spectral
	reconstructions.  The useful question is whether their visible structures
	remain stable when the sampled SDR constraints are enriched at fixed
	\((N_\sigma,J_{\max})\).  For this purpose we record three simple support
	metrics in addition to the target-value drift:
	\begin{align}
		f_{b/R_S<3}
		&=
		\frac{\sum_{i:\,b_i/R_S(\sig_i)<3}\rphysi}{\sum_i \rphysi},
		\label{S-eq:support-metric-lowb}\\
		\langle b/R_S\rangle_\rho
		&=
		\frac{\sum_i (b_i/R_S(\sig_i))\,\rphysi}{\sum_i \rphysi},
		\label{S-eq:support-metric-meanb}\\
		N_{\rphys\ge1.8}
		&=
		\#\{i:\rphysi\ge1.8\}.
		\label{S-eq:support-metric-saturated}
	\end{align}
	Here the sums run over residual bins in the support CSV for the given
	boundary optimizer.  They are raw bin-density sums: no energy-quadrature
	weights, partial-wave factors \(n_J^{(6)}\), or dispersive kernels are included.
	The first quantity measures the fraction of this raw density at low impact
	parameter, the second is a raw-density-weighted impact-parameter average, and
	the third counts bins close to the reflective
	unitarity endpoint.  These observables are intentionally crude; their purpose is
	to check qualitative stability of the Regge/low-\(b\) picture, not to
	define a continuum limit.

	A  comparison including the complete eikonal contribution at \(\GN=4\pi^2\), \(X=20\),
	\(Y_{\max}\) is:
	\begin{table}[t]
		\centering
		\small\setlength{\tabcolsep}{3pt}
		\caption{ Support measures including the complete eikonal contribution for the same boundary point on two
			finite grids.}
		\begin{tabular}{crrrr}
			\toprule
			\((N_\sigma,J_{\max},N_\lambda)\)&\(N_{\rm supp}\)&
			\(N_{\rphys\ge1.8}\)&\(f_{b/R_S<3}\)&
			\(\langle b/R_S\rangle_\rho\)\\
			\midrule
			\((600,240,80)\)&2604&2525&0.97833&1.45315\\
			\((1200,400,120)\)&5221&5102&0.97854&1.44623\\
			\bottomrule
		\end{tabular}
		\label{S-tab:support-metric-table}
	\end{table}
	The number of occupied bins grows with the spectral grid and is not itself a
	continuum observable.  The density-weighted quantities are the useful
	comparison: the low-impact fraction changes by \(2.2\times10^{-4}\), and the
	weighted mean impact parameter changes by about \(0.5\%\).  Together with the
	coefficient plateau in \cref{S-sec:nlambda-drift}, this shows that the principal
	low-impact  support pattern is stable across these two finite-grid
	 solutions.  It does not establish uniqueness of the optimizer or pointwise
	convergence of every occupied bin.

	\subsection{No-eikonal cutoff scan}
	\label{S-sec:noeik-scan}

	In the control without the prescribed eikonal contribution (Sec.~\ref{sec:no-eikonal-controls}
	of the main text), at \(\GN=4\pi^2\) and \(X=6.5\), increasing the spectral
	cutoff with only sparse constraint enrichment gives a rapidly moving
	\(Y_{\max}\), whereas enriching the spectral grid and the set of sampled values
	together gives the much more stable sequence in \cref{S-tab:noeik-grid-balance}.
	The final balanced step changes \(Y_{\max}\) by about \(3\%\), although the
	support still reaches the imposed cutoff.
\begin{table*}[t]
		\centering
		\small\setlength{\tabcolsep}{3pt}
		\begin{tabular}{c c r c r}
			\toprule
			\((N_\sigma,J_{\max})\) & \(N_\lambda\) sparse & \(Y_{\max}\) sparse
			& \(N_\lambda\) balanced & \(Y_{\max}\) balanced\\
			\midrule
			\((300,80)\)   & 15 & 171.859 & 30  & 7.399\\
			\((600,160)\)  & 30 & 199.625 & 60  & 8.940\\
			\((900,240)\)  & 45 & 215.485 & 90  & 9.457\\
			\((1200,320)\) & 60 & 229.475 & 120 & 9.715\\
			\bottomrule
		\end{tabular}
		\caption{No-eikonal cutoff  scan.  In every run the support reaches
			\(\sigma_{\max}=N_\sigma\).  The sparse sequence exposes excess freedom from
			enlarging the spectral grid faster than the sampled constraints; the
			balanced sequence is the relevant finite-cutoff control.}
		\label{S-tab:noeik-grid-balance}
	\end{table*}

	Why does the spectrum reorganize so strongly?  The finite-\(\lambda\) sum rule
	makes the reason transparent.  With the eikonal contribution supplied,
	\begin{equation}
		\frac{\lambda\Ktwo[\rho_{\rm res}](\lambda)}{8\pi \GN}
		=1+2\lambda X+\lambda^2Y-F_{\rm eik}(\lambda) ,
		\label{S-eq:lambda-budget-eikonal}
	\end{equation}
	whereas the no-eikonal control obeys
	\begin{equation}
		\frac{\lambda\Ktwo[\rho](\lambda)}{8\pi \GN}
		=1+2\lambda X+\lambda^2Y .
		\label{S-eq:lambda-budget-noeikonal}
	\end{equation}
	In the first equation the residual contribution supplies only the difference
	between the target and the complete eikonal contribution.  Its limiting constant is
	\(1-\alpha_E\) for the chosen finite phase window, but the full
	\(\lambda\)-dependence is retained.  In the second equation the finite-grid
	density itself must supply the unit constant.  A regional decomposition locates that contribution mainly on
	the low-energy high-spin branch in Fig.~\ref{fig:noeik-control-x6p5} of the main text; the visible
	low-impact band contributes very little to the leading pole budget.  The
	finite LP therefore does not derive the eikonal contribution.  Supplying it selects
	the standard asymptotic realization before asking how the spectrum completes
	the smaller-impact-parameter region.

\section{The absorptive core: selection, controls and disk shapes}
\label{S-sec:supp-absorb}

	This section documents the calculations behind Sec.~\ref{sec:absorptive-density-pilot}
	of the main text. On the original \((600,240,80)\) grid at \(X=20\), we
	start from a stored upper-boundary solution with residual density
	\(\rho^{\rm ref}\) and coefficient \(Y_{\rm ref}=Y_{\max}\) and solve for
	the change of the density on a set \(F\) of free bins,
	\begin{equation}
		\sum_{i\in F}A_{\alpha i}\left(\rho_i-\rho_i^{\rm ref}\right)
		-\lambda_\alpha^2\left(Y-Y_{\rm ref}\right)=0,
		\qquad
		0\le\rho_i\le u_i ,
		\label{S-eq:delta-form}
	\end{equation}
	so that the prescribed source cancels and the reference itself is an exactly
	feasible point.  This matters numerically: the 80 rows are close to linearly
	dependent (their singular values span seventeen orders of magnitude after
	column scaling), and a formulation with an independently computed right-hand
	side is reported infeasible by the solver.  The matrix \(A\) is rebuilt with
	the production kernel and reproduces the stored selected solutions to
	\(8\times10^{-9}\).  Columns and rows are scaled, and the programs are solved
	with HiGHS.

	\subsection{Selection of an absorptive spectrum}

	The selection minimizes the mean distance of the total density from
	\(\rho=1\) over a region \(R\), subject to \cref{S-eq:delta-form} and
	\(Y\ge Y_{\rm ref}-\epsilon|Y_{\rm ref}|\) with \(\epsilon=0.5\%\).  The stored
	solutions use the squared distance (a quadratic program solved with Clarabel).
	At \(\GN=4\pi^2\) the region is \(b<2R_S\), the reference has the eikonal contribution
	switched on smoothly between \(2R_S\) and \(3R_S\), all bins outside the region
	are frozen, and the sum rules are allowed a normalized slack of \(10^{-2}\); the
	residual on 1601 unsampled values of \(\lambda\) is \(0.078\), against \(0.020\)
	for the reference. Figure~\ref{fig:core-vs-band}(a) of the main text shows
	this selection. At \(\GN=2\pi^2\)
	and \(\GN=\pi^2\) the region is \(b<10R_S\), the eikonal contribution starts at \(10R_S\), and
	the sum rules are exact; the residuals on unsampled \(\lambda\) are \(0.086\)
	and \(0.072\).

	For Fig.~\ref{fig:absorptive-core-selection}(b), at \(X=20\) we let
	\(Y\) fall by at most \(0.5\%\) of \(|Y_{\max}|\). We minimize the weighted
	mean of \((\rho_{\rm tot}-1)^2\), equivalently maximizing the weighted mean of
	\(\mathcal A_0=1-(\rho_{\rm tot}-1)^2\). The impact-parameter weight is
	one through \(b=2.5R_S\) and drops smoothly to zero at \(3R_S\), as
	\(u^2(3-2u)\) with \(u=(3-b/R_S)/0.5\) clipped to \([0,1]\).
	Panel (b) shows the absorption proxy through \(\sigma=500\) and \(J=200\).

	We keep the original equal-width bins in \(1/\sigma\) through the
	node \(\sigma=20\), then use \(\sigma=21,24,\ldots,198\) and
	\(200,210,\ldots,1200\), giving 732 nodes. Each bin is weighted by its
	width in \(1/\sigma\), divided by \(\pi\), so the measure is preserved
	under refinement. We use even spins through \(J=240\) below \(\sigma=200\)
	and \(J=480\) thereafter, retain the original 80 sampled sum rules, and
	let residual bins with positive capacity inside \(10R_S\) readjust.

	The reference and selected spectra have \(Y_{\max}=-134.0578259\)
	and \(Y=-134.7281150\). Clarabel returns \texttt{AlmostSolved}.
	The maximum change to a sampled sum rule is \(3.57\times10^{-7}\)
	after enforcing the density box. The largest relative residual is
	\(1.31\%\) over the sampled \(\lambda\) interval and \(2.01\%\) when
	the small-\(\lambda\) endpoint is included, as for the reference. See
	\nolinkurl{codesforarxiv/fig9_v23/README.md} for the bin construction,
	solver settings and archived data.

	For the selection in Fig.~\ref{fig:absorptive-core-selection}(d) we
	use the same grid, weights and absolute loss allowance
	\(\Delta Y\le0.6702891\) at \(X=25.3\).
	The reference has \(Y_{\max}=-188.8884721\) and the selected spectrum has
	\(Y=-189.2879016\). A small adjustment outside \(3R_S\) leaves the objective
	unchanged and brings the maximum sampled row shift to \(2.67\times10^{-7}\).
	Panel (c) shows the reference residual density and panel (d) shows
	\(\mathcal A_0=1-(\rho_{\rm tot}-1)^2\) for the selected spectrum.
	Both panels show spins through \(J=200\).
	The two outer crossings and the first outward crossing of
	\(\mathcal A_0=0.9\) are interpolated between adjacent even-spin bins and
	fitted over \(610\le\sigma\le1200\). The arrays and plotting scripts are in
	\nolinkurl{codesforarxiv/fig9_v23/higher_x_band_check/README.md}.

	The controls replace the squared distance by the absolute distance, which
	makes the selection a linear program with exact rows, and vary the region.
	One finding is that with exact rows and all exterior bins frozen a region as
	small as \(b<2R_S\) admits no change at all: the reference is the only
	feasible point.  This is why the original \(4\pi^2\) run needed a slack.  In
	the \(L^1\) controls at \(\GN=4\pi^2\) we therefore let the bins between the
	region and \(10R_S\) readjust, with the objective still restricted to the
	region.  \Cref{S-tab:core-controls-full} lists the exponent of the core edge on
	four windows and its position at four energies.  Regions bounded by a fixed
	impact parameter, whose shape in the \((\sigma,J)\) plane is \(\sigma^{1/2}\),
	give the same edge as regions bounded by a multiple of \(R_S\).  The entry
	\(b<4\,\ME^{-1}\) at \(\GN=4\pi^2\) shows what a region that is too small does: above
	\(\sigma\simeq200\) the core runs into its boundary and the fitted exponent on
	the highest window drops.

	\begin{table*}[t]
		\centering
		\small\setlength{\tabcolsep}{3pt}
		\caption{Edge of the absorptive core (outward crossing of \(\mathcal A_0=1/2\) from \(J=0\)) for the stored
			\(L^2\) selections and the \(L^1\) controls, all at \(X=20\) and \(\epsilon=0.5\%\) on the \((600,240,80)\)
			grid.  Columns: exponent \(p\) of \(J_{\rm edge}+3/2\propto\sigma^p\) on four windows of \(\sigma\), and
			\(b_{\rm edge}/R_S\) at four energies.  Regions are given in units of \(R_S\) or of
			\(M_{\rm EFT}^{-1}\).  At \(\GN=\pi^2\) the core is one to three spins wide below \(\sigma\simeq80\), and only
			the windows starting at \(\sigma=20\) are meaningful.}
		
		\begin{tabular}{llcccccccc}
			\toprule
			\(\GN\) & norm, region & \(p_{10\text{--}80}\) & \(p_{20\text{--}200}\) & \(p_{20\text{--}600}\) & \(p_{5\text{--}600}\)
			& \(\sigma=5\) & 20 & 75 & 300\\
			\midrule
			\(4\pi^2\) & \(L^2\), \(b<2R_S\), slack & 0.70 & 0.69 & 0.69 & 0.72 & 0.64 & 0.66 & 0.70 & 0.75\\
			\(4\pi^2\) & \(L^1\), \(b<2R_S\) & 0.65 & 0.72 & 0.80 & 0.75 & 0.64 & 0.66 & 0.61 & 0.75\\
			\(4\pi^2\) & \(L^1\), \(b<5.3\,\ME^{-1}\) & 0.70 & 0.75 & 0.72 & 0.74 & 0.64 & 0.66 & 0.70 & 0.78\\
			\(4\pi^2\) & \(L^1\), \(b<4\,\ME^{-1}\) & 0.72 & 0.67 & 0.62 & 0.71 & 0.64 & 0.66 & 0.70 & 0.58\\
			\midrule
			\(2\pi^2\) & \(L^2\), \(b<10R_S\) & 0.77 & 0.73 & 0.80 & 0.67 & 0.81 & 0.58 & 0.67 & 0.69\\
			\(2\pi^2\) & \(L^1\), \(b<10R_S\) & 0.79 & 0.73 & 0.72 & 0.63 & 0.81 & 0.58 & 0.56 & 0.64\\
			\(2\pi^2\) & \(L^1\), \(b<21\,\ME^{-1}\) & 0.79 & 0.73 & 0.72 & 0.63 & 0.81 & 0.58 & 0.56 & 0.64\\
			\(2\pi^2\) & \(L^1\), \(b<12\,\ME^{-1}\) & 0.77 & 0.72 & 0.73 & 0.64 & 0.81 & 0.58 & 0.67 & 0.69\\
			\midrule
			\(\pi^2\) & \(L^2\), \(b<10R_S\) & 0.61 & 0.71 & 0.71 & 0.54 & -- & 0.40 & 0.57 & 0.55\\
			\(\pi^2\) & \(L^1\), \(b<10R_S\) & 0.52 & 0.76 & 0.74 & 0.54 & -- & 0.40 & 0.44 & 0.49\\
			\(\pi^2\) & \(L^1\), \(b<16\,\ME^{-1}\) & 0.53 & 0.75 & 0.73 & 0.54 & -- & 0.40 & 0.44 & 0.49\\
			\(\pi^2\) & \(L^1\), \(b<8\,\ME^{-1}\) & 0.53 & 0.75 & 0.73 & 0.54 & -- & 0.40 & 0.44 & 0.49\\
			\bottomrule
		\end{tabular}
		\label{S-tab:core-controls-full}
	\end{table*}

	\subsection{The whole-leaf bound, disk shapes and energy windows}
	\label{S-sec:whole-leaf-bound}

	For the whole-leaf extension of Fig.~\ref{fig:inverse-absorptive-radius}
	we use the same \((600,240,80)\) grid and vary every residual bin with
	positive capacity. With a trial disk imposed, we first bound the
	feasible range of \(X\) and cover it with intervals. The chord between
	two feasible points on the upper boundary is a lower bound \(L(X)\) on
	\(Y_{\max}(X)\), by convexity of the original feasible region. On each
	interval, a dual bound on the maximum of \(Y-L(X)\), checked against
	all bins, bounds the loss \(Y_{\max}(X)-Y\) from below.

	At \(X=-10.8708\), the witness for a disk of radius \(0.97817045R_S\)
	has a certified loss of at most \(0.668128\). Raising the radius past
	that breakpoint adds the \(J=0\) bin at \(\sigma=600/485\simeq1.2371\).
	The 27 intervals for the enlarged disk give a loss of at least
	\(0.670623\) everywhere, above the allowance \(0.669939\). This
	certifies the limiting radius on the grid. The code, feasible spectrum,
	and interval certificates are indexed in
	\nolinkurl{codesforarxiv/leaf_radius_bound_v23/README.md}.

	\paragraph{Shape and energy dependence.}

	The bound of the main text uses a disk \(b/R_S<B_*/R_S\) at all energies.  To see
	what limits it we replaced the disk by the regions
	\(J+3/2\le C\sigma^{p}\), at \(\GN=2\pi^2\), with the bins at \(b<10R_S\) free
	and the exterior frozen, and found the largest \(C\) for which
	\(\mathcal A_0\ge\eta\) can be imposed throughout for a loss
	\(\epsilon\le0.5\%\).  \Cref{S-tab:disk-shapes} gives the result.  The value
	\(p=2/3\) is the disk of the main text and gives \(B_*=0.83R_S\) for
	\(\eta=0.5\), close to the \(0.86R_S\) found at \(\GN=4\pi^2\) with all bins free.
	For \(p\le2/3\) the largest admissible region passes close to
	\(b\simeq2\,M_{\rm EFT}^{-1}\) at \(\sigma\simeq2\): the constraint is set
	by the lowest energies.  Steeper shapes give up some radius there and gain
	it at high energy.  The number of bins in the region is largest for
	\(p=0.75\)--\(0.85\) (515 for \(\eta=0.5\), against 421 for the fixed impact
	parameter \(p=1/2\) and 430 for \(p=1\)), which is close to the exponent
	\(0.70\)--\(0.75\) of the selected core.

	\begin{table*}[t]
		\centering
		\small\setlength{\tabcolsep}{3pt}
		\caption{Largest regions \(J+3/2\le C\sigma^p\) in which \(\mathcal A_0\ge\eta\) can be imposed for a loss
			\(\epsilon\le0.5\%\) (\(\GN=2\pi^2\), \(X=20\), bins at \(b<10R_S\) free).  For each threshold: the
			coefficient \(C\), the number of bins in the region, its boundary at \(\sigma=2\) in units of
			\(M_{\rm EFT}^{-1}\), and its boundary at \(\sigma=20\) and \(300\) in units of \(R_S\).}
		
		\begin{tabular}{c|ccccc|ccccc}
			\toprule
			& \multicolumn{5}{c|}{\(\eta=0.5\)} & \multicolumn{5}{c}{\(\eta=0.9\)}\\
			\(p\) & \(C\) & bins & \(b(2)\) & \(b/R_S(20)\) & \(b/R_S(300)\) & \(C\) & bins & \(b(2)\) & \(b/R_S(20)\) & \(b/R_S(300)\)\\
			\midrule
			\(0.4\) & 1.156 & 372 & 2.16 & 0.49 & 0.24 & 1.041 & 284 & 1.94 & 0.44 & 0.22\\
			\(0.5\) & 1.076 & 421 & 2.15 & 0.62 & 0.39 & 0.938 & 318 & 1.88 & 0.54 & 0.34\\
			\(0.6\) & 0.966 & 470 & 2.07 & 0.75 & 0.63 & 0.817 & 351 & 1.75 & 0.63 & 0.53\\
			\(2/3\) & 0.874 & 494 & 1.96 & 0.83 & 0.83 & 0.719 & 362 & 1.61 & 0.68 & 0.68\\
			\(0.75\) & 0.736 & 515 & 1.75 & 0.89 & 1.12 & 0.586 & 371 & 1.39 & 0.71 & 0.89\\
			\(0.85\) & 0.551 & 514 & 1.41 & 0.90 & 1.49 & 0.425 & 364 & 1.08 & 0.70 & 1.14\\
			\(1\) & 0.281 & 430 & 0.79 & 0.72 & 1.78 & 0.212 & 306 & 0.60 & 0.54 & 1.34\\
			\bottomrule
		\end{tabular}
		\label{S-tab:disk-shapes}
	\end{table*}

	The same conclusion follows from imposing the disk only in a window of
	energies (\cref{S-tab:disk-windows}).  The bound in units of \(R_S\) rises from
	\(1.2\) for \(4\le\sigma<12\) to \(2.3\) above \(\sigma\simeq40\), where it
	levels off.

	\begin{table}[t]
		\centering
		\small\setlength{\tabcolsep}{3pt}
		\caption{The bound when the disk \(b/R_S<B_*/R_S\) with \(\mathcal A_0\ge0.5\) is imposed only in a window of
			energies (\(\GN=2\pi^2\), \(X=20\), \(\epsilon=0.5\%\)): radius in units of \(R_S\), the same radius in
			units of \(M_{\rm EFT}^{-1}\) at the geometric mean energy of the window, and the number of bins in the
			disk.}
		
		\begin{tabular}{cccc}
			\toprule
			window & \(B_*/R_S\) & \(B_*M_{\rm EFT}\) & bins\\
			\midrule
			\(4\le\sigma<12\) & 1.21 & 3.5 & 199\\
			\(12\le\sigma<36\) & 1.48 & 5.2 & 182\\
			\(36\le\sigma<110\) & 2.26 & 9.5 & 195\\
			\(110\le\sigma<330\) & 2.34 & 11.8 & 162\\
			\bottomrule
		\end{tabular}
		\label{S-tab:disk-windows}
	\end{table}

	\subsection{The loss in \(Y\) for a disk of the classical radius}
	\label{S-sec:supp-floor}

	The comparison with the classical trapped surface in
	Sec.~\ref{sec:ceiling-floor} of the main text uses the same linear
	program, written relative to the stored optimum at \(X=20\), with the
	condition imposed on the total density and the residual bins inside
	\(10R_S\) free.  Freeing the bins out to \(20R_S\), or every bin of the
	grid, changes the loss in the fourth digit (\(0.8588\%\), \(0.8587\%\)
	and \(0.8587\%\) at \(\GN=2\pi^2\), \(\sigma_{\min}=36\)), and the same holds about the matched references of
	\cref{S-sec:supp-matched}, where frozen and free losses differ by at most \(0.1\%\) at the lowest and
	highest energies.

	An exclusion found with the exterior frozen can weaken when the exterior is freed, so the exclusions that
	matter were certified directly.  For any multipliers \(\mu_\alpha\) on the 80 sampled rows with
	\(\sum_\alpha\mu_\alpha\lambda_\alpha^2=1\), every density on the grid that obeys the rows satisfies
	\begin{equation}
		Y-Y_{\max}\le\sum_i\max_{\rho_i\in{\rm box}_i}\,(\mu^{T}A)_i\,(\rho_i-\rho_i^{\rm ref}),
		\label{S-eq:dual-certificate}
	\end{equation}
	where the sum runs over all bins and the boxes include the disk condition.  We take \(\mu\) from the
	program with every bin free (interior-point method), form \(\mu^{T}A\) with the unscaled rows in
	double-double arithmetic, and add a rounding budget, which is below \(10^{-9}\) of the loss in every case.
	\Cref{S-tab:dual-certificate} lists the result.  The certified bound equals the loss found by the solver
	to the digits shown, so the half-percent exclusions at \(E_{\min}R_S=29.8\) (\(4\pi^2\)) and \(34.7\)
	(\(2\pi^2\)) hold on the whole grid and do not rest on the convergence of the solver.  The multipliers
	of the program with the exterior frozen certify nothing outside the free bins, as expected.  The table
	also corrects the statement above for one case.  At \(\GN=2\pi^2\) the frozen and free losses agree
	(to \(0.02\%\)), but at \(\GN=4\pi^2\), \(X=20\), where the eikonal contribution begins at \(3R_S\), freeing every bin lowers
	the loss by \(2.9\)--\(11.4\%\).  The frozen-exterior values at \(\GN=4\pi^2\) are therefore
	not used: for this coupling \cref{S-tab:price-floor}, the losses and crossings quoted in the main text and the points of Fig.~\ref{fig:ceiling-floor}(b) of the main text are computed with every bin free.

	\begin{table}[t]
		\centering
		\small\setlength{\tabcolsep}{3pt}
		\caption{Loss in \(Y\), in percent of \(Y_{\max}\) at \(X=20\), of the disk \(b<1.47R_S\) with
			\(\mathcal A_0\ge0.9\) at \(\sigma\ge\sigma_{\min}\): with the bins outside \(10R_S\) frozen, with
			every bin free, and the lower bound certified by \cref{S-eq:dual-certificate}.}
		
		\begin{tabular}{cccccc}
			\toprule
			\(\GN\) & \(\sigma_{\min}\) & \(E_{\min}R_S\) & frozen & free & certified\\
			\midrule
			\(4\pi^2\) & \(7\) & 9.8 & 2.031 & 1.822 & 1.822\\
			\(4\pi^2\) & \(12\) & 13.9 & 1.330 & 1.260 & 1.260\\
			\(4\pi^2\) & \(20\) & 19.6 & 0.902 & 0.864 & 0.864\\
			\(4\pi^2\) & \(36\) & 29.8 & 0.549 & 0.530 & 0.530\\
			\(4\pi^2\) & \(63\) & 43.8 & 0.335 & 0.325 & 0.325\\
			\(4\pi^2\) & \(110\) & 64.7 & 0.197 & 0.191 & 0.191\\
			\(4\pi^2\) & \(200\) & 91.0 & 0.123 & 0.120 & 0.120\\
			\(2\pi^2\) & \(7\) & 7.8 & 2.835 & 2.835 & 2.835\\
			\(2\pi^2\) & \(63\) & 34.7 & 0.542 & 0.542 & 0.542\\
			\bottomrule
		\end{tabular}
		\label{S-tab:dual-certificate}
	\end{table}
  At \(\GN=4\pi^2\) the program reproduces
	the radii of Fig.~\ref{fig:inverse-absorptive-radius} of the main text:
	\(0.71\) and \(0.85R_S\) at all energies and \(1.01\) and \(1.34R_S\) for
	\(10\le ER_S\le44\), against \(0.71\), \(0.86\), \(1.02\) and \(1.35\); with every bin free it gives \(0.71\), \(0.86\),
	\(1.02\) and \(1.35\).
	\Cref{S-tab:price-floor} lists the losses.  The sampled sum rules are
	satisfied to \(10^{-6}\) in every solution and the residual at unsampled
	\(\lambda\) changes by less than \(10^{-9}\).  At \(\GN=4\pi^2\) the
	programs of \cref{S-tab:price-floor} were solved with every bin free by the interior-point method (with
	the exterior frozen the simplex method fails for \(\sigma_{\min}\le4\)).  At \(\sigma_{\min}=1\) the
	losses are \(6.3\%\) and \(2.7\%\), the values quoted in Sec.~\ref{sec:absorptive-density-pilot} of
	the main text for the disk imposed at all energies.  For the Myers--Perry radius
	\(1.09R_S\) the product \(\epsilon\,E_{\min}R_S\) at \(\mathcal A_0\ge0.9\)
	stays between \(0.082\) and \(0.092\) at \(\GN=4\pi^2\) and between \(0.09\) and
	\(0.11\) at \(\GN=2\pi^2\) for \(\sigma_{\min}\ge7\).

	\begin{table*}[t]
		\centering
		\small\setlength{\tabcolsep}{3pt}
		\caption{Loss \(\epsilon\) in \(Y\), in percent of \(Y_{\max}\), of a disk \(b<1.47R_S\) with
			\(\mathcal A_0\ge0.9\) (\(\epsilon_{0.9}\)) or \(0.5\) (\(\epsilon_{0.5}\)) at all energies
			\(\sigma_{\min}\le\sigma\le600\), at \(X=20\) on the \((600,240,80)\) grid.  \(E_{\min}R_S\) is the value
			at the lowest energy of the disk.  At \(\GN=4\pi^2\) every bin of the grid is free; at the other two
			couplings the bins outside \(10R_S\) are frozen, which changes nothing there (see text).}
		
		\begin{tabular}{c ccc ccc ccc}
			\toprule
			 & \multicolumn{3}{c}{\(\GN=4\pi^2\)} & \multicolumn{3}{c}{\(\GN=2\pi^2\)} & \multicolumn{3}{c}{\(\GN=\pi^2\)}\\
			\(\sigma_{\min}\) & \(E_{\min}R_S\) & \(\epsilon_{0.9}\) & \(\epsilon_{0.5}\) & \(E_{\min}R_S\) & \(\epsilon_{0.9}\) & \(\epsilon_{0.5}\) & \(E_{\min}R_S\) & \(\epsilon_{0.9}\) & \(\epsilon_{0.5}\)\\
			\midrule
			\(1\) & 2.7 & 6.30 & 2.68 & 2.1 & 13.43 & 5.73 & 1.7 & 41.28 & 17.63\\
			\(2\) & 4.2 & 4.06 & 1.73 & 3.4 & 5.58 & 2.38 & 2.7 & 21.28 & 9.10\\
			\(4\) & 6.7 & 2.61 & 1.11 & 5.3 & 4.00 & 1.71 & 4.2 & 14.30 & 6.11\\
			\(7\) & 9.8 & 1.82 & 0.78 & 7.8 & 2.84 & 1.21 & 6.2 & 9.90 & 4.23\\
			\(12\) & 13.9 & 1.26 & 0.54 & 11.1 & 1.96 & 0.84 & 8.8 & 7.22 & 3.09\\
			\(20\) & 19.6 & 0.86 & 0.37 & 15.6 & 1.38 & 0.59 & 12.4 & 5.25 & 2.25\\
			\(36\) & 29.8 & 0.53 & 0.23 & 23.7 & 0.86 & 0.37 & 18.8 & 3.36 & 1.44\\
			\(63\) & 43.8 & 0.32 & 0.14 & 34.7 & 0.54 & 0.23 & 27.6 & 2.15 & 0.92\\
			\(110\) & 64.7 & 0.19 & 0.08 & 51.4 & 0.33 & 0.14 & 40.8 & 1.30 & 0.56\\
			\(200\) & 91.0 & 0.12 & 0.05 & 72.2 & 0.21 & 0.09 & 57.3 & 0.83 & 0.36\\
			\bottomrule
		\end{tabular}
		\label{S-tab:price-floor}
	\end{table*}

	\subsection{Leaf geometry and matched positions}
	\label{S-sec:supp-matched}

	Percentages of \(Y_{\max}\) at \(X=20\) do not agree between the three couplings of
	\cref{S-tab:price-floor}: at equal \(E_{\min}R_S\) the entry at \(\GN=\pi^2\) is three times the other two.
	The absolute loss \(\Delta Y=\epsilon|Y_{\max}|\) does not show this.  It falls by \(1.8\) from
	\(4\pi^2\) to \(2\pi^2\) and by \(1.65\) from \(2\pi^2\) to \(\pi^2\), uniformly in energy, while
	\(|Y_{\max}|\) falls by \(2.1\) and \(4.9\).  The reason is that \(X=20\) is a different place on each
	leaf.  To locate it we solved, with every residual bin free and \(X\) added as a variable, for the largest
	and smallest \(X\) (the tips \(X_{\rm R}\), \(X_{\rm L}\)) and for the upper and lower boundary at the
	matched values \(X_f\) of Eq.~\eqref{eq:leaf-fraction} of the main text (\cref{S-tab:leaf-geometry}).
	These are 21 programs of about \(7\times10^4\) variables, solved with the interior-point method of HiGHS in
	11 to 14 minutes each; the sampled rows hold to machine precision and the residual at unsampled \(\lambda\)
	moves by at most \(2\%\) of the scale of the equation at \(\GN=4\pi^2\) and by less than \(1\%\) at the
	other two couplings.  The upper boundary passes through \(Y_{\max}=0\) between the positions listed, so a
	loss quoted as a fraction of \(Y_{\max}\) has no meaning across positions.

	\begin{table*}[t]
		\centering
		\small\setlength{\tabcolsep}{3pt}
		\caption{Leaf geometry on the \((600,240,80)\) grid: the two tips, the leaf fraction of the point
			\(X=20\), and the upper and lower boundary at the matched fractions \(f=0.43\) and \(0.28\).
			The reference at \(\GN=4\pi^2\) has its eikonal contribution from \(b=3R_S\), the other two from \(11R_S\).}
		
		\begin{tabular}{c ccc ccc ccc}
			\toprule
			& & & & \multicolumn{3}{c}{\(f=0.43\)} & \multicolumn{3}{c}{\(f=0.28\)}\\
			\(\GN\) & \(X_{\rm L}\) & \(X_{\rm R}\) & \(f(X{=}20)\) & \(X_f\) & \(Y_{\max}\) & \(Y_{\min}\) & \(X_f\) & \(Y_{\max}\) & \(Y_{\min}\)\\
			\midrule
			\(4\pi^2\) & -11.70 & 25.58 & 0.85 & 4.25 & -1.8 & -74.3 & -1.42 & 37.0 & -28.4\\
			\(2\pi^2\) & -15.11 & 66.93 & 0.43 & 20.00 & -62.6 & -243.4 & 7.52 & 26.7 & -136.0\\
			\(\pi^2\) & -14.92 & 111.70 & 0.28 & 39.27 & -139.4 & -392.4 & 20.00 & -12.8 & -243.6\\
			\bottomrule
		\end{tabular}
		\label{S-tab:leaf-geometry}
	\end{table*}

	The loss required by the disk \(b<1.47R_S\) was then recomputed about the upper-boundary solutions at the matched
	positions (\cref{S-tab:price-matched}).  At \(f=0.43\) a single power law,
	\(\Delta Y=0.58\,(20/E_{\min}R_S)^{1.10}\), describes the three couplings for \(E_{\min}R_S\ge8\) with a
	scatter of \(-16\%\) to \(+14\%\) (rms \(9\%\)); the separate fits give \(0.60\), \(0.63\) and
	\(0.52\) at \(E_{\min}R_S=20\) with exponents \(1.08\), \(1.20\) and \(1.09\).  At \(f=0.28\) the
	pooled fit is \(0.46\,(20/E_{\min}R_S)^{0.97}\) with a scatter of \(\pm40\%\) (rms \(19\%\)); the
	curve at \(\GN=4\pi^2\) is flatter there (exponent \(0.85\)).  The ratio between the two thresholds is \(2.33\)--\(2.34\)
	throughout, equal to the ratio of the amounts \(1-\sqrt{1-\eta}\) by which they tighten the density box.  At \(f=0.85\), the position of the \(4\pi^2\) point and close to the positive tip, 17 of the 56
	programs failed and that comparison was not used.  Freeing more bins about a matched reference changes the loss in the fourth digit, as at \(X=20\) (\(0.5866\), \(0.5863\), \(0.5863\) for bins inside \(10R_S\), inside \(20R_S\) and all bins, at \(\GN=\pi^2\), \(\sigma_{\min}=36\)).

	\begin{table*}[t]
		\centering
		\small\setlength{\tabcolsep}{3pt}
		\caption{Loss \(\Delta Y\) required by a disk \(b<1.47R_S\) with \(\mathcal A_0\ge0.9\) or \(0.5\) at all
			energies \(\sigma_{\min}\le\sigma\le600\), at the matched leaf fraction \(f=0.43\)
			(\(X=4.25\), \(20\) and \(39.27\)).}
		
		\begin{tabular}{c ccc ccc ccc}
			\toprule
			 & \multicolumn{3}{c}{\(\GN=4\pi^2\)} & \multicolumn{3}{c}{\(\GN=2\pi^2\)} & \multicolumn{3}{c}{\(\GN=\pi^2\)}\\
			\(\sigma_{\min}\) & \(E_{\min}R_S\) & \(\Delta Y_{0.9}\) & \(\Delta Y_{0.5}\) & \(E_{\min}R_S\) & \(\Delta Y_{0.9}\) & \(\Delta Y_{0.5}\) & \(E_{\min}R_S\) & \(\Delta Y_{0.9}\) & \(\Delta Y_{0.5}\)\\
			\midrule
			\(2\) & 4.2 & 3.334 & 1.418 & 3.4 & 3.489 & 1.490 & 2.7 & 3.281 & 1.403\\
			\(4\) & 6.7 & 2.015 & 0.858 & 5.3 & 2.506 & 1.070 & 4.2 & 2.276 & 0.974\\
			\(7\) & 9.8 & 1.343 & 0.573 & 7.8 & 1.774 & 0.758 & 6.2 & 1.610 & 0.689\\
			\(12\) & 13.9 & 0.892 & 0.381 & 11.1 & 1.226 & 0.524 & 8.8 & 1.197 & 0.513\\
			\(20\) & 19.6 & 0.595 & 0.254 & 15.6 & 0.865 & 0.370 & 12.4 & 0.890 & 0.381\\
			\(36\) & 29.8 & 0.365 & 0.156 & 23.7 & 0.537 & 0.230 & 18.8 & 0.587 & 0.251\\
			\(63\) & 43.8 & 0.244 & 0.104 & 34.7 & 0.339 & 0.145 & 27.6 & 0.386 & 0.165\\
			\(110\) & 64.7 & 0.169 & 0.072 & 51.4 & 0.204 & 0.087 & 40.8 & 0.240 & 0.103\\
			\(200\) & 91.0 & 0.123 & 0.053 & 72.2 & 0.131 & 0.056 & 57.3 & 0.155 & 0.066\\
			\bottomrule
		\end{tabular}
		\label{S-tab:price-matched}
	\end{table*}

\section{Inward continuation of the eikonal contribution}
\label{S-sec:supp-inward}

	\subsection{The inward-continued eikonal contribution}

	The inward-continuation control of Sec.~\ref{sec:oscillation} of the main
	text moves the prescribed eikonal contribution beyond the conservative
	large-impact-parameter trust region.  We use
	\(b/R_S\ge0.9,\ 0.75,\ 0.5,\ 0.25\), leave a residual variable in every spectral bin,
	and impose the cap on the total density,
	\begin{equation}
		\rho_{\rm tot}^{\rm phys}
		=\rho_{\rm eik}^{\rm phys}+\rho_{\rm res}^{\rm phys},
		\qquad 0\leq\rho_{\rm tot}^{\rm phys}\leq2 .
		\label{S-eq:extended-eikonal-total-cap}
	\end{equation}
	This is a model deformation, not a claim that the pointwise Einstein eikonal
	formula is controlled at small \(b/R_S\).  It lets us ask whether the low-impact component disappears, moves, or reorganizes once the
	prescribed eikonal contribution competes with it directly.

	A  calculation including the complete eikonal contribution with the same mask in the source and the cap
	uses the refined residual grid
	\( (N_\sigma,J_{\max},N_\lambda)=(1200,220,100),
	\) with 1600 refined-energy points.\footnote{For \(J\leq20\), the interval \(1\leq\sigma\leq20\) is replaced by 1600 points approximately uniform in \(E=\ME\sqrt{\sigma}\), with quadrature weights recomputed from the cell boundaries; the original harmonic grid is retained elsewhere. This refinement is applied inside the LP itself, not merely during plotting, and prevents the rapidly varying low-spin residual profiles from being undersampled. It therefore makes the observed extrema,  variation of the algebraic density map, and histogram broadening reliable against the coarse-grid aliasing that can hide oscillations.}
	We define \(\xi\equiv(b/R_S)_{\min}\), so the four runs use
	\(\xi=0.9,0.75,0.5,0.25\).  The eikonal contribution is evaluated on a separate
	\(N_\mu=6400\) grid, with exact finite spins through \(J=320\) and 4000-node
	leading and regular high-spin continuum integrals.  We take
	\(\chi_{\max}=1000\).\footnote{At fixed \(J\), lowering \(\xi\) admits higher energies, and since \(\chi \propto\sigma^2\), the phase at the support boundary grows as \(\chi_{\rm edge}\propto \xi^{-3}\). Thus a small \(\chi_{\max}\) would artificially remove the high-phase, rapidly oscillating region newly exposed at small \(\xi\); \(\chi_{\max}=1000\) avoids this truncation for the low-spin sector studied here.}

	All four LPs solve to optimality.  On the 100 \(\lambda\)-values imposed in
	the optimization, the largest absolute \(\Ktwo\) equality residual is
	\(7.97\times10^{-9}\).  The total-density cap is satisfied up to a maximal
	numerical excess of \(4.45\times10^{-16}\).  The corresponding values of
	\(Y_{\max}\) are
	\(
	-33.496835,
	-33.499802,
	-33.499347,
	-33.501440.
	\) The occupied residual-bin count changes only from 7979 to 7977, and the
	support Jaccard relative to the \(b/R_S\ge0.9\) solution remains above
	0.998997.  The relative \(L^1\) distances from the
	\(\xi=0.9\) density are $(0, 0.0762, 0.1742, 0.2752)$ for
	\(\xi=0.9,0.75,0.5,0.25\), respectively.  The successive changes are
	\(7.62\%\), \(10.61\%\), and \(12.23\%\).

	\begin{figure*}[t]
		\centering
		\includegraphics[width=0.83\textwidth]
		{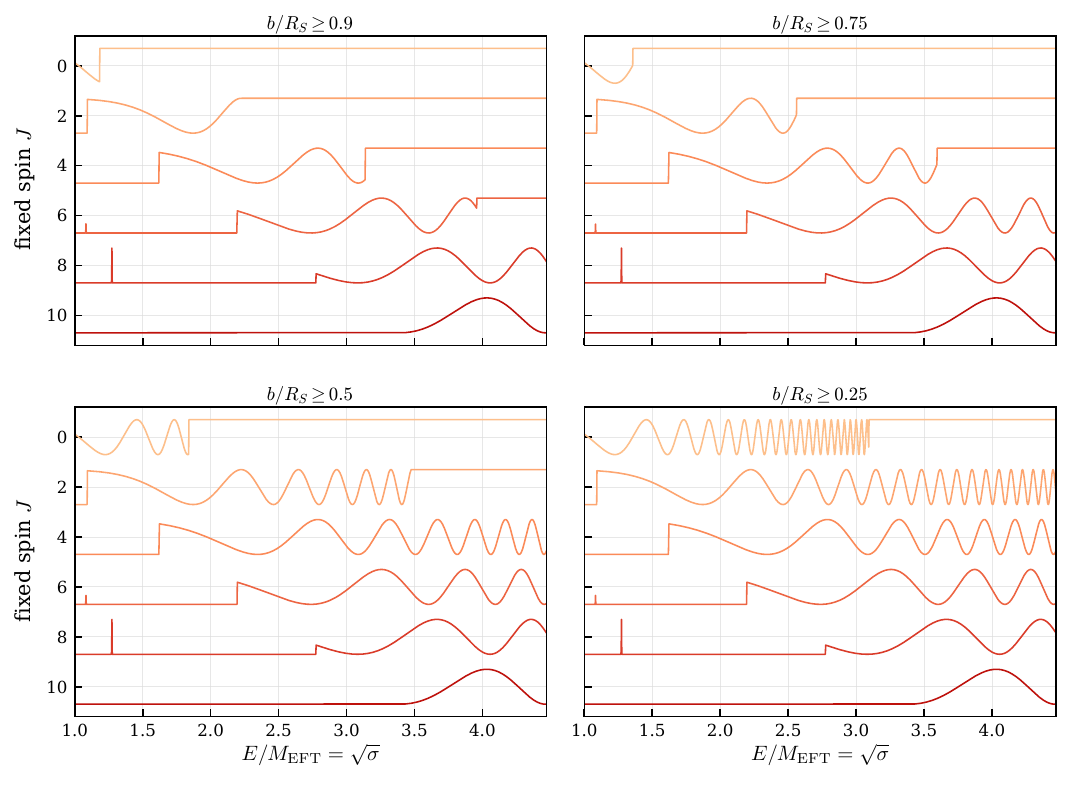}
		\caption{Fixed-spin residual-density profiles for the four
			 values of \(\xi=(b/R_S)_{\min}\) at \(X=6.5\). Each curve is
			vertically offset by its spin \(J=0,2,\ldots,10\), with displacement
			proportional to \(\rho_{\rm res}^{\rm phys}-1\).   The modulation grows strongly as the $b/R_S$ eikonal contribution is extended inward.  These are LP densities, not reconstructed elastic phases.}
		\label{S-fig:ext-eik-oscillation}
	\end{figure*}

	\Cref{S-fig:ext-eik-oscillation} shows the six lowest spins.  As the eikonal contribution is continued from
	\(b/R_S\ge0.9\) through \(0.75\), \(0.5\) and \(0.25\), the summed total variation of the six
	density profiles rises from \(45.0\) to \(58.0\), \(96.2\) and \(208.6\), and the number of maxima
	on the refined grid from \(21\) to \(26\), \(46\) and \(103\).  We do not attach a physical
	interpretation to this modulation: the residual density is added to a prescribed eikonal density
	and is not the absorptive part of an independently unitary partial wave.

\section{Born partial waves, absorptive density and the eikonal}
\label{S-sec:supp-born}\label{S-app:eikonal-background}
The normalized angular block is \(C_J^{(\nu)}(z)/C_J^{(\nu)}(1)\), with
\(\nu=(D-3)/2\). Its coefficient is
\begin{equation}
n_J^{(D)}=\frac{(4\pi)^{D/2}(D+2J-3)\Gamma(D+J-3)}
{\pi\,\Gamma((D-2)/2)\Gamma(J+1)}.
\end{equation}
The tree graviton pole in Eq.~\eqref{eq:eft-expansion} gives
\(2\pi G_Ns(3+z^2)^2/(1-z^2)\) at \(z=1+2t/s\), in physical variables.
Its partial waves are real. For \(S_J=1+if_J=e^{i\chi_J}\),
\(f_J=\sin\chi_J+i(1-\cos\chi_J)\). The density
\(\rho_J=1-\operatorname{Re}S_J=1-\cos\chi_J\) starts at \(\chi_J^2/2\),
while the real Born wave starts at \(\chi_J\).
The forward tree amplitude \(8\pi G_Ns^2/q^2\), with \(q^2=-t\), gives
the transverse Fourier transform~\cite{S-tHooftGravitonDominance,S-ACV,S-Weinberg}
\begin{equation}
\chi_D(s,b)=\frac{G_Ns\,\Gamma((D-4)/2)}
{\pi^{(D-4)/2}b^{D-4}}.
\end{equation}
For \(D=6\) this is Eq.~\eqref{eq:chi-def} with \(s=\ME^2\sigma\)
and \(b=2(J+3/2)/(\ME\sqrt\sigma)\). The physical transfer in the
coefficient sum rules tends to \(-\lambda\ME^2\) at large \(\sigma\).
Integration over all large energies generates an order-\(G_N\) term
even though each eikonal bin starts at order-\(G_N^2\).

\section{Algebra of the eikonal sum rule}
\label{S-sec:supp-algebra}\label{S-app:algebra}
The scaled SDR derivatives are \(\mathcal C_x=2K_{g_2}\)
and \(\mathcal C_y=-K_{g_3}\). The combination
\(\mathcal K_2=\mathcal C_x-\lambda\mathcal C_y\) cancels angular
derivatives since \(t_\lambda\) depends on \(x,y\) through \(y+\lambda x\).
At \(x=y=0\),
\[
-(\partial_x-\lambda\partial_y)K_0
=\frac{2\sigma+3\lambda}{\sigma^4},\qquad
z_\lambda=\sqrt{\frac{\sigma-3\lambda}{\sigma+\lambda}} .
\]
The extra \(1/\sigma\) from the partial-wave expansion gives
Eq.~\eqref{eq:k2-explicit-kernel}. In this normalization
\(\mathcal K_2=8\pi\GN/\lambda+2g_2\ME^6+\lambda g_3\ME^8\)
to the order used here. A scalar pole with scaled mass \(\widehat m=m/\ME\)
and scaled residue \(\widehat h\) gives
\(\mathcal K_2=\widehat h(2\widehat m^2+3\lambda)/\widehat m^8\).

In the large-spin integral put \(L=J+3/2\), \(\bar b=b\ME=2L/\sqrt\sigma\)
and \(w=\bar b\sqrt\lambda\). The even-spin measure satisfies
\(\tfrac12 n_J^{(6)}dL\to4\pi^2\bar b^3\sigma^2d\bar b\).
The angular block becomes \(2J_1(w)/w\). Changing variables to
\(\sigma=\pi\chi\bar b^2/\GN\) leaves, in \(d\chi\,d\bar b\), the
leading and regular integrands
\[
(1-\cos\chi)\frac{2J_1(w)}{w}
\left\{\frac{\lambda\bar b}{\pi\chi^2},
\ \frac{3\GN\lambda^2}{2\pi^2\bar b\chi^3}\right\}.
\]
These give the \(J_0\) and \(H\) brackets in
Eqs.~\eqref{eq:continuum-eikonal-leading} and
\eqref{eq:continuum-eikonal-regular}, with the numerical identity
\begin{equation}
H(x)=\frac{J_1(x)}{2x}
+\frac12\left[\int_0^x\frac{1-J_0(t)}{t}\,dt-\log\frac{x}{2}-\gamma_E\right].
\label{S-eq:continuum-eikonal-h-stable}
\end{equation}
Differentiating gives \(-J_1(x)/x^2\), and \(H(\infty)=0\) fixes the
constant. At \(\lambda\to0\) the regular term vanishes and the leading
term tends to the pole fraction \(\alpha_E\) of
Eq.~\eqref{eq:alpha-window-main}. For the split coefficient sum rules, the leading large-\(\sigma\)
pieces show how the pole enters each projector separately. Let
\(Q(\sigma,\lambda)=\mathcal A^{(s)}_\rho(\sigma,-\lambda)\)
and take \(\Sigma\) high enough to use the large-spin approximation.
The pole-sensitive parts are
\begin{align}
\mathcal C_x^{>\Sigma} &=
\frac2\pi\int_\Sigma^\infty\frac{d\sigma}{\sigma^3}
       (Q-\lambda\partial_\lambda Q),\\
\mathcal C_y^{>\Sigma} &=
-\frac2\pi\int_\Sigma^\infty\frac{d\sigma}{\sigma^3}
       \partial_\lambda Q.
\end{align}
The second term in \(\mathcal C_x^{>\Sigma}\) cancels against
\(-\lambda\mathcal C_y^{>\Sigma}\). With the eikonal density,
the integral of \(Q\) in the first equation generates
\(8\pi\GN\alpha_E/\lambda\), and \(-\partial_\lambda\) then gives
\(8\pi\GN\alpha_E/\lambda^2\) in \(\mathcal C_y\).
The two terms in \(\mathcal C_x\) each give
\(8\pi\GN\alpha_E/\lambda\). This uses the large-spin Bessel kernel
before taking \(\lambda\to0\).

The exact-spin quadrature through
\(J_*\) and the continuum integral above \(J_*\) enter the sum rule once.

\section{Reduced costs and cap marginals}
\label{S-sec:supp-reduced}\label{S-app:reduced-cost}
For \(A\rphys-\lambda^2Y=r_X\), \(0\le\rphysi\le u_i=2-\reiki\),
the bounded dual maximizes
\[
r_X\cdot h-\sum_i u_i[A_i\cdot h]_+,\qquad
-\sum_\alpha\lambda_\alpha^2h_\alpha=\pm1 .
\]
Here \([z]_+=\max(z,0)\). The signs give \(Y_{\min}\) and
\(-Y_{\max}\), respectively. In the solver variable
\(\widetilde\rho_i=\bar s_i\rphysi\), \(\bar s_i=s_i/(8\pi\GN)\)
for column scale \(s_i\) on the pole-normalized density.
At fixed \(X\) in the minimization of \(-Y\), let
\(m_i^{\rm low},m_i^{\rm up}\) be the derivatives of the optimum
with respect to the lower and upper bounds of \(\widetilde\rho_i\).
The turn-on cost for an empty bin and cap value for a saturated bin are
\[
c_i=[\bar s_i m_i^{\rm low}]_+,\qquad
v_i=[-\bar s_i m_i^{\rm up}]_+ .
\]
The first is the loss in \(Y_{\max}\) from forcing density into the bin.
The second is the gain from raising its cap.

\section{EFT matching below \texorpdfstring{\(\ME\)}{M EFT}}
\label{S-sec:supp-eft}\label{S-app:eft-matching}
Let \(m=M_{\min}/\ME\in(0,1)\). The fixed input contains the cut
below \(\sigma=m^2\) and the known low-spin discontinuity and large-spin
tail for \(m^2\le\sigma<1\). Denote its contribution to the sum rule by
\(B_{\rm calc}^{\rm low}\). The density \(\rho_{\rm low}\) runs only
over unresolved spins in \(m^2\le\sigma<1\). The unknown UV density
\(\rho_{\rm UV}\) begins at \(\sigma=1\) and shares its cap with the
prescribed eikonal density. The amplitude contains the pole and local terms, the fixed low-energy cut,
the high-energy eikonal term, and the dispersive integrals of these two
unresolved densities over their respective energy ranges. This counts each
contribution to the cut once. The matching equation is
\begin{equation}
\begin{split}
\frac{\lambda\mathcal K_2[\rho_{\rm low}+\rho_{\rm UV}]}{8\pi\GN}
+F_{\rm eik}(\lambda)+B_{\rm calc}^{\rm low}(\lambda)
&=1+2\lambda X+\lambda^2Y .
\end{split}
\end{equation}
A positive unresolved low-energy term can be bounded by
\(0\le\rho_{J,\rm low}^{\rm phys}\le\gamma_J(\sigma)\) using the
first omitted absorptive term. A signed truncation error needs a
two-sided envelope around the calculated density. Above the threshold,
\(0\le\rho_{J,\rm eik}^{\rm phys}+\rho_{J,\rm UV}^{\rm phys}\le2\).

\clearpage
\section{Reproducibility}
\label{S-sec:supp-repro}\label{S-app:reproducibility}
The code, grids and checks are indexed in the arXiv ancillary file
\mbox{\nolinkurl{20261124_reproduction_readme_v1.md}}.
For
Fig.~\ref{fig:absorptive-core-selection}, the Fig.~9 package in S5\,A
contains the selected and reference arrays and the check of the core and
outer crossings. The package in S5\,B contains the density witness and
27 interval certificates for the whole-leaf extension of
Fig.~\ref{fig:inverse-absorptive-radius}. Its audit applies the saved
multipliers to the original, unscaled constraints, checks the feasible
points defining the chords and verifies coverage of the \(X\) interval.

The strong-coupling eikonal integration uses 6400 energy nodes through
\(J_*=320\) and 4000 nodes in the continuum tail. In radius searches we reduce the 80 sampled
equalities to rank 45 at a singular-value cutoff of \(10^{-12}\), check
monotonicity in radius, threshold and loss, and report unresolved trials
as one-sided feasible bounds.

\end{document}